\documentclass[11pt]{article}
\usepackage{epsfig} 
\usepackage{graphics}
\newcommand{\sect}[1]{ \section{#1} \setcounter{equation}{0} }

\newcommand{\third}{\mbox{\small{$\frac{1}{3}$}}} 
 
\newcommand{\pitwo}{\mbox{\small{$\frac{\pi}{2}$}}}

\newcommand{\pisix}{\mbox{\small{$\frac{\pi}{6}$}}}
\newcommand{\Nc}{N_{\!c}}
\newcommand{\Nf}{N_{\!f}}

\newcommand{\Nda}{{N^d_{\!A}}}
\newcommand{\Noda}{{N^o_{\!A}}}

\newcommand{\MSbar}{\overline{\mbox{MS}}}
\newcommand{\MSbars}{\overline{\mbox{\footnotesize{MS}}}}
\newcommand{\RI}{\mbox{RI${}^\prime$}}
\newcommand{\RIs}{\mbox{\footnotesize{RI${}^\prime$}}}
\newcommand{\RIc}{\widetilde{\mbox{RIc${}^\prime$}}}
\newcommand{\RIcs}{\widetilde{\mbox{\footnotesize{RIc${}^\prime$}}}}
\newcommand{\mMOM}{\mbox{mMOM}}
\newcommand{\mMOMs}{\mbox{\footnotesize{mMOM}}}

\newcommand{\lins}{\mbox{\footnotesize{lin}}}

\newcommand{\CFs}{\mbox{\footnotesize{CF}}}

\newcommand{\MAGs}{\mbox{\footnotesize{MAG}}}
\newcommand{\MOMc}{\mbox{MOMc}}
\newcommand{\MOMcs}{\mbox{\footnotesize{MOMc}}}
\newcommand{\MOMg}{\mbox{MOMg}}
\newcommand{\MOMgs}{\mbox{\footnotesize{MOMg}}}
\newcommand{\MOMq}{\mbox{MOMq}}

\newcommand{\MOMis}{\mbox{\footnotesize{MOMi}}}

\begin{document}

\title{The Crewther relation, schemes, gauges and fixed points}

\author{J.A. Gracey \& R.H. Mason, \\ Theoretical Physics Division, \\ 
Department of Mathematical Sciences, \\ University of Liverpool, \\ P.O. Box 
147, \\ Liverpool, \\ L69 3BX, \\ United Kingdom.} 

\date{}

\maketitle 

\vspace{5cm} 
\noindent 
{\bf Abstract.} We investigate the Crewther relation at high loop order in a 
variety of renormalization schemes and gauges. By examining the properties of 
the relation in schemes other than modified minimal subtraction ($\MSbar$) at 
the fixed points of Quantum Chromodynamics we propose a generalization of the 
Crewther relation that extends the $\MSbar$ construction of Broadhurst and 
Kataev. A derivation based on the properties of the renormalization group 
equation is provided for the generalization which is tested in various 
scenarios.

\vspace{-16.0cm}
\hspace{13.9cm}
{\bf LTH 1344}

\newpage 

\sect{Introduction.}

In \cite{1} Crewther made an interesting and remarkable connection between the
Adler $D$-function and the Bjorken sum rule in non-abelian gauge theories. 
Using the then available explicit perturbative expressions for those seemingly 
unrelated quantities Crewther showed that their product was a constant and 
there was no $O(a)$ correction where $a$ is related to the strong coupling
constant. When the higher order corrections subsequently became available in 
the modified minimal subtraction ($\MSbar$) scheme \cite{2,3,4} it was apparent
that Crewther's observation of constancy of the product was not retained at 
these orders. It was demonstrated in \cite{5} however that the non-zero $a$ 
dependence could be written as the product of two functions of $a$. One of the 
functions was the $\beta$-function of Quantum Chromodynamics (QCD). While the 
presence of a non-zero term appeared to undermine the ethos that Crewther's 
observation was an exact result, it was in fact in keeping with a more general 
property which is that of conformal symmetry, \cite{6,7,8}. Indeed this 
non-zero $O(a^2)$ term is referred to as the conformal symmetry breaking term, 
\cite{5}, since the $\beta$-function vanishes when conformal symmetry is 
present. Later the improvement in loop technology meant the $O(a^4)$ terms of 
both the Adler $D$-function and Bjorken sum rules became available in \cite{9}.
It was shown in \cite{9} that the resulting $O(a^4)$ correction to the 
conformal symmetry breaking term could also be accommodated within the product 
of the two loop $\beta$-function and a correction to the other function in the 
breaking term as a function of $a$. Thus the conformal aspect of the Crewther 
relation was reinforced. 

In briefly summarizing the background to the development of Crewther's 
observation it is important to note that the so-called conformal symmetry 
breaking term was always determined in the $\MSbar$ scheme in the first 
instance, \cite{5,9}. More recently there have been several detailed studies of
the Crewther construction in other renormalization schemes in \cite{10,11,12}. 
One of the motivations in carrying out such investigations is to ascertain 
whether or not the conformal symmetry breaking term is always the product of 
the $\beta$-function and another function of $a$ in schemes other than 
$\MSbar$. Amongst the schemes examined in \cite{10,11,12} were the $V$ scheme 
and the minimal momentum ($\mMOM$) scheme. The former is based on the static 
quark potential, \cite{13,14}, while the latter is defined with respect to the 
ghost-gluon vertex of QCD, \cite{15}. In particular it preserves the 
non-renormalization property of that vertex in the Landau gauge, discovered by 
Taylor in \cite{16}, but extended to an arbitrary linear covariant gauge. The 
renormalization group functions of the $\mMOM$ scheme are known to five loops, 
\cite{15,17,18,19}. Although the Crewther relation $\mMOM$ scheme study only 
required four loop information one of its main conclusions was that at leading 
order the factorization of the conformal symmetry breaking term only occurred 
for specific values of the covariant gauge fixing parameter $\alpha$, 
\cite{12}. These were $\alpha$~$=$~$0$, $-$~$1$ and $-$~$3$. It was also
recognized that the special cases of $\alpha$~$=$~$0$ and $-$~$3$ arose in 
other schemes, \cite{12}. At the subsequent order the factorization only 
occurred in the Landau gauge where $\alpha$ vanishes. While disappointing, 
detailed investigations of the structure of where the factorization, \cite{12},
failed did provide significant insight into and key clues as to how to 
reconcile the hope that the Crewther construction could accommodate other 
schemes. Indeed this has to be the case if the Crewther relation is to be 
regarded as a fundamental property of a non-abelian gauge theory. 

One interesting aspect of the emergence of the particular case of 
$\alpha$~$=$~$-$~$3$ is that this value has already been noted as having 
special significance in situations primarily associated with phenomenology or 
infrared dynamics in QCD. For instance, one of the earliest occurrences was in 
\cite{20} where it emerged in a study of the Wilson loop and its 
renormalization properties. This connection was noted again later in the case 
of Wilson operators in deep inelastic scattering in \cite{21}. In a different 
context, \cite{22}, Schwinger-Dyson methods were used to examine models of 
chiral symmetry breaking and this choice of the gauge parameter was shown to be
necessary for a scale invariant solution. In other words when the ans\"{a}tzen 
for the form factors have a power law behaviour then the underlying equations 
can be solved to determine when chiral symmetry breaking occurs. Separately in 
\cite{23,24}, where the worldline formalism was used to construct gauge 
invariant variables to study parton distribution functions, the choice of 
$\alpha$~$=$~$-$~$3$ was required to ensure there were no divergences 
associated with rapidity. The same choice of $\alpha$ was also needed in the 
related area of deep inelastic scattering in \cite{25} where renormalon chains 
of bubble graphs were examined in the large $\Nf$ expansion as that particular
value of $\alpha$ identified an abelian characteristic of the underlying 
renormalization in a similar vein to \cite{20,21}. At leading order the 
approximation was relatively accurate for $\alpha$~$=$~$-$~$3$ but corrections 
to this value were also established to preserve the abelian property, 
\cite{26}. This leading order value also appeared in an exploration of 
Yang-Mills theory at low energy. In particular in \cite{27,28,29} an effective 
Lagrangian was constructed in terms of dimension six gluonic operators that was
a solution to the infrared Schwinger-Dyson equations. Indeed it was argued in 
\cite{27,28,29} that to have a scaling solution for the gluon propagator which 
would lead to a linearly rising static quark potential associated with 
confinement required $\alpha$~$=$~$-$~$3$. Similarly the role this particular 
value plays in the behaviour of the effective coupling constant derived in 
quark-quark scattering was discussed in \cite{30}. Specifically at 
$\alpha$~$=$~$-$~$3$ the effective coupling decreases as a function of the 
exchanged momentum consistent with asymptotic freedom. A more phenomenological
origin for this gauge parameter choice was noted in \cite{31}. Moreover the
appearance of $\alpha$~$=$~$-$~$3$ was not restricted to strong dynamics. It 
was shown in \cite{32} that the $Z$ boson is multiplicatively renormalizable
for this gauge parameter choice in the electroweak sector of the Standard 
Model. In support of the infrared connection the critical properties of the QCD
renormalization group equations in \cite{33} indicated that there was a fixed 
point similar to that of Banks and Zaks, \cite{34}, but with a value of 
$\alpha$ close to $-$~$3$. That it was not exactly a negative integer was 
primarily because that critical point analysis was carried out at four loops 
which was an order much higher than the previous work on this special value. 
This would suggest that while the infrared dynamics observations are pointing 
to the same underlying property quantum corrections will necessarily have to be
taken into account too. Our Crewther study will also lend weight to that point 
of view. Indeed the connection provided in \cite{34} also directs our attention
to perhaps the most significant clue to accommodate schemes for the Crewther 
relation. This is the fact that in examining the critical properties of the 
renormalization group functions of QCD treated in the $(a,\alpha)$ plane there 
is a fixed point in the neighbourhood of $\alpha$~$=$~$-$~$3$ in the conformal 
window. While it is not precisely this value it is however suggestive of a 
connection to the Crewther relation with the discrepancy somehow being 
accounted for by higher order corrections lurking within a reconciliation of 
the factorization issue with the conformal symmetry property. 

It is therefore the aim of this article to extend the Crewther relation in some
way that it is inclusive of schemes beyond the $\MSbar$ one yet at the same
time obeying the conformal aspects on top of respecting the renormalization
group. In our investigation in order to gain as wide a picture as possible we 
will consider not only the mMOM scheme but a similar one used in lattice gauge 
theories, termed the modified Regularization Invariant ($\RI$) scheme 
\cite{35,36}, as well as the kinematic momentum subtraction (MOM) schemes 
developed by Celmaster and Gonsalves, \cite{37,38}. In addition we will examine
the relation in two other gauges which are nonlinear covariant gauge fixings.
The reason for this is that while the Lorenz suite of covariant gauges are 
canonically used in computations, with $\alpha$ arbitrary or zero, gauge 
independent results should be blind to the choice of the particular gauge 
fixing functional. Our aim is to construct a generalization of the conformal 
symmetry breaking term that is consistent with the properties of the 
renormalization group equation and preserves the conformal symmetry property as
well as putting the origin of the special case of $\alpha$~$\approx$~$-$~$3$ of
the linear covariant gauge fixing on a solid footing. To achieve all these aims
required both numerical examination of the perturbative expansions of the 
product of the Adler $D$-function and Bjorken sum rule as well as the analytic 
expressions. These will be studied at the highest available loop order in each 
of the different gauges and schemes. Indeed the way the article is structured 
it will follow the same development ethos of \cite{5} where numerical and 
analytic evidence was presented that led to and justified the factorization 
property of the original conformal symmetry breaking term of \cite{5}. By 
assemblying this evidence a clear picture emerges which allows us to formulate 
the generalization before providing an analytic justification of our 
observations. This is then robustly and satisfactorily tested across the 
various schemes and gauges. It would be remiss of us at this point not to 
mention the related work in this area provided in \cite{39}. There the 
principle of maximal conformality was used to provide a way of setting the 
renormalization scale for phenomenological applications from the Crewther 
relation. 

The article is organized as follows. We recall the core observations 
surrounding the Crewther relation in Section $2$ as well as details of the
factorization property in the $\mMOM$ scheme. The focus of Section $3$ is to
present numerical evidence that the conformal symmetry breaking term vanishes
at a large number of fixed points of the QCD renomalization group equations for
nonzero $\alpha$. Based on those observations the consequences for the Crewther
relation in the $\mMOM$ scheme are studied analytically in Section $4$ where a 
generalization of \cite{5} is introduced. The consistency of this
generalization is then tested in other schemes and gauges in Section $5$ as
well as several other schemes in the linear covariant gauge in Section $6$. 
The formalization of the generalized Crewther relation within the 
renormalization group is presented in Section $7$. Section $8$ by contrast is
devoted to explaining the relation of the gauge parameter critical point values
to the special cases of the gauge parameter that were indicated in the $\mMOM$ 
scheme in \cite{12}. An overview of the field theory origin of certain
properties of the Crewther relation that we observe is provided in Section $9$.
Conclusions are provided in Section $10$ ahead of two appendices. Expressions 
illustrating the structure of key functions in the Crewther relation in a MOM 
scheme for a general Lie group are recorded in Appendix A. Finally Appendix B 
presents the renormalization group functions for a new renormalization scheme 
that is introduced in Section $6$.

\sect{Background.}

It is instructive to first review the construction given in \cite{5} which 
revealed Crewther's observation that the product of the Bjorken sum rule and 
Adler $D$-function is not a constant. Instead there is a discrepancy which
depends on the $\beta$-function in the $\MSbar$ scheme. We recall the sum rule 
and $D$-function to $O(a^3)$ are, \cite{2,3,4},
\begin{eqnarray}
C_{{\mbox{\footnotesize{Bjr}}}}^{\MSbars}(a) &=&
1 - 3 C_F a
+ \left[
\frac{21}{2} C_F^2
- 23 C_F C_A
+ 8 \Nf T_F C_F
\right] a^2
\nonumber \\
&& + \left[
\frac{440}{3} \zeta_5 C_F C_A^2
+ \frac{1241}{9} C_F^2 C_A
+ \frac{7070}{27} \Nf T_F C_F C_A
+ 48 \zeta_3 \Nf T_F C_F C_A
\right. \nonumber \\
&& \left. ~~~
- \frac{10874}{27} C_F C_A^2
- \frac{920}{27} \Nf^2 T_F^2 C_F
- \frac{176}{3} \zeta_3 C_F^2 C_A
- \frac{160}{3} \zeta_5 \Nf T_F C_F C_A
\right. \nonumber \\
&& \left. ~~~
- \frac{133}{9} \Nf T_F C_F^2
- \frac{80}{3} \zeta_3 \Nf T_F C_F^2
- \frac{3}{2} C_F^3
\right] a^3 ~+~ O(a^4)
\label{bjms}
\end{eqnarray}
and 
\begin{eqnarray}
C_{{\mbox{\footnotesize{Adl}}}}^{\MSbars}(a) &=&
\left[
1 + 3 C_F a
+ \left[
\frac{123}{2} C_F C_A
- \frac{3}{2} C_F^2
- 44 \zeta_3 C_F C_A
- 22 \Nf T_F C_F
+ 16 \zeta_3 \Nf T_F C_F
\right] a^2
\right. \nonumber \\
&&
\left. + \left[
\frac{160}{3} \zeta_5 \Nf T_F C_F C_A
+ \frac{4832}{27} \Nf^2 T_F^2 C_F
+ \frac{7168}{9} \zeta_3 \Nf T_F C_F C_A
+ \frac{90445}{54} C_F C_A^2
\right. \right. \nonumber \\
&& \left. \left. ~~~
- \frac{31040}{27} \Nf T_F C_F C_A
- \frac{10948}{9} \zeta_3 C_F C_A^2
- \frac{1216}{9} \zeta_3 \Nf^2 T_F^2 C_F
- \frac{440}{3} \zeta_5 C_F C_A^2
\right. \right. \nonumber \\
&& \left. \left. ~~~
- \frac{69}{2} C_F^3
- 572 \zeta_3 C_F^2 C_A
- 320 \zeta_5 \Nf T_F C_F^2
- 127 C_F^2 C_A
- 29 \Nf T_F C_F^2
\right. \right. \nonumber \\
&& \left. \left. ~~~
+ 304 \zeta_3 \Nf T_F C_F^2
+ 880 \zeta_5 C_F^2 C_A
\right] a^3 ~+~ O(a^4)
\right] d_R 
\label{adms}
\end{eqnarray}
in the $\MSbar$ scheme where $C_F$, $C_A$ and $T_F$ are the usual colour group
Casimirs, $\Nf$ is the number of (massless) quark flavours, $d_R$ is the 
dimension of the fundamental representation and $\zeta_n$ is the Riemann zeta 
function. Although the $O(a^4)$ terms of both are available \cite{9}, and 
involve higher rank $4$ colour Casimirs, the $O(a^3)$ expressions are 
sufficient for the moment to illustrate the properties of the discrepancy. We 
note that our focus will be on the flavour non-singlet construction throughout 
the article. For clarity we define the product in the same way as \cite{5} 
through
\begin{equation}
C_{{\mbox{\footnotesize{Bjr}}}}^{\MSbars}(a) 
C_{{\mbox{\footnotesize{Adl}}}}^{\MSbars}(a) ~=~ d_R \left[ 1 ~+~
\Delta_{\mbox{\footnotesize{csb}}}^{\MSbars}(a) \right]
\label{crewms}
\end{equation}
where ${\mbox{csb}}$ on $\Delta_{\mbox{\footnotesize{csb}}}^{\MSbars}(a)$ 
labels the conformal symmetry breaking term, \cite{5}. The scheme that the 
variables are in will be indicated by the superscript label throughout. It is 
clear from the product of (\ref{bjms}) and (\ref{adms}) that 
$\Delta_{\mbox{\footnotesize{csb}}}^{\MSbars}(a)$ is $O(a^2)$. The subsequent 
$O(a^2)$ term is partly comprised of the sum of the two $O(a^2)$ terms of 
(\ref{bjms}) and (\ref{adms}) which results in five terms where the coefficient 
of the $C_F^2$ term is $9$. The remaining $O(a^2)$ part arises from the product
of the one loop terms of each of (\ref{bjms}) and (\ref{adms}) that precisely 
cancels the other $C_F^2$ contribution. In total four terms remain. One pair 
has rational coefficients and the other pair involves $\zeta_3$. While this 
summarizes the underlying algebra the major result of \cite{5} was to observe 
that the remaining terms factorize with one of the factors equivalent to the 
one loop $\beta$-function coefficient of \cite{40,41}. Consequently it was 
proposed that $\Delta_{\mbox{\footnotesize{csb}}}^{\MSbars}(a)$ took the all 
orders form
\begin{equation}
\Delta_{\mbox{\footnotesize{csb}}}^{\MSbars}(a) ~=~ 
\frac{\beta^{\MSbars}(a)}{a} K_a^{\MSbars}(a) ~.
\label{deltms}
\end{equation}
We recall explicit computation produced, \cite{5},
\begin{eqnarray}
K_a^{\MSbars}(a) &=& 
\left[
12 \zeta_3 C_F
- \frac{21}{2} C_F
\right] a
\nonumber \\
&&
+ \left[
\frac{326}{3} \Nf T_F C_F
+ \frac{397}{6} C_F^2
- 240 \zeta_5 C_F^2
+ 136 \zeta_3 C_F^2
+ \frac{884}{3} \zeta_3 C_F C_A
- \frac{629}{2} C_F C_A
\right. \nonumber \\
&& \left. ~~~
- \frac{304}{3} \zeta_3 \Nf T_F C_F
\right] a^2
\nonumber \\
&&
+ \left[
\frac{6496}{9} \zeta_3 \Nf^2 T_F^2 C_F
+ \frac{11900}{3} \zeta_5 C_F C_A^2
- \frac{406043}{36} C_F C_A^2
- \frac{40336}{9} \zeta_3 \Nf T_F C_F C_A
\right. \nonumber \\
&& \left. ~~~
- \frac{24880}{3} \zeta_5 C_F^2 C_A
- \frac{9824}{9} \Nf^2 T_F^2 C_F
- \frac{8000}{3} \zeta_5 \Nf T_F C_F C_A
- \frac{7729}{18} \Nf T_F C_F^2
\right. \nonumber \\
&& \left. ~~~
+ \frac{2471}{12} C_F^3
+ \frac{16570}{3} \zeta_3 C_F^2 C_A
+ \frac{67520}{9} \Nf T_F C_F C_A
+ \frac{72028}{9} \zeta_3 C_F C_A^2
\right. \nonumber \\
&& \left. ~~~
+ \frac{99757}{36} C_F^2 C_A
- 5720 \zeta_5 C_F^3
- 3668 \zeta_3 \Nf T_F C_F^2
- 1232 \zeta_3^2 C_F C_A^2
- 840 \zeta_7 C_F^2 C_A
\right. \nonumber \\
&& \left. ~~~
- 128 \zeta_3^2 \Nf T_F C_F C_A
+ 320 \zeta_5 \Nf^2 T_F^2 C_F
+ 488 \zeta_3 C_F^3
+ 576 \zeta_3^2 \Nf T_F C_F^2
\right. \nonumber \\
&& \left. ~~~
+ 4000 \zeta_5 \Nf T_F C_F^2
+ 5040 \zeta_7 C_F^3
\right] a^3 ~+~ O(a^4) 
\end{eqnarray}
which we include as a reference point for later.

One subsequent question that was studied after the establishment of 
(\ref{crewms}) and (\ref{deltms}) was whether or not this relation took the
same form in renormalization schemes other than $\MSbar$. This was examined at 
length in the $\mMOM$ scheme in several articles, \cite{10,11,12}. The scheme 
is based on preserving the non-renormalization property of the ghost-gluon 
vertex in the Landau gauge, that Taylor observed in \cite{16}, in an arbitrary 
linear covariant gauge. QCD has been renormalized to high loop order in $\mMOM$
in \cite{15,17,18,19}. In order to carry out a Crewther relation study the 
expressions for the Adler $D$-function and Bjorken sum rule had first to be 
established in the $\mMOM$ scheme, \cite{12}. This was achieved by mapping the 
$\MSbar$ coupling constant dependence in 
$C_{{\mbox{\footnotesize{Bjr}}}}^{\MSbars}(a)$ and
$C_{{\mbox{\footnotesize{Adl}}}}^{\MSbars}(a)$ to the $\mMOM$ coupling constant 
using the explicit relations between the variables in both schemes that are
available in \cite{15,17} for instance. As this mapping is gauge parameter 
dependent it therefore produces expressions that are $\alpha$ dependent, 
\cite{12}. In order to facilitate our subsequent analysis we note, \cite{12},
\begin{eqnarray}
C_{{\mbox{\footnotesize{Bjr}}}}^{\mMOMs}(a,\alpha) &=&
1 - 3 C_F a
+ \left[
\frac{3}{2} \alpha C_F C_A
- \frac{107}{12} C_F C_A
+ \frac{3}{4} \alpha^2 C_F C_A
+ \frac{4}{3} \Nf T_F C_F
+ \frac{21}{2} C_F^2
\right] a^2
\nonumber \\
&&
+ \left[
24 \zeta_3 \Nf T_F C_F C_A
- \frac{20585}{144} C_F C_A^2
- \frac{208}{9} \Nf T_F C_F^2
- \frac{176}{3} \zeta_3 C_F^2 C_A
\right. \nonumber \\
&& \left. ~~~
- \frac{160}{3} \zeta_5 \Nf T_F C_F C_A
- \frac{117}{8} \zeta_3 C_F C_A^2
- \frac{40}{3} \Nf^2 T_F^2 C_F
- \frac{21}{2} \alpha C_F^2 C_A
\right. \nonumber \\
&& \left. ~~~
- \frac{21}{4} \alpha^2 C_F^2 C_A
- \frac{9}{8} \alpha^2 \zeta_3 C_F C_A^2
- \frac{4}{3} \alpha \Nf T_F C_F C_A
- \frac{3}{2} C_F^3
\right. \nonumber \\
&& \left. ~~~
- \frac{2}{3} \alpha^2 \Nf T_F C_F C_A
+ \frac{9}{16} \alpha^3 C_F C_A^2
+ \frac{33}{4} \alpha \zeta_3 C_F C_A^2
+ \frac{64}{3} \zeta_3 \Nf T_F C_F^2
\right. \nonumber \\
&& \left. ~~~
+ \frac{215}{48} \alpha C_F C_A^2
+ \frac{349}{48} \alpha^2 C_F C_A^2
+ \frac{440}{3} \zeta_5 C_F C_A^2
+ \frac{832}{9} \Nf T_F C_F C_A
\right. \nonumber \\
&& \left. ~~~
+ \frac{1415}{36} C_F^2 C_A
\right] a^3 ~+~ O(a^4)
\label{mmombjr}
\end{eqnarray}
and
\begin{eqnarray}
C_{{\mbox{\footnotesize{Adl}}}}^{\mMOMs}(a,\alpha) &=&
\left[
1 + 3 C_F a
\right. \nonumber \\
&& \left.
+ \left[
\frac{569}{12} C_F C_A
- \frac{46}{3} \Nf T_F C_F
- \frac{3}{2} C_F^2
- \frac{3}{2} \alpha C_F C_A
- \frac{3}{4} \alpha^2 C_F C_A
- 44 \zeta_3 C_F C_A
\right. \right. \nonumber \\
&& \left. \left. ~~~
+ 16 \zeta_3 \Nf T_F C_F
\right] a^2
\right. \nonumber \\
&& \left.
+ \left[
\frac{50575}{48} C_F C_A^2
- \frac{18929}{24} \zeta_3 C_F C_A^2
- \frac{2063}{48} \alpha C_F C_A^2
- \frac{2033}{3} \Nf T_F C_F C_A
\right. \right. \nonumber \\
&& \left. \left. ~~~
- \frac{1355}{12} C_F^2 C_A
- \frac{1273}{48} \alpha^2 C_F C_A^2
- \frac{440}{3} \zeta_5 C_F C_A^2
- \frac{69}{2} C_F^3
- \frac{9}{16} \alpha^3 C_F C_A^2
\right. \right. \nonumber \\
&& \left. \left. ~~~
+ \frac{3}{2} \alpha C_F^2 C_A
+ \frac{3}{4} \alpha^2 C_F^2 C_A
+ \frac{23}{3} \alpha^2 \Nf T_F C_F C_A
+ \frac{46}{3} \alpha \Nf T_F C_F C_A
\right. \right. \nonumber \\
&& \left. \left. ~~~
+ \frac{58}{3} \Nf T_F C_F^2
+ \frac{143}{4} \alpha \zeta_3 C_F C_A^2
+ \frac{160}{3} \zeta_5 \Nf T_F C_F C_A
+ \frac{185}{8} \alpha^2 \zeta_3 C_F C_A^2
\right. \right. \nonumber \\
&& \left. \left. ~~~
+ \frac{1424}{3} \zeta_3 \Nf T_F C_F C_A
- 572 \zeta_3 C_F^2 C_A
- 320 \zeta_5 \Nf T_F C_F^2
- 64 \zeta_3 \Nf^2 T_F^2 C_F
\right. \right. \nonumber \\
&& \left. \left. ~~~
- 16 \alpha \zeta_3 \Nf T_F C_F C_A
- 8 \alpha^2 \zeta_3 \Nf T_F C_F C_A
+ 96 \Nf^2 T_F^2 C_F
+ 256 \zeta_3 \Nf T_F C_F^2
\right. \right. \nonumber \\
&& \left. \left. ~~~
+ 880 \zeta_5 C_F^2 C_A
\right] a^3 ~+~ O(a^4)
\right] 
\label{mmomadl}
\end{eqnarray}
where we include the explicit gauge parameter dependence in the $\mMOM$ scheme
in the arguments of the functions. A general observation is that (\ref{crewms})
has to be replaced by the more accommodating formal relation
\begin{equation}
C_{{\mbox{\footnotesize{Bjr}}}}^{\cal S}(a,\alpha) 
C_{{\mbox{\footnotesize{Adl}}}}^{\cal S}(a,\alpha) ~=~ d_R \left[ 1 ~+~
\Delta_{\mbox{\footnotesize{csb}}}^{\cal S}(a,\alpha) \right]
\end{equation}
where ${\cal S}$ labels the scheme which for the moment is $\mMOM$. In
\cite{12} the objective was to write 
$a \Delta_{\mbox{\footnotesize{csb}}}^{\mMOMs}(a,\alpha)$ in the form
$\beta^{\mMOMs}(a,\alpha) K_a^{\mMOMs}(a,\alpha)$ which was partially 
successful. By this we mean that while the coefficient of the leading term of  
$K_a^{\mMOMs}(a,\alpha)$ had the same value as the $\MSbar$ scheme the next
order term, being $\alpha$ dependent, could not be accommodated within 
(\ref{deltms}) except for several specific values of $\alpha$. These were
$\alpha$~$=$~$0$, $-$~$1$ and $-$~$3$, \cite{12}. At the subsequent loop order 
neither of the last two values preserved the structure of (\ref{deltms}). 
Indeed the situation for the failure of $\alpha$~$=$~$-$~$3$ was examined in
depth. In particular where the breakdown arose was distilled to a set of terms 
with identifiable colour factors. In summarizing the state of play for the 
$\mMOM$ scheme, \cite{12}, it might appear that the relation (\ref{deltms}) 
limits Crewther's original relation, \cite{1}, and its extension to include a 
term proportional to the $\beta$-function to a subset of schemes. However we do
not believe this to be a satisfactory situation.

Instead while the establishment of (\ref{crewms}) and (\ref{deltms}) is an
important observation, the properties of the $\MSbar$ scheme perhaps overlook
aspects of the renormalization group functions in other schemes that actually
gave a direction as to how to resolve the difficulty with the $\mMOM$ scheme.
There will be several threads to our argument. One of the $\beta$-function
properties has already been incorporated into (\ref{mmombjr}) and
(\ref{mmomadl}) which is that in general the $\beta$-function is gauge 
dependent. In the $\MSbar$ scheme the $\beta$-function is independent of the 
gauge parameter, \cite{42}. In addition the gauge parameter is in essence a
second coupling constant in a gauge theory even though its origin is in the 
quadratic part of the Lagrangian rather than in a higher order interaction. The
renormalization group function associated with the renormalization of $\alpha$,
that is not usually referred to as a $\beta$-function as such, is the anomalous
dimension of the gauge parameter which we will denote by 
$\gamma_\alpha(a,\alpha)$ here. Indeed the relevance of this anomalous 
dimension can be seen in the renormalization group equation, derived from 
ensuring that a finite renormalized $n$-point Green's function 
$\Gamma_{(n)}(\mu,a,\alpha)$ is independent of the renormalization scale $\mu$
associated with a regularization, since
\begin{equation}
\left[ \mu \frac{\partial ~}{\partial \mu} ~+~ 
\beta(a,\alpha) \frac{\partial ~}{\partial a} ~+~ 
\alpha \gamma_\alpha(a,\alpha) \frac{\partial ~}{\partial \alpha} ~-~ 
n \gamma_\phi(a,\alpha) \right] \Gamma_{(n)}(\mu,a,\alpha) ~=~ 0 ~.
\label{rgeeqn}
\end{equation}
We have appended an $\alpha$ dependence in the $\beta$-function since in
general it will depend on the gauge parameter. It is only in a subset of
schemes such as $\MSbar$ and $\RI$ that the $\beta$-function is independent of 
$\alpha$. Here $\gamma_\phi(a,\alpha)$ represents the anomalous dimensions of 
all the fields of the $n$-function and we have simplified to the massless case.
While this is a standard expression for a gauge theory if one interprets 
$\alpha$ as a second coupling constant then the second and third terms of 
(\ref{rgeeqn}) could be rewritten as the differential operator
\begin{equation}
\beta_1(g_i) \frac{\partial ~}{\partial g_1} ~+~
\beta_2(g_i) \frac{\partial ~}{\partial g_2}
\label{betacomb}
\end{equation}
where $g_1$~$=$~$a$, $g_2$~$=$~$\alpha$ and
\begin{equation}
\beta_1(a,\alpha) ~\equiv~ \beta(a,\alpha) ~~~,~~~
\beta_2(a,\alpha) ~\equiv~ \alpha \gamma_\alpha(a,\alpha) ~.
\label{betai}
\end{equation}
The final thread of our argument rests in the observation of \cite{5} that 
$\Delta_{\mbox{\footnotesize{csb}}}^{\MSbars}(a)$ vanishes at a fixed point of 
the underlying theory. This is clearly the case in $\MSbar$ as the 
$\beta$-function zeros determine the critical points of the renormalization
group flow. In other schemes where the $\beta$-function is gauge parameter
dependent the situation is more involved. For instance, in \cite{43} the fixed 
point properties of QCD were studied to several loop orders in schemes other 
than $\MSbar$ such as the MOM schemes of Celmaster and Gonsalves, \cite{37,38}. 
In the analysis of \cite{43} the gauge parameter was treated as a second
coupling constant and the critical point values of $a$ and $\alpha$, where the 
functions of (\ref{betai}) are zero, were determined. On top of the Gaussian 
and Banks-Zaks critical points, \cite{34,44}, that were derived from the 
vanishing of $\beta^{\MSbars}(a)$, there are fixed points for both $a$ and 
$\alpha$ non-zero simultaneously. More recently the analysis of \cite{43} has 
been extended to several further loop orders, \cite{45}, for the $\MSbar$, 
$\mMOM$ and the MOM schemes of \cite{37,38} in the linear covariant gauge as 
well as the nonlinear gauges considered here. Not only is the non-trivial 
infrared stable fixed point observed in \cite{43} preserved to higher order it 
has analogues in the nonlinear gauges as well. Therefore considering all these 
ingredients together it should be the case that the product of 
$C_{{\mbox{\footnotesize{Bjr}}}}^{\mMOMs}(a,\alpha)$ and
$C_{{\mbox{\footnotesize{Adl}}}}^{\mMOMs}(a,\alpha)$ gives $d_R$ not only at 
the Gaussian and Banks-Zaks fixed points but also at the other non-trivial 
critical points.

{\begin{table}[ht]
\begin{center}
 \begin{tabular}{ |c|c||c|c|}
 \hline \multicolumn{4}{|c|}{$\mMOM$ $2$ loop} \\
 \hline \hline
\rule{0pt}{12pt}
  $a_{\infty}$&$\alpha_{\infty}$ &$O(a^3)$&$O(a^4)$\\ \hline
  $0.0033112583$ &$0.0000000000$ & $2.9999991596$ & $3.0000039877$ \\ \hline
  $0.0032001941$ &$-3.0301823312$ & $2.9999982468$ & $3.0000012469$ \\ \hline
 $9.1803474173$ &$2.4636080795$ &  $1271156.8083213258$ & $17202735.3015072510$ \\ \hline
\hline \multicolumn{4}{|c|}{$\mMOM$ $3$ loop} \\
 \hline \hline
\rule{0pt}{12pt}
  $a_{\infty}$& $\alpha_{\infty}$ &$O(a^3)$&$O(a^4)$\\ \hline
  $0.0031177883$ & $0.0000000000$ & $2.9999963264$ & $3.0000001212$ \\ \hline
  $0.0031380724$ & $-3.0274210489$ & $2.9999973439$ & $3.0000001217$ \\ \hline
 $0.1279084604$ & $1.9051106246$ & $6.2952539870$ & $10.1893903424$ \\ \hline
 \hline \multicolumn{4}{|c|}{$\mMOM$ $4$ loop} \\
 \hline \hline
\rule{0pt}{12pt}
  $a_{\infty}$&$\alpha_{\infty}$ &$O(a^3)$&$O(a^4)$\\ \hline
  $0.0031213518$ & $0.0000000000$ &$2.9999963720$ & $3.0000001843$ \\ \hline
  $0.0031430130$ &  $-3.0273541344$ &$2.9999974127$ & $3.0000002080$ \\ \hline
 $0.1162651496$ & $0.5286066929$ &  $5.3930704057$ & $11.8942763573$ \\ \hline
 $0.1902883419$ & $0.0000000000$ &  $13.5399867931$ & $66.1969134786$ \\ \hline
 \hline \multicolumn{4}{|c|}{$\mMOM$ $5$ loop} \\
 \hline \hline
\rule{0pt}{12pt}
  $a_{\infty}$& $\alpha_{\infty}$ &$O(a^3)$&$O(a^4)$\\ \hline
  $0.0031220809$ &$0.0000000000$ & $2.9999963814$ & $3.0000001972$ \\ \hline
  $0.0031434144$ & $-3.0273765993$ & $2.9999974183$ & $3.0000002151$ \\ \hline
 $0.0502252330$ & $-3.8653031470$ &  $3.1912609578$ & $3.2787374506$ \\ \hline
 $0.0577103776$ & $0.0000000000$ &  $3.2818695828$ & $3.7273436677$ \\ \hline
\end{tabular}
 \end{center}
\begin{center}
\caption{Values of $C_{{\mbox{\footnotesize{Bjr}}}}^{\mMOMs}(a,\alpha) 
C_{{\mbox{\footnotesize{Adl}}}}^{\mMOMs}(a,\alpha)$ at critical points at
successive loop orders for $\Nf$~$=$~$16$.}
\label{adbjmmfp}
\end{center}
\end{table}}

{\begin{table}[ht]
\begin{center}
 \begin{tabular}{ |c|c||c|c|}
 \hline \multicolumn{4}{|c|}{\rule{0pt}{12pt} $\MSbar$ 2 loop} \\
 \hline \hline
\rule{0pt}{12pt}
 $a_{\infty}$& $\alpha_{\infty}$ & $O(a^3)$&$O(a^4)$\\ \hline
  $0.0033112583$ & $0.0000000000$ & $3.0000005633$ & $3.0000037708$ \\ \hline
 $0.0033112583$ & $-2.9529847269$ &  $3.0000005633$ & $3.0000037708$ \\ \hline
 $0.0033112583$ & $-203.8803486064$ & $3.0000005633$ & $3.0000037708$ \\ \hline 
 \hline \multicolumn{4}{|c|}{\rule{0pt}{12pt} $\MSbar$ 3 loop} \\
 \hline \hline
\rule{0pt}{12pt}
  $a_{\infty}$& $\alpha_{\infty}$ &$O(a^3)$&$O(a^4)$\\ \hline
  $0.0031618421$ & $0.0000000000$ & $2.9999981298$ & $3.0000007963$ \\ \hline
 $0.0031618421$ & $-2.9458392416$ &  $2.9999981298$ & $3.0000007963$ \\ \hline
 \hline \multicolumn{4}{|c|}{\rule{0pt}{12pt} $\MSbar$ 4 loop} \\
 \hline \hline
\rule{0pt}{12pt}
  $a_{\infty}$& $\alpha_{\infty}$ &$O(a^3)$&$O(a^4)$\\ \hline
  $0.0031699203$ & $0.0000000000$ & $2.9999982498$ & $3.0000009437$ \\ \hline
  $0.0031699203$ & $-2.9459051005$ & $2.9999982498$ & $3.0000009437$ \\ \hline
$0.0031699203$ & $-126.8989470199$ &  $2.9999982498$ & $3.0000009437$ \\ \hline
 $0.0917000502$ & $0.0000000000$ &  $4.1865556391$ & $6.0730678402$ \\ \hline
 $0.0917000502$ & $2.6061754737$ &  $4.1865556391$ & $6.0730678402$ \\ \hline
 $0.0917000502$ & $-4.8367657559$ &  $4.1865556391$ & $6.0730678402$ \\ \hline
 \hline \multicolumn{4}{|c|}{\rule{0pt}{12pt} $\MSbar$ 5 loop} \\
 \hline \hline
\rule{0pt}{12pt}
  $a_{\infty}$& $\alpha_{\infty}$ &$O(a^3)$&$O(a^4)$\\ \hline
  $0.0031708402$ & $0.0000000000$ & $2.9999982636$ & $3.0000009605$ \\ \hline
 $0.0031708402$ & $-2.9459382941$ & $2.9999982636$ & $3.0000009605$ \\ \hline
 $0.0468980276$ & $0.0000000000$ &  $3.1531010299$ & $3.2821634001$ \\ \hline
 $0.0468980276$ & $-3.7013593081$ & $3.1531010299$ & $3.2821634001$ \\ \hline
\end{tabular}
 \end{center}
\begin{center}
\caption{Values of $C_{{\mbox{\footnotesize{Bjr}}}}^{\MSbars}(a)
C_{{\mbox{\footnotesize{Adl}}}}^{\MSbars}(a)$ at critical points at successive 
loop orders for $\Nf$~$=$~$16$.}
\label{adbjmsfp}
\end{center}
\end{table}}

\sect{Numerical evidence.}

In order to test the idea given in \cite{5} that 
$C_{{\mbox{\footnotesize{Bjr}}}}^{\mMOMs}(a,\alpha) 
C_{{\mbox{\footnotesize{Adl}}}}^{\mMOMs}(a,\alpha)$ reflects some conformal
property of the underlying field theory at a fixed point, we have carried out a
numerical investigation in the first instance using the data given in 
\cite{45}. In particular the critical point values of the coupling constant and
gauge parameter are known to various loop orders in different renormalization 
schemes and gauges. The numerical analysis of this section will be for the 
$SU(3)$ colour group. Also as the Crewther relation is a purely four 
dimensional one the fixed points we consider include all real solutions with
positive coupling constant where the functions of (\ref{betai}) vanish. To 
remain within the perturbative domain of applicability our analysis was 
restricted to values of $\Nf$ close to the top of the conformal window which is
at $\Nf$~$=$~$16$ for $SU(3)$, \cite{34,44}. As a guide to this reasoning we 
have recorded the successive loop order evaluation of 
$C_{{\mbox{\footnotesize{Bjr}}}}^{\mMOMs}(a,\alpha) 
C_{{\mbox{\footnotesize{Adl}}}}^{\mMOMs}(a,\alpha)$ at criticality in Table
\ref{adbjmmfp} for $\Nf$~$=$~$16$. In the table the various three or four 
critical values of $a$ and $\alpha$, denoted by $a_\infty$ and $\alpha_\infty$ 
in the same notation as \cite{43}, are given at two to five loops respectively 
in four sub-sectors as indicated by the sub-heading. The critical values are 
recorded to higher precision than \cite{45}. In each sub-sector the first fixed
point is the Banks-Zaks solution, \cite{34,44}, which is a saddle point in the 
$(a,\alpha)$ plane. The second fixed point is the infrared stable one whose 
critical coupling value tends to that of the Banks-Zaks one as the loop order 
increases. The remaining one or two fixed point solutions are artefacts of 
solving polynomial equations at higher loop order. While their critical 
coupling values are outside the range of perturbative validity they are 
included as they are relevant to subsequent arguments. The data in the final 
two columns are the $O(a^3)$ and $O(a^4)$ values of 
$C_{{\mbox{\footnotesize{Bjr}}}}^{\mMOMs}(a,\alpha) 
C_{{\mbox{\footnotesize{Adl}}}}^{\mMOMs}(a,\alpha)$ at criticality.

{\begin{table}[ht]
\begin{center}
 \begin{tabular}{ |c|c||c|c|}
 \hline \multicolumn{4}{|c|}{$\MOMg$ 3 loops} \\
 \hline \hline
\rule{0pt}{12pt}
  $a_{\infty}$&$\alpha_{\infty}$ &$O(a^3)$&$O(a^4)$\\ \hline
  $0.0029141531$ & $0.0000000000$ & $2.9999877567$ & $2.9999976812$ \\ \hline
 $0.0029387804$ & $-3.0259847376$ &  $2.9999876182$ & $2.9999981783$ \\ \hline
 \hline \multicolumn{4}{|c|}{$\MOMc$ 3 loops} \\
 \hline \hline
\rule{0pt}{12pt}
  $a_{\infty}$& $\alpha_{\infty}$ &$O(a^3)$&$O(a^4)$\\ \hline
  $0.0031361417$ & $0.0000000000$ & $2.9999980055$ & $2.9999998120$ \\ \hline
 $0.0031893967$ & $-3.0263835218$ &  $2.9999980117$ & $3.0000010487$ \\ \hline
  $0.0108889865$ & $-9.6139587420$ & $3.0017410493$ & $3.0006478718$ \\ \hline
 \hline \multicolumn{4}{|c|}{$\MOMq$ 3 loops} \\
 \hline \hline
\rule{0pt}{12pt}
  $a_{\infty}$& $\alpha_{\infty}$ &$O(a^3)$&$O(a^4)$\\ \hline
  $0.0031361417$ &$0.0000000000$ & $2.9999974139$ & $2.9999999355$ \\ \hline
  $0.0031893967$ &$-3.0263835218$ & $3.0000000826$ & $2.9999999820$ \\ \hline
  $0.0108889865$ &$-9.6139587420$ & $3.0015252800$ & $3.0013631672$ \\ \hline
\end{tabular}
 \end{center}
\begin{center}
\caption{Values of $C_{{\mbox{\footnotesize{Bjr}}}}^{\MOMis}(a,\alpha)
C_{{\mbox{\footnotesize{Adl}}}}^{\MOMis}(a,\alpha)$ at critical points at successive
loop orders for $\Nf$~$=$~$16$.}
\label{adbjmomfp}
\end{center}
\end{table}}

{\begin{table}[ht]
\begin{center}
 \begin{tabular}{ |c|c||c|c|}
 \hline \multicolumn{4}{|c|}{Curci-Ferrari $\MOMg$ 3 loops} \\
 \hline \hline
\rule{0pt}{12pt}
 $a_{\infty}$& $\alpha_{\infty}$ & $O(a^3)$&$O(a^4)$\\ \hline
 $0.0029141531$ & $0.0000000000$ & $2.9999877567$ & $2.9999976812$ \\ \hline
 $0.0031077787$ & $-5.8860065418$ & $2.9999745145$ & $2.9999979271$ \\ \hline
 $0.0069944637$ & $-6.2196239630$ & $2.9999956606$ & $3.0006212984$ \\ \hline
 \hline \multicolumn{4}{|c|}{Curci-Ferrari $\MOMc$ 3 loops} \\
 \hline \hline
\rule{0pt}{12pt}
 $a_{\infty}$& $\alpha_{\infty}$ & $O(a^3)$&$O(a^4)$\\ \hline
 $0.0031822955$ & $0.0000000000$ & $2.9999996960$ & $3.0000000816$ \\ \hline
 $0.0033905650$ & $-5.8484574473$ & $3.0000096730$ & $2.9999996739$ \\ \hline
 $0.0256527015$ & $-17.6663992530$ & $3.0598055895$ & $2.4987327333$ \\ \hline
 $0.0776442642$ & $2.1961076699$ & $3.7711417659$ & $2.5429487508$ \\ \hline
 \hline \multicolumn{4}{|c|}{Curci-Ferrari $\MOMq$ 3 loops} \\
 \hline \hline
\rule{0pt}{12pt}
 $a_{\infty}$&$\alpha_{\infty}$ & $O(a^3)$&$O(a^4)$\\ \hline
 $0.0031361417$ & $0.0000000000$ & $2.9999974139$ & $2.9999999354$ \\ \hline
 $0.0032120881$ & $-5.8504247987$ & $3.0000013385$ & $2.9999998776$ \\ \hline
 $0.0068658178$ & $-12.1450763308$ & $3.0002807117$ & $3.0002920650$ \\ \hline
\end{tabular}
 \end{center}
\begin{center}
\caption{Values of $C_{{\mbox{\footnotesize{Bjr}}}}^{\MOMis}(a,\alpha)
C_{{\mbox{\footnotesize{Adl}}}}^{\MOMis}(a,\alpha)$ at critical points at successive
loop orders for $\Nf$~$=$~$16$ in the Curci-Ferrari gauge.}
\label{adbjcfmomfp}
\end{center}
\end{table}}
{\begin{table}[ht]
\begin{center}
 \begin{tabular}{ |c|c||c|c|}
 \hline \multicolumn{4}{|c|}{MAG $\MOMg$ 3 loops} \\
 \hline \hline
\rule{0pt}{12pt}
 $a_{\infty}$& $\alpha_{\infty}$ & $O(a^3)$&$O(a^4)$\\ \hline
 $0.0028654266$ & $-0.4168022707$ &  $2.9999860489$ & $2.9999968722$ \\ \hline
 $0.0029254124$ & $-5.5625872573$ & $2.9999889816$ & $2.9999980333$ \\ \hline
 \hline \multicolumn{4}{|c|}{MAG $\MOMc$ 3 loops} \\
 \hline \hline
\rule{0pt}{12pt}
 $a_{\infty}$& $\alpha_{\infty}$ & $O(a^3)$&$O(a^4)$\\ \hline
 $0.0031683606$ & $-0.4162766338$ & $2.9999989914$ & $3.0000000637$ \\ \hline
 $0.0031577946$ & $-5.5686496127$ & $2.9999979720$ & $3.0000000132$ \\ \hline
 $0.0014328961$ & $-60.7998590768$ & $3.0000187570$ & $2.9999749282$ \\ \hline
 \hline \multicolumn{4}{|c|}{MAG $\MOMq$ 3 loops} \\
 \hline \hline
\rule{0pt}{12pt}
 $a_{\infty}$& $\alpha_{\infty}$ & $O(a^3)$&$O(a^4)$\\ \hline
 $0.0031256674$ & $-0.4162158544$ & $2.9999968662$ & $2.9999999505$ \\ \hline
 $0.0031921140$ & $-5.5677230790$ & $3.0000001817$ & $3.0000000315$ \\ \hline
\end{tabular}
 \end{center}
\begin{center}
\caption{Values of $C_{{\mbox{\footnotesize{Bjr}}}}^{\MOMis}(a,\alpha)
C_{{\mbox{\footnotesize{Adl}}}}^{\MOMis}(a,\alpha)$ at critical points at successive
loop orders for $\Nf$~$=$~$16$ in the MAG.}
\label{adbjmagmomfp}
\end{center}
\end{table}}

Several features emerge from the table. The first is that for the first two 
critical points the combination of the Adler function and Bjorken sum rule 
produces a value very close to the expected value of $3$. Moreover there is a 
clear but slow convergence to $3$ as the loop order increases. We qualify the 
lack of precise agreement by noting that in computing the product there will be
errors from the truncation of the series. With the available $O(a^4)$ 
expressions for both series the product will have terms up to $O(a^8)$. However
this would not be the true $O(a^8)$ expression for
$C_{{\mbox{\footnotesize{Bjr}}}}^{\mMOMs}(a,\alpha) 
C_{{\mbox{\footnotesize{Adl}}}}^{\mMOMs}(a,\alpha)$ since the as yet 
undetermined $O(a^5)$ term of one factor would lead to an $O(a^8)$ contribution
from the known $O(a^3)$ term of the other factor for instance. For the purposes
of the tables we do not include such incomplete higher order contributions. 
Therefore the most reliable indication of the value of  
$C_{{\mbox{\footnotesize{Bjr}}}}^{\mMOMs}(a,\alpha) 
C_{{\mbox{\footnotesize{Adl}}}}^{\mMOMs}(a,\alpha)$ is that from the five loop
critical points at both $O(a^3)$ and $O(a^4)$. A guide to this is that the
final two fixed points produce a value significantly closer to $3$ than might
have seemed possible from the corresponding lower loop data even at four loops.
That the value for these two five loop fixed points is not as accurate as the 
Banks-Zaks or the infrared stable ones resides in the fact that one is still
outside the domain of perturbative reliability. More importantly for the two 
that are clearly within the domain the critical value of
$C_{{\mbox{\footnotesize{Bjr}}}}^{\mMOMs}(a,\alpha) 
C_{{\mbox{\footnotesize{Adl}}}}^{\mMOMs}(a,\alpha)$ remains close to $3$. We
have analysed the small deviation from $3$ for these two fixed points in a
variety of ways and concluded that the discrepancy is due to the terms beyond
$O(a^4)$ that result from the product of two perturbative functions of $a$.
Perhaps one clear indication of this is to repeat the same exercise that 
produced Table \ref{adbjmmfp} but for the $\MSbar$ scheme as that case led to 
the expressions for $\Delta_{\mbox{\footnotesize{csb}}}^{\MSbars}(a)$ in 
(\ref{crewms}). The results are recorded in Table \ref{adbjmsfp} to the same 
loop order. What is apparent is that the same numerical accuracy emerges for 
$C_{{\mbox{\footnotesize{Bjr}}}}^{\MSbars}(a) 
C_{{\mbox{\footnotesize{Adl}}}}^{\MSbars}(a)$ at each loop order for the 
Banks-Zaks and infrared stable fixed points as the corresponding cases in the
$\mMOM$ scheme. So the deviation from the value of $3$ in the $\mMOM$ scheme is
on a similar footing to $\MSbar$. While this reinforces the notion that the
Crewther relation has a connection to conformal properties it does not resolve
how the observations of \cite{12} can be accommodated.

We conclude this section by recording the situation in other schemes as well as
gauges. In \cite{45} the fixed point properties of QCD in the MOM schemes of 
Celmaster and Gonsalves were studied in the linear covariant gauge as well as 
two nonlinear gauges. These were the Curci-Ferrari gauge and the Maximal 
Abelian Gauge (MAG) whose more technical properties we discuss later but note 
that for both gauges the perturbative renormalization group functions are only 
available to three loops. Our results for the linear gauge are given in Table 
\ref{adbjmomfp} while those for the other gauges are noted in Tables 
\ref{adbjcfmomfp} and \ref{adbjmagmomfp} respectively where $\MOMg$, $\MOMc$ 
and $\MOMq$ denote the three distinct MOM schemes based on the triple gluon, 
ghost-gluon and quark-gluon $3$-point vertices respectively. For brevity we 
have solely recorded the situation at three loops. Again a very similar state 
to the five loop analysis arises. In each case and particularly at three loops 
the product of the Adler function and the Bjorken sum rule in effect evaluate 
to $3$. Moreover it reinforces the observation that the Crewther relation has 
to accommodate properties consistent with the critical points of the underlying
theory for a variety of schemes and gauges.

\sect{$\mMOM$ scheme.}

In light of these arguments and numerical evidence as well as the association
of $\beta(a,\alpha)$ and $\gamma_\alpha(a,\alpha)$ with the underlying
$\beta$-functions of gauge fixed QCD we propose that the extension of the 
Crewther relation of \cite{5} to accommodate renormalization schemes other than 
$\MSbar$ is 
\begin{equation}
\Delta_{\mbox{\footnotesize{csb}}}^{\cal S}(a_{\cal S},\alpha_{\cal S}) ~=~ 
\frac{\beta^{\cal S}(a_{\cal S},\alpha_{\cal S})}{a_{\cal S}} 
K_a^{\cal S}(a_{\cal S},\alpha_{\cal S}) ~+~ 
\alpha_{\cal S} \gamma_\alpha^{\cal S}(a_{\cal S},\alpha_{\cal S}) 
K_\alpha^{\cal S}(a_{\cal S},\alpha_{\cal S}) 
\label{deltgen}
\end{equation}
where ${\cal S}$ labels the scheme. Again we have included $\alpha$ in the 
argument of the $\beta$-function to be as general as possible. While 
(\ref{deltgen}) reflects the combination (\ref{betacomb}) with the 
identification of (\ref{betai}) in (\ref{rgeeqn}), unlike its $\MSbar$ 
counterpart two functions $K_a(a,\alpha)$ and $K_\alpha(a,\alpha)$ are now 
required to satisfy the vanishing of 
$\Delta_{\mbox{\footnotesize{csb}}}(a,\alpha)$ at all the critical points of 
theory in a scheme ${\cal S}$. In other words if the origin of the Crewther
relation rests in the fact that it reflects conformal symmetry, which is 
related to vanishing $\beta$-functions, then this property should not be a pure
$\MSbar$ one. The combination of the Bjorken sum rule and Adler $D$-function 
that is the Crewther relation should vanish at the fixed points of another 
scheme as suggested by our numerical investigation. So this can be resolved for
other schemes provided the relevant $\beta$-functions are present as proposed 
in (\ref{deltgen}).

In keeping with the ethos of \cite{5} where the coefficients of the 
renormalization group functions of (\ref{deltgen}) were established by explicit
calculation we now focus in this section on the $\mMOM$ scheme to study the 
$K$-functions partly to be able to compare and contrast with \cite{10,11,12}. 
Therefore we have constructed the following functions in the $\mMOM$ scheme 
\begin{eqnarray}
K_a^{\mMOMs}(a,\alpha) &=& 
\left[
12 \zeta_3 C_F
- \frac{21}{2} C_F
\right] a
\nonumber \\
&&
+ \left[
\frac{21}{4} C_F C_A \alpha^2
+ \frac{21}{2} C_F C_A \alpha
- \frac{2591}{12} C_F C_A
+ \frac{397}{6} C_F^2
- 240 \zeta_5 C_F^2
\right. \nonumber \\
&& \left. ~~~
- 48 \zeta_3 \Nf T_F C_F
- 12 \zeta_3 C_F C_A \alpha
- 6 \zeta_3 C_F C_A \alpha^2
+ 62 \Nf T_F C_F
+ 136 \zeta_3 C_F^2
\right. \nonumber \\
&& \left. ~~~
+ 182 \zeta_3 C_F C_A
\right] a^2
\nonumber \\
&&
+ \left[
\frac{2471}{12} C_F^3
- \frac{1840145}{288} C_F C_A^2
- \frac{14740}{3} \zeta_5 C_F^2 C_A
- \frac{8000}{3} \zeta_5 \Nf T_F C_F C_A
\right. \nonumber \\
&& \left. ~~~
- \frac{5092}{9} \Nf T_F C_F^2
- \frac{4792}{3} \zeta_3 \Nf T_F C_F^2
- \frac{4312}{3} \zeta_3 \Nf T_F C_F C_A
\right. \nonumber \\
&& \left. ~~~
- \frac{2961}{16} \zeta_3 C_F C_A^2 \alpha^2
- \frac{2113}{2} \zeta_3^2 C_F C_A^2
- \frac{1568}{3} \Nf^2 T_F^2 C_F
- \frac{397}{4} C_F^2 C_A \alpha
\right. \nonumber \\
&& \left. ~~~
- \frac{397}{8} C_F^2 C_A \alpha^2
- \frac{63}{4} \zeta_3 C_F C_A^2 \alpha^3
- \frac{31}{2} \Nf T_F C_F C_A \alpha^2
+ \frac{27}{2} \zeta_3^2 C_F C_A^2 \alpha^2
\right. \nonumber \\
&& \left. ~~~
+ \frac{155}{3} \Nf T_F C_F C_A \alpha
+ \frac{391}{8} \zeta_3 C_F C_A^2 \alpha
+ \frac{561}{32} C_F C_A^2 \alpha^3
+ \frac{4051}{96} C_F C_A^2 \alpha
\right. \nonumber \\
&& \left. ~~~
+ \frac{6251}{32} C_F C_A^2 \alpha^2
+ \frac{11900}{3} \zeta_5 C_F C_A^2
+ \frac{71251}{18} \Nf T_F C_F C_A
+ \frac{132421}{72} C_F^2 C_A
\right. \nonumber \\
&& \left. ~~~
+ \frac{152329}{48} \zeta_3 C_F C_A^2
- 5720 \zeta_5 C_F^3
- 840 \zeta_7 C_F^2 C_A
- 204 \zeta_3 C_F^2 C_A \alpha
\right. \nonumber \\
&& \left. ~~~
- 102 \zeta_3 C_F^2 C_A \alpha^2
- 99 \zeta_3^2 C_F C_A^2 \alpha
- 40 \zeta_3 \Nf T_F C_F C_A \alpha
\right. \nonumber \\
&& \left. ~~~
+ 12 \zeta_3 \Nf T_F C_F C_A \alpha^2
+ 160 \zeta_3^2 \Nf T_F C_F C_A
+ 180 \zeta_5 C_F^2 C_A \alpha^2
\right. \nonumber \\
&& \left. ~~~
+ 224 \zeta_3 \Nf^2 T_F^2 C_F
+ 320 \zeta_5 \Nf^2 T_F^2 C_F
+ 360 \zeta_5 C_F^2 C_A \alpha
+ 488 \zeta_3 C_F^3
\right. \nonumber \\
&& \left. ~~~
+ 2400 \zeta_5 \Nf T_F C_F^2
+ 3608 \zeta_3 C_F^2 C_A
+ 5040 \zeta_7 C_F^3
\right] a^3 ~+~ O(a^4)
\end{eqnarray}
and
\begin{eqnarray}
K_\alpha^{\mMOMs}(a,\alpha) &=&
\left[
\frac{21}{4} C_F C_A
+ \frac{21}{4} C_F C_A \alpha
- 6 \zeta_3 C_F C_A
- 6 \zeta_3 C_F C_A \alpha
\right] a^2
\nonumber \\
&&
+ \left[
\frac{189}{32} \alpha^2 C_F C_A^2
+ \frac{341}{3} \Nf T_F C_F C_A
+ \frac{2035}{48} \alpha C_F C_A^2
+ \frac{2109}{8} \zeta_3 C_F C_A^2
\right. \nonumber \\
&& \left. ~~~
+ 9 \zeta_3^2 C_F C_A^2 \alpha
+ 120 \zeta_5 C_F^2 C_A
+ 120 \zeta_5 C_F^2 C_A \alpha
- \frac{9773}{32} C_F C_A^2
\right. \nonumber \\
&& \left. ~~~
- \frac{443}{8} \zeta_3 C_F C_A^2 \alpha
- \frac{397}{12} C_F^2 C_A
- \frac{397}{12} C_F^2 C_A \alpha
- \frac{27}{4} \zeta_3 C_F C_A^2 \alpha^2
\right. \nonumber \\
&& \left. ~~~
- 88 \zeta_3 \Nf T_F C_F C_A
- 68 \zeta_3 C_F^2 C_A
- 68 \zeta_3 C_F^2 C_A \alpha
- 33 \zeta_3^2 C_F C_A^2
\right] a^3
\nonumber \\
&& +~ O(a^4) ~.
\end{eqnarray}
These expressions were arrived at perturbatively by solving for the two unknown
$K$-functions at each order in the coupling constant. This approach was taken
from the point of view of seeing whether such an ansatz would admit an explicit
solution and is a valid method to apply. What it did reveal was that one 
property of the $K$-functions is they are not strictly unique. This is in the 
sense that one could in principle arrange terms in one of the $K$-functions to 
be absorbed into the other in such a way that the vanishing of 
$\Delta_{\mbox{\footnotesize{csb}}}(a,\alpha)$ at all the fixed points is
preserved. One noteworthy property of $K_\alpha^{\mMOMs}(a,\alpha)$ is that the
leading term, which is always $O(a^2)$ for this function, vanishes at 
$\alpha$~$=$~$-$~$1$ but not for the $O(a^3)$ term. This particular gauge 
parameter value was singled out for specific comment in \cite{12} as being of 
special interest. Equally it was noted that its special status was not 
preserved at next order. Similar observations were also made for the case of
$\alpha$~$=$~$-$~$3$ and we will discuss both cases in a later section.

\sect{MOM schemes.} 

While the Crewther relation has been studied previously in various schemes but 
at length in the $\mMOM$ scheme in \cite{12}, it is instructive to study 
several specific schemes as well as gauges other than the commonly used linear 
covariant gauge. This is the topic for this section and we examine the MOM 
schemes of \cite{37,38}. In particular we will focus on the $\MOMg$ scheme 
associated with the triple gluon vertex as the mathematical content of the 
$K$-functions of each of the three MOM schemes is similar. The $\MOMg$ 
renormalization group functions of \cite{37,38} and the other two schemes were 
extended to three loops in \cite{46} numerically as well as analytically for an
arbitrary linear gauge in \cite{47}. More recently the four loop Landau gauge 
renormalization group functions were determined in \cite{48}. One reason we 
have not used the results of \cite{48} in our calculations is partly as they 
are recorded for a specific gauge. However, another is that those higher order 
corrections will not contribute to the $K$-functions. This can be seen by 
simply examining the orders of $a$ in the two terms of (\ref{deltgen}) and the 
fact that the Adler and Bjorken expressions are only available to $O(a^4)$. As 
the mapping between the coupling constants and gauge parameters in the $\MSbar$
and $\MOMg$ schemes are known, \cite{47}, we have found 
$C_{{\mbox{\footnotesize{Bjr}}}}^{\MOMgs}(a,\alpha)$ and 
$C_{{\mbox{\footnotesize{Adl}}}}^{\MOMgs}(a,\alpha)$ to $O(a^4)$. Consequently 
we have been able to determine the two $K$-functions parallel to those of the 
$\mMOM$ scheme. For instance the $SU(3)$ Yang-Mills expressions are
\begin{eqnarray}
\left. K_a^{\MOMgs}(a,\alpha) \right|^{SU(3)}_{\Nf=0} &=& 
\left[
16 \zeta_3
- 14
\right] a
\nonumber \\
&&
+ \left[
\frac{6448}{9} \zeta_3
- \frac{14158}{27}
- \frac{1280}{3} \zeta_5
- \frac{644}{27} \pi^2
- \frac{368}{9} \psi^\prime(\third) \zeta_3
- \frac{56}{9} \pi^2 \alpha^2
\right. \nonumber \\
&& \left. ~~~
- \frac{32}{3} \psi^\prime(\third) \zeta_3 \alpha^2
+ \frac{28}{3} \psi^\prime(\third) \alpha^2
+ \frac{64}{9} \zeta_3 \pi^2 \alpha^2
+ \frac{322}{9} \psi^\prime(\third)
+ \frac{736}{27} \zeta_3 \pi^2
\right. \nonumber \\
&& \left. ~~~
- 72 \zeta_3 \alpha
- 42 \alpha^2
- 42 \psi^\prime(\third) \alpha
- 32 \zeta_3 \pi^2 \alpha
- 8 \zeta_3 \alpha^3
+ 7 \alpha^3
+ 28 \pi^2 \alpha
\right. \nonumber \\
&& \left. ~~~
+ 48 \zeta_3 \alpha^2
+ 48 \psi^\prime(\third) \zeta_3 \alpha
+ 63 \alpha
\right] a^2
\nonumber \\
&&
+ \left[
\frac{1}{4} \psi^{\prime\prime\prime}(\third) \zeta_3 \alpha^3
+ \frac{7}{2} \psi^{\prime\prime\prime}(\third) \zeta_3 \alpha^2
- \frac{127914947}{2592}
- \frac{696830}{243} \pi^2
\right. \nonumber \\
&& \left. ~~~
- \frac{372568}{81} \zeta_3 \pi^2 \alpha
- \frac{307715}{54} \psi^\prime(\third) \alpha
- \frac{210179}{324} \pi^2 \alpha^2
\right. \nonumber \\
&& \left. ~~~
- \frac{177919}{243} \psi^\prime(\third) \pi^2 \alpha
- \frac{147991}{96} \alpha^2
- \frac{144856}{27} \psi^\prime(\third) \zeta_3
- \frac{133352}{729} \zeta_3 \pi^4 \alpha
\right. \nonumber \\
&& \left. ~~~
- \frac{54362}{243} \zeta_3 \pi^4
- \frac{50834}{81} \psi^\prime(\third)^2 \zeta_3 \alpha
- \frac{29440}{27} \zeta_5 \pi^2
- \frac{25921}{216} \psi^\prime(\third)^2
\right. \nonumber \\
&& \left. ~~~
- \frac{24514}{3} \zeta_3 \alpha
- \frac{21805}{216} \psi^\prime(\third)^2 \alpha^2
- \frac{14812}{81} \psi^\prime(\third) \zeta_3 \pi^2
\right. \nonumber \\
&& \left. ~~~
- \frac{12460}{81} \psi^\prime(\third) \zeta_3 \pi^2 \alpha^2
- \frac{11663}{9} \psi^\prime(\third) \zeta_3 \alpha^2
- \frac{8918}{243} \pi^4 \alpha^2
- \frac{5349}{32} \alpha^4
\right. \nonumber \\
&& \left. ~~~
- \frac{4399}{72} \alpha^3
- \frac{2989}{32} \psi^{\prime\prime\prime}(\third)
- \frac{2560}{9} \zeta_5 \pi^2 \alpha^2
- \frac{1286}{81} \zeta_3 \pi^4 \alpha^3
- \frac{1205}{6} \zeta_3 \alpha^3
\right. \nonumber \\
&& \left. ~~~
- \frac{1115}{18} \pi^2 \alpha^3
- \frac{1078}{27} \psi^\prime(\third) \pi^2 \alpha^3
- \frac{763}{12} \pi^2 \alpha^4
- \frac{308}{9} \psi^\prime(\third)^2 \zeta_3 \alpha^3
\right. \nonumber \\
&& \left. ~~~
- \frac{250}{3} \psi^\prime(\third) \zeta_3 \alpha^3
- \frac{153}{2} \zeta_3 \alpha^5
- \frac{147}{32} \alpha^6
- \frac{112}{9} \psi^\prime(\third) \zeta_3 \pi^2 \alpha^4
- \frac{98}{27} \pi^4 \alpha^4
\right. \nonumber \\
&& \left. ~~~
- \frac{49}{4} \psi^\prime(\third) \alpha^5
- \frac{49}{6} \psi^\prime(\third)^2 \alpha^4
- \frac{49}{16} \psi^{\prime\prime\prime}(\third) \alpha^2
- \frac{28}{3} \zeta_3 \pi^2 \alpha^5
\right. \nonumber \\
&& \left. ~~~
- \frac{7}{32} \psi^{\prime\prime\prime}(\third) \alpha^3
+ \frac{21}{4} \zeta_3 \alpha^6
+ \frac{28}{3} \psi^\prime(\third)^2 \zeta_3 \alpha^4
+ \frac{49}{6} \pi^2 \alpha^5
+ \frac{63}{2} \psi^{\prime\prime\prime}(\third) \alpha
\right. \nonumber \\
&& \left. ~~~
+ \frac{98}{9} \psi^\prime(\third) \pi^2 \alpha^4
+ \frac{112}{27} \zeta_3 \pi^4 \alpha^4
+ \frac{218}{3} \zeta_3 \pi^2 \alpha^4
+ \frac{427}{4} \psi^{\prime\prime\prime}(\third) \zeta_3
\right. \nonumber \\
&& \left. ~~~
+ \frac{500}{9} \zeta_3 \pi^2 \alpha^3
+ \frac{539}{18} \psi^\prime(\third)^2 \alpha^3
+ \frac{763}{8} \psi^\prime(\third) \alpha^4
+ \frac{867}{4} \zeta_3 \alpha^4
+ \frac{1071}{16} \alpha^5
\right. \nonumber \\
&& \left. ~~~
+ \frac{1115}{12} \psi^\prime(\third) \alpha^3
+ \frac{1232}{27} \psi^\prime(\third) \zeta_3 \pi^2 \alpha^3
+ \frac{1280}{3} \psi^\prime(\third) \zeta_5 \alpha^2
\right. \nonumber \\
&& \left. ~~~
+ \frac{3115}{27} \psi^\prime(\third)^2 \zeta_3 \alpha^2
+ \frac{3703}{27} \psi^\prime(\third)^2 \zeta_3
+ \frac{4501}{324} \pi^4 \alpha^3
+ \frac{10192}{243} \zeta_3 \pi^4 \alpha^2
\right. \nonumber \\
&& \left. ~~~
+ \frac{14720}{9} \psi^\prime(\third) \zeta_5
+ \frac{19415}{8} \zeta_3 \alpha^2
+ \frac{21805}{162} \psi^\prime(\third) \pi^2 \alpha^2
+ \frac{22400}{3} \zeta_7
\right. \nonumber \\
&& \left. ~~~
+ \frac{23326}{27} \zeta_3 \pi^2 \alpha^2
+ \frac{25921}{162} \psi^\prime(\third) \pi^2
+ \frac{116683}{729} \pi^4 \alpha
+ \frac{177919}{324} \psi^\prime(\third)^2 \alpha
\right. \nonumber \\
&& \left. ~~~
+ \frac{186284}{27} \psi^\prime(\third) \zeta_3 \alpha
+ \frac{190267}{972} \pi^4
+ \frac{203336}{243} \psi^\prime(\third) \zeta_3 \pi^2 \alpha
\right. \nonumber \\
&& \left. ~~~
+ \frac{210179}{216} \psi^\prime(\third) \alpha^2
+ \frac{289169}{48} \alpha
+ \frac{289712}{81} \zeta_3 \pi^2
+ \frac{307715}{81} \pi^2 \alpha
\right. \nonumber \\
&& \left. ~~~
+ \frac{348415}{81} \psi^\prime(\third)
+ \frac{485200}{27} \zeta_5
+ \frac{9840343}{216} \zeta_3
- 23325 \zeta_3^2
- 1920 \zeta_5 \alpha^2
\right. \nonumber \\
&& \left. ~~~
- 1920 \psi^\prime(\third) \zeta_5 \alpha
- 109 \psi^\prime(\third) \zeta_3 \alpha^4
- 36 \zeta_3^2 \alpha^3
- 36 \psi^{\prime\prime\prime}(\third) \zeta_3 \alpha
+ 9 \zeta_3^2 \alpha^2
\right. \nonumber \\
&& \left. ~~~
+ 14 \psi^\prime(\third) \zeta_3 \alpha^5
+ 320 \zeta_5 \alpha^3
+ 1280 \zeta_5 \pi^2 \alpha
+ 2052 \zeta_3^2 \alpha
+ 2880 \zeta_5 \alpha
\right] a^3 \nonumber \\
&& +~ O(a^4)
\end{eqnarray}
and
\begin{eqnarray}
\left. K_\alpha^{\MOMgs}(a,\alpha) \right|^{SU(3)}_{\Nf=0} &=& 
\left[
\frac{63}{2}
- \frac{56}{9} \pi^2 \alpha
- \frac{32}{3} \psi^\prime(\third) \zeta_3 \alpha
+ \frac{21}{2} \alpha^2
+ \frac{28}{3} \psi^\prime(\third) \alpha
+ \frac{64}{9} \zeta_3 \pi^2 \alpha
- 42 \alpha
\right. \nonumber \\
&& \left. ~
- 36 \zeta_3
- 21 \psi^\prime(\third)
- 16 \zeta_3 \pi^2
- 12 \zeta_3 \alpha^2
+ 14 \pi^2
+ 24 \psi^\prime(\third) \zeta_3
+ 48 \zeta_3 \alpha
\right] a^2
\nonumber \\
&&
+ \left[
684 \zeta_3^2
+ 960 \zeta_5
- \frac{528991}{108} \psi^\prime(\third)
- \frac{298948}{81} \zeta_3 \pi^2
- \frac{207487}{324} \psi^\prime(\third) \alpha
\right. \nonumber \\
&& \left. ~~~
- \frac{179000}{729} \zeta_3 \pi^4
- \frac{177037}{243} \psi^\prime(\third) \pi^2
- \frac{50582}{81} \psi^\prime(\third)^2 \zeta_3
- \frac{26252}{81} \zeta_3 \pi^2 \alpha
\right. \nonumber \\
&& \left. ~~~
- \frac{12853}{3} \zeta_3
- \frac{5120}{27} \zeta_5 \pi^2 \alpha
- \frac{2464}{27} \zeta_3 \pi^4 \alpha
- \frac{2331}{8} \alpha^3
- \frac{2009}{9} \psi^\prime(\third) \pi^2 \alpha
\right. \nonumber \\
&& \left. ~~~
- \frac{1295}{6} \zeta_3 \alpha^2
- \frac{917}{36} \alpha^2
- \frac{574}{3} \psi^\prime(\third)^2 \zeta_3 \alpha
- \frac{448}{27} \psi^\prime(\third) \zeta_3 \pi^2 \alpha^3
\right. \nonumber \\
&& \left. ~~~
- \frac{392}{81} \pi^4 \alpha^3
- \frac{385}{3} \pi^2 \alpha^3
- \frac{255}{2} \zeta_3 \alpha^4
- \frac{245}{12} \psi^\prime(\third) \alpha^4
- \frac{147}{16} \alpha^5
\right. \nonumber \\
&& \left. ~~~
- \frac{140}{9} \zeta_3 \pi^2 \alpha^4
- \frac{98}{9} \psi^\prime(\third)^2 \alpha^3
- \frac{49}{24} \psi^{\prime\prime\prime}(\third) \alpha
- \frac{35}{2} \pi^2 \alpha^2
- \frac{7}{32} \psi^{\prime\prime\prime}(\third) \alpha^2
\right. \nonumber \\
&& \left. ~~~
+ \frac{1}{4} \psi^{\prime\prime\prime}(\third) \zeta_3 \alpha^2
+ \frac{7}{3} \psi^{\prime\prime\prime}(\third) \zeta_3 \alpha
+ \frac{21}{2} \zeta_3 \alpha^5
+ \frac{21}{2} \psi^{\prime\prime\prime}(\third)
\right. \nonumber \\
&& \left. ~~~
+ \frac{70}{3} \psi^\prime(\third) \zeta_3 \alpha^4
+ \frac{105}{4} \psi^\prime(\third) \alpha^2
+ \frac{112}{9} \psi^\prime(\third)^2 \zeta_3 \alpha^3
+ \frac{133}{4} \pi^4 \alpha^2
\right. \nonumber \\
&& \left. ~~~
+ \frac{147}{2} \psi^\prime(\third)^2 \alpha^2
+ \frac{245}{18} \pi^2 \alpha^4
+ \frac{385}{2} \psi^\prime(\third) \alpha^3
+ \frac{392}{27} \psi^\prime(\third) \pi^2 \alpha^3
\right. \nonumber \\
&& \left. ~~~
+ \frac{440}{3} \zeta_3 \pi^2 \alpha^3
+ \frac{448}{81} \zeta_3 \pi^4 \alpha^3
+ \frac{1280}{3} \zeta_5 \pi^2
+ \frac{1785}{16} \alpha^4
+ \frac{2009}{12} \psi^\prime(\third)^2 \alpha
\right. \nonumber \\
&& \left. ~~~
+ \frac{2156}{27} \pi^4 \alpha
+ \frac{2296}{9} \psi^\prime(\third) \zeta_3 \pi^2 \alpha
+ \frac{2560}{9} \psi^\prime(\third) \zeta_5 \alpha
+ \frac{11215}{12} \zeta_3 \alpha
\right. \nonumber \\
&& \left. ~~~
+ \frac{13126}{27} \psi^\prime(\third) \zeta_3 \alpha
+ \frac{20909}{144} \alpha
+ \frac{78711}{16}
+ \frac{149474}{27} \psi^\prime(\third) \zeta_3
\right. \nonumber \\
&& \left. ~~~
+ \frac{156625}{729} \pi^4
+ \frac{177037}{324} \psi^\prime(\third)^2
+ \frac{202328}{243} \psi^\prime(\third) \zeta_3 \pi^2
+ \frac{207487}{486} \pi^2 \alpha
\right. \nonumber \\
&& \left. ~~~
+ \frac{528991}{162} \pi^2
- 1280 \zeta_5 \alpha
- 640 \psi^\prime(\third) \zeta_5
- 220 \psi^\prime(\third) \zeta_3 \alpha^3
\right. \nonumber \\
&& \left. ~~~
- 98 \psi^\prime(\third) \pi^2 \alpha^2
- 84 \psi^\prime(\third)^2 \zeta_3 \alpha^2
- 38 \zeta_3 \pi^4 \alpha^2
- 36 \zeta_3^2 \alpha^2
\right. \nonumber \\
&& \left. ~~~
- 30 \psi^\prime(\third) \zeta_3 \alpha^2
- 12 \psi^{\prime\prime\prime}(\third) \zeta_3
+ 6 \zeta_3^2 \alpha
+ 20 \zeta_3 \pi^2 \alpha^2
+ 112 \psi^\prime(\third) \zeta_3 \pi^2 \alpha^2
\right. \nonumber \\
&& \left. ~~~
+ 320 \zeta_5 \alpha^2
+ 333 \zeta_3 \alpha^3
\right] a^3 ~+~ O(a^4) 
\label{momgkfn}
\end{eqnarray}
where $\psi(z)$ is the Euler psi function. Its appearance is directly related
to the symmetric point configuration of the triple gluon vertex and also arises
in the other two MOM schemes based on the ghost-gluon ($\MOMc$) and quark-gluon
($\MOMq$) vertices. In \cite{47} the expressions for the renormalization group 
functions additionally involved harmonic polynomials $H^{(2)}_i$ as well as 
\begin{equation}
s_n(z) ~=~ \frac{1}{\sqrt{3}} \mbox{Im} \, \left[ \mbox{Li}_n \left(
\frac{e^{iz}}{\sqrt{3}} \right) \right]
\end{equation}
where $\mbox{Li}_n(z)$ is the polylogarithm function. In \cite{48} it was shown
using the PSLQ algorithm that the actual number basis for MOM schemes was 
rationals, $\pi$, $\zeta_2$, $\zeta_3$, $\zeta_4$, $\zeta_5$, 
$\psi^\prime(\third)$ and $\psi^{\prime\prime\prime}(\third)$. The absence of
harmonic polylogarithms and the polylogarithm function was because they were
not independent of this basis. Indeed we have demonstrated this by verifying 
that the relation 
\begin{eqnarray}
s_2 (\pisix) &=& \frac{1}{11664}
\left[ 324 \sqrt{3} \ln(3) \pi - 27 \sqrt{3} \ln^2(3) \pi + 29 \sqrt{3} \pi^3
- 1944 \psi^\prime(\third)
\right. \nonumber \\
&& \left. ~~~~~~~~~
+ 23328 s_2 (\pitwo) + 19440 s_3 (\pisix) - 15552 s_3 (\pitwo) 
+ 1296 \pi^2 \right]
\label{harmpolrel}
\end{eqnarray}
holds numerically. We have incorporated (\ref{harmpolrel}) in our computations
and therefore any MOM scheme expressions will only involve the basis of 
\cite{48}. We have also established relations similar to the $\MOMg$ ones of 
(\ref{momgkfn}) for the other MOM schemes. Those results as well as the $\MOMg$
ones are available for an arbitrary colour group and $\Nf$~$\neq$~$0$ in the
data file associated with this article. As there was a special value for
$\alpha$ in the $\mMOM$ case that means the $K_\alpha$ function was absent at
leading order we have examined $K_\alpha^{\MOMgs}(a,\alpha)$ for $SU(3)$ to see
if the same or another value of $\alpha$ arises to produce a zero leading order
term of the $\MOMg$ $K_\alpha$ function. For $\Nf$~$\neq$~$0$ this term is 
$\Nf$ independent but unlike the $\mMOM$ case it is quadratic in $\alpha$.
Although there are analytic solutions for $\alpha$ when the leading term
vanishes the numerical values are $\alpha$~$=$~$-$~$1.617608$ and $2.492398$.
The fact that neither values are integer is solely due to the presence of
$\psi(\third)$, $\zeta_3$ and $\pi$ contributions which result from the
renormalization prescription defining each MOM scheme and the particular
external momentum configuration of the $3$-point vertices where the
renormalization constants are defined. For the $\MOMc$ and $\MOMq$ schemes the 
respective values of $\alpha$ are $-$~$2.358904$ and $-$~$7.145900$ as both 
leading order $K_\alpha$ functions are linear in the gauge parameter.  

Having concentrated on the linear covariant gauge to this point we turn our
attention to two nonlinear gauges which are the Curci-Ferrari gauge \cite{49}
and the MAG, \cite{50,51,52}. However both gauges are not unrelated. The MAG is
a gauge fixing where the gluon fields are partitioned into those whose 
associated group generator form an abelian subgroup of the non-abelian gauge 
group and the remainder. The former are referred to as the diagonal gluons and 
there are $\Nda$ such fields while there are $\Noda$ remaining off-diagonal 
fields where $\Nda$~$=$~$\Nc-1$ and $\Noda$~$=$~$\Nc(\Nc-1)$ for the group 
$SU(\Nc)$. The gauge fixing functional takes a different form in both sectors 
where the diagonal gluons are fixed in the Landau gauge but the $\Noda$ gluons 
have a nonlinear functional that involves the ghosts associated with the 
off-diagonal gluons. These ghost fields differ from those of the abelian 
subgroup gluons. The MAG has several similarities with the background field 
gauge in that the diagonal gluons behave like background field gluons. The 
Curci-Ferrari gauge is different from the MAG in that the gluons are not 
distinguished in the gauge fixing but do share a similar quadratic ghost term
in the gauge fixing functional. Both gauges have quartic ghost interactions 
that have no analogue in the linear covariant gauge fixing. What does connect 
the MAG to the Curci-Ferrari gauge, however, is the fact that taking the formal 
$\Nda$~$\to$~$0$ limit produces the Curci-Ferrari gauge. This is of practical 
benefit as a check on any computations involving the MAG since the 
Curci-Ferrari gauge expressions have to be reproduced as $\Nda$~$\to$~$0$. 

Although both the Bjorken sum rule and the Adler $D$-function were originally
computed perturbatively to high loop order in the linear gauge, since the
operators involved in the underlying field theory formalism are gauge
independent the $\MSbar$ expressions for both quantities will be the same in
either the Curci-Ferrari gauge or MAG. Therefore to determine the extension
to the Crewther relation we have carried out the mapping of the $\MSbar$
coupling constant to the Curci-Ferrari coupling in the $\MOMg$ scheme as well
as also the $\MOMc$ and $\MOMq$ schemes. As a check on this procedure,
however, we carried out the explicit direct computation of the Adler 
$D$-function in both the Curci-Ferrari gauge and the MAG to high loop order.
This could only be achieved by using the {\sc Forcer} package, \cite{53,54},
written in the symbolic manipulation language {\sc Form}, \cite{55,56}. The 
{\sc Forcer} algorithm evaluates massless $2$-point Feynman graphs up to four 
loops very efficiently. In carrying out this exercise it is important to 
recognize that in the Curci-Ferrari gauge there are several extra Feynman 
graphs to be evaluated for the $D$-function than the linear gauge due to the 
extra quartic ghost interaction. By contrast in the case of the MAG there are 
considerably more diagrams to evaluate compared to the other two gauges not 
only due to quartic ghost interactions but because there are twice as many 
gluon and ghost fields. These have intricately connected interactions. On top 
of this one has to implement a routine to handle the colour group factors 
associated with gluon and ghost fields as they lie in two sectors. So the group
generators and structure functions obey a more complicated algebra. Useful in 
this context were the identities used in the three loop $\MSbar$ 
renormalization of the MAG, \cite{57}. The overall outcome was that the 
$\MSbar$ Adler $D$-function resulted as expected. Once that was established for
both gauges it was a simple exercise to change the renormalization of the 
$D$-function to another scheme such as the $\MOMg$ one for both gauges. 
Consequently this checked that the direct application of the mapping from the 
linear gauge expression was consistent. The various coupling constant mappings 
were orginally determined in \cite{58}. 

Having summarized the background to both gauges we have managed to construct
the respective $K$-functions in keeping with (\ref{deltgen}). The expressions
for the Curci-Ferrari (CF) gauge are not dissimilar to those of the linear 
covariant gauge since
\begin{eqnarray}
\left. K_{a\,\CFs}^{\MOMgs}(a,\alpha) \right|^{SU(3)}_{\Nf=0} &=& 
\left[
16 \zeta_3
- 14
\right] a
\nonumber \\
&&
+ \left[
28 \pi^2 \alpha
+ 48 \zeta_3 \alpha^2
+ 48 \psi^{\prime}(\third) \zeta_3 \alpha
- \frac{14158}{27}
- \frac{1280}{3} \zeta_5
- \frac{644}{27} \pi^2
\right. \nonumber \\
&& \left. ~~~
- \frac{368}{9} \psi^{\prime}(\third) \zeta_3
- \frac{56}{9} \pi^2 \alpha^2
- \frac{32}{3} \psi^{\prime}(\third) \zeta_3 \alpha^2
+ \frac{28}{3} \psi^{\prime}(\third) \alpha^2
+ \frac{64}{9} \zeta_3 \pi^2 \alpha^2
\right. \nonumber \\
&& \left. ~~~
+ \frac{322}{9} \psi^{\prime}(\third)
+ \frac{736}{27} \zeta_3 \pi^2
+ \frac{6448}{9} \zeta_3
- 72 \zeta_3 \alpha
- 42 \alpha^2
- 42 \psi^{\prime}(\third) \alpha
\right. \nonumber \\
&& \left. ~~~
- 32 \zeta_3 \pi^2 \alpha
- 8 \zeta_3 \alpha^3
+ 7 \alpha^3
+ 63 \alpha
\right] a^2
\nonumber \\
&&
+ \left[
\frac{22400}{3} \zeta_7
- \frac{127914947}{2592}
- \frac{696830}{243} \pi^2
- \frac{562472}{81} \zeta_3 \pi^2 \alpha
\right. \nonumber \\
&& \left. ~~~
- \frac{491041}{54} \psi^{\prime}(\third) \alpha
- \frac{402143}{243} \psi^{\prime}(\third) \pi^2 \alpha
- \frac{389608}{729} \zeta_3 \pi^4 \alpha
\right. \nonumber \\
&& \left. ~~~
- \frac{238097}{324} \pi^2 \alpha^2
- \frac{164461}{96} \alpha^2
- \frac{144856}{27} \psi^{\prime}(\third) \zeta_3
- \frac{114898}{81} \psi^{\prime}(\third)^2 \zeta_3 \alpha
\right. \nonumber \\
&& \left. ~~~
- \frac{54362}{243} \zeta_3 \pi^4
- \frac{32389}{216} \psi^{\prime}(\third)^2 \alpha^2
- \frac{29948}{3} \zeta_3 \alpha
- \frac{29440}{27} \zeta_5 \pi^2
\right. \nonumber \\
&& \left. ~~~
- \frac{25921}{216} \psi^{\prime}(\third)^2
- \frac{18508}{81} \psi^{\prime}(\third) \zeta_3 \pi^2 \alpha^2
- \frac{14812}{81} \psi^{\prime}(\third) \zeta_3 \pi^2
\right. \nonumber \\
&& \left. ~~~
- \frac{13643}{243} \pi^4 \alpha^2
- \frac{12761}{9} \psi^{\prime}(\third) \zeta_3 \alpha^2
- \frac{5415}{32} \alpha^4
- \frac{2989}{32} \psi^{\prime\prime\prime}(\third)
\right. \nonumber \\
&& \left. ~~~
- \frac{2560}{9} \zeta_5 \pi^2 \alpha^2
- \frac{2236}{81} \zeta_3 \pi^4 \alpha^3
- \frac{1862}{27} \psi^{\prime}(\third) \pi^2 \alpha^3
- \frac{875}{12} \pi^2 \alpha^4
\right. \nonumber \\
&& \left. ~~~
- \frac{841}{3} \zeta_3 \alpha^3
- \frac{545}{12} \psi^{\prime}(\third) \alpha^3
- \frac{532}{9} \psi^{\prime}(\third)^2 \zeta_3 \alpha^3
- \frac{380}{9} \zeta_3 \pi^2 \alpha^3
\right. \nonumber \\
&& \left. ~~~
- \frac{153}{2} \zeta_3 \alpha^5
- \frac{147}{32} \alpha^6
- \frac{112}{9} \psi^{\prime}(\third) \zeta_3 \pi^2 \alpha^4
- \frac{98}{27} \pi^4 \alpha^4
- \frac{63}{16} \psi^{\prime\prime\prime}(\third) \alpha^2
\right. \nonumber \\
&& \left. ~~~
- \frac{49}{4} \psi^{\prime}(\third) \alpha^5
- \frac{49}{6} \psi^{\prime}(\third)^2 \alpha^4
- \frac{28}{3} \zeta_3 \pi^2 \alpha^5
- \frac{7}{16} \psi^{\prime\prime\prime}(\third) \alpha^3
\right. \nonumber \\
&& \left. ~~~
+ \frac{1}{2} \psi^{\prime\prime\prime}(\third) \zeta_3 \alpha^3
+ \frac{9}{2} \psi^{\prime\prime\prime}(\third) \zeta_3 \alpha^2
+ \frac{21}{4} \zeta_3 \alpha^6
+ \frac{28}{3} \psi^{\prime}(\third)^2 \zeta_3 \alpha^4
\right. \nonumber \\
&& \left. ~~~
+ \frac{49}{6} \pi^2 \alpha^5
+ \frac{63}{2} \psi^{\prime\prime\prime}(\third) \alpha
+ \frac{98}{9} \psi^{\prime}(\third) \pi^2 \alpha^4
+ \frac{112}{27} \zeta_3 \pi^4 \alpha^4
\right. \nonumber \\
&& \left. ~~~
+ \frac{190}{3} \psi^{\prime}(\third) \zeta_3 \alpha^3
+ \frac{250}{3} \zeta_3 \pi^2 \alpha^4
+ \frac{427}{4} \psi^{\prime\prime\prime}(\third) \zeta_3
+ \frac{545}{18} \pi^2 \alpha^3
\right. \nonumber \\
&& \left. ~~~
+ \frac{825}{4} \zeta_3 \alpha^4
+ \frac{875}{8} \psi^{\prime}(\third) \alpha^4
+ \frac{931}{18} \psi^{\prime}(\third)^2 \alpha^3
+ \frac{1071}{16} \alpha^5
\right. \nonumber \\
&& \left. ~~~
+ \frac{1280}{3} \psi^{\prime}(\third) \zeta_5 \alpha^2
+ \frac{2128}{27} \psi^{\prime}(\third) \zeta_3 \pi^2 \alpha^3
+ \frac{2917}{36} \alpha^3
+ \frac{3703}{27} \psi^{\prime}(\third)^2 \zeta_3
\right. \nonumber \\
&& \left. ~~~
+ \frac{3913}{162} \pi^4 \alpha^3
+ \frac{4627}{27} \psi^{\prime}(\third)^2 \zeta_3 \alpha^2
+ \frac{5279}{2} \zeta_3 \alpha^2
+ \frac{14720}{9} \psi^{\prime}(\third) \zeta_5
\right. \nonumber \\
&& \left. ~~~
+ \frac{15592}{243} \zeta_3 \pi^4 \alpha^2
+ \frac{25522}{27} \zeta_3 \pi^2 \alpha^2
+ \frac{25921}{162} \psi^{\prime}(\third) \pi^2
\right. \nonumber \\
&& \left. ~~~
+ \frac{32389}{162} \psi^{\prime}(\third) \pi^2 \alpha^2
+ \frac{190267}{972} \pi^4
+ \frac{238097}{216} \psi^{\prime}(\third) \alpha^2
\right. \nonumber \\
&& \left. ~~~
+ \frac{281236}{27} \psi^{\prime}(\third) \zeta_3 \alpha
+ \frac{289712}{81} \zeta_3 \pi^2
+ \frac{340907}{729} \pi^4 \alpha
+ \frac{348415}{81} \psi^{\prime}(\third)
\right. \nonumber \\
&& \left. ~~~
+ \frac{402143}{324} \psi^{\prime}(\third)^2 \alpha
+ \frac{433885}{48} \alpha
+ \frac{459592}{243} \psi^{\prime}(\third) \zeta_3 \pi^2 \alpha
+ \frac{485200}{27} \zeta_5
\right. \nonumber \\
&& \left. ~~~
+ \frac{491041}{81} \pi^2 \alpha
+ \frac{9840343}{216} \zeta_3
- 23325 \zeta_3^2
- 1920 \zeta_5 \alpha^2
\right. \nonumber \\
&& \left. ~~~
- 1920 \psi^{\prime}(\third) \zeta_5 \alpha
- 125 \psi^{\prime}(\third) \zeta_3 \alpha^4
- 72 \zeta_3^2 \alpha^3
- 54 \zeta_3^2 \alpha^2
\right. \nonumber \\
&& \left. ~~~
- 36 \psi^{\prime\prime\prime}(\third) \zeta_3 \alpha
+ 14 \psi^{\prime}(\third) \zeta_3 \alpha^5
+ 320 \zeta_5 \alpha^3
+ 1280 \zeta_5 \pi^2 \alpha
+ 2052 \zeta_3^2 \alpha
\right. \nonumber \\
&& \left. ~~~
+ 2880 \zeta_5 \alpha
\right] a^3 ~+~ O(a^4)
\end{eqnarray}
and
\begin{eqnarray}
\left. K_{\alpha\,\CFs}^{\MOMgs}(a,\alpha) \right|^{SU(3)}_{\Nf=0} &=& 
\left[
\frac{63}{2}
+ \frac{64}{9} \zeta_3 \pi^2 \alpha
- \frac{56}{9} \pi^2 \alpha
- \frac{32}{3} \psi^{\prime}(\third) \zeta_3 \alpha
+ \frac{21}{2} \alpha^2
+ \frac{28}{3} \psi^{\prime}(\third) \alpha
- 42 \alpha
\right. \nonumber \\
&& \left. ~
- 36 \zeta_3
- 21 \psi^{\prime}(\third)
- 16 \zeta_3 \pi^2
- 12 \zeta_3 \alpha^2
+ 14 \pi^2
+ 24 \psi^{\prime}(\third) \zeta_3
+ 48 \zeta_3 \alpha
\right] a^2
\nonumber \\
&&
+ \left[
\frac{1}{2} \psi^{\prime\prime\prime}(\third) \zeta_3 \alpha^2
- \frac{1149479}{108} \psi^{\prime}(\third)
- \frac{620324}{81} \zeta_3 \pi^2
- \frac{612664}{729} \zeta_3 \pi^4
\right. \nonumber \\
&& \left. ~~~
- \frac{556493}{243} \psi^{\prime}(\third) \pi^2
- \frac{158998}{81} \psi^{\prime}(\third)^2 \zeta_3
- \frac{52297}{81} \psi^{\prime}(\third) \alpha
\right. \nonumber \\
&& \left. ~~~
- \frac{26576}{81} \zeta_3 \pi^2 \alpha
- \frac{22049}{3} \zeta_3
- \frac{5120}{27} \zeta_5 \pi^2 \alpha
- \frac{2512}{27} \zeta_3 \pi^4 \alpha
- \frac{2079}{8} \alpha^3
\right. \nonumber \\
&& \left. ~~~
- \frac{2009}{9} \psi^{\prime}(\third) \pi^2 \alpha
- \frac{715}{3} \zeta_3 \alpha^2
- \frac{574}{3} \psi^{\prime}(\third)^2 \zeta_3 \alpha
- \frac{448}{27} \psi^{\prime}(\third) \zeta_3 \pi^2 \alpha^3
\right. \nonumber \\
&& \left. ~~~
- \frac{392}{81} \pi^4 \alpha^3
- \frac{255}{2} \zeta_3 \alpha^4
- \frac{245}{12} \psi^{\prime}(\third) \alpha^4
- \frac{147}{16} \alpha^5
- \frac{140}{9} \zeta_3 \pi^2 \alpha^4
\right. \nonumber \\
&& \left. ~~~
- \frac{116}{3} \zeta_3 \pi^4 \alpha^2
- \frac{98}{9} \psi^{\prime}(\third)^2 \alpha^3
- \frac{21}{4} \psi^{\prime}(\third) \alpha^2
- \frac{21}{8} \psi^{\prime\prime\prime}(\third) \alpha
\right. \nonumber \\
&& \left. ~~~
- \frac{7}{16} \psi^{\prime\prime\prime}(\third) \alpha^2
+ \frac{7}{2} \pi^2 \alpha^2
+ \frac{21}{2} \zeta_3 \alpha^5
+ \frac{21}{2} \psi^{\prime\prime\prime}(\third)
+ \frac{70}{3} \psi^{\prime}(\third) \zeta_3 \alpha^4
\right. \nonumber \\
&& \left. ~~~
+ \frac{112}{9} \psi^{\prime}(\third)^2 \zeta_3 \alpha^3
+ \frac{147}{2} \psi^{\prime}(\third)^2 \alpha^2
+ \frac{196}{9} \alpha^2
+ \frac{203}{6} \pi^4 \alpha^2
+ \frac{245}{18} \pi^2 \alpha^4
\right. \nonumber \\
&& \left. ~~~
+ \frac{357}{2} \psi^{\prime}(\third) \alpha^3
+ \frac{392}{27} \psi^{\prime}(\third) \pi^2 \alpha^3
+ \frac{448}{81} \zeta_3 \pi^4 \alpha^3
+ \frac{1280}{3} \zeta_5 \pi^2
\right. \nonumber \\
&& \left. ~~~
+ \frac{1785}{16} \alpha^4
+ \frac{2009}{12} \psi^{\prime}(\third)^2 \alpha
+ \frac{2198}{27} \pi^4 \alpha
+ \frac{2296}{9} \psi^{\prime}(\third) \zeta_3 \pi^2 \alpha
\right. \nonumber \\
&& \left. ~~~
+ \frac{2560}{9} \psi^{\prime}(\third) \zeta_5 \alpha
+ \frac{3076}{3} \zeta_3 \alpha
+ \frac{13288}{27} \psi^{\prime}(\third) \zeta_3 \alpha
+ \frac{14105}{144} \alpha
\right. \nonumber \\
&& \left. ~~~
+ \frac{104594}{243} \pi^2 \alpha
+ \frac{310162}{27} \psi^{\prime}(\third) \zeta_3
+ \frac{481037}{48}
+ \frac{536081}{729} \pi^4
\right. \nonumber \\
&& \left. ~~~
+ \frac{556493}{324} \psi^{\prime}(\third)^2
+ \frac{635992}{243} \psi^{\prime}(\third) \zeta_3 \pi^2
+ \frac{1149479}{162} \pi^2
- 1280 \zeta_5 \alpha
\right. \nonumber \\
&& \left. ~~~
- 640 \psi^{\prime}(\third) \zeta_5
- 204 \psi^{\prime}(\third) \zeta_3 \alpha^3
- 119 \pi^2 \alpha^3
- 98 \psi^{\prime}(\third) \pi^2 \alpha^2
\right. \nonumber \\
&& \left. ~~~
- 84 \psi^{\prime}(\third)^2 \zeta_3 \alpha^2
- 72 \zeta_3^2 \alpha^2
- 36 \zeta_3^2 \alpha
- 12 \psi^{\prime\prime\prime}(\third) \zeta_3
- 4 \zeta_3 \pi^2 \alpha^2
\right. \nonumber \\
&& \left. ~~~
+ 3 \psi^{\prime\prime\prime}(\third) \zeta_3 \alpha
+ 6 \psi^{\prime}(\third) \zeta_3 \alpha^2
+ 112 \psi^{\prime}(\third) \zeta_3 \pi^2 \alpha^2
+ 136 \zeta_3 \pi^2 \alpha^3
\right. \nonumber \\
&& \left. ~~~
+ 297 \zeta_3 \alpha^3
+ 320 \zeta_5 \alpha^2
+ 684 \zeta_3^2
+ 960 \zeta_5
\right] a^3 ~+~ O(a^4)
\end{eqnarray}
for $SU(3)$ Yang-Mills. For the MAG a similar picture emerges since
\begin{eqnarray}
\left. K_{a\,\MAGs}^{\MOMgs}(a,\alpha) \right|^{SU(3)}_{\Nf=0} &=& 
\left[
16 \zeta_3
- 14
\right] a
\nonumber \\
&&
+ \left[
\frac{6448}{9} \zeta_3
- \frac{14158}{27}
- \frac{1280}{3} \zeta_5
- \frac{704}{27} \zeta_3 \pi^2 \alpha
- \frac{512}{27} \zeta_3 \pi^2
- \frac{308}{9} \psi^{\prime}(\third) \alpha
\right. \nonumber \\
&& \left. ~~~
- \frac{224}{9} \psi^{\prime}(\third)
- \frac{56}{27} \pi^2 \alpha^2
- \frac{32}{9} \psi^{\prime}(\third) \zeta_3 \alpha^2
+ \frac{28}{9} \psi^{\prime}(\third) \alpha^2
+ \frac{64}{27} \zeta_3 \pi^2 \alpha^2
\right. \nonumber \\
&& \left. ~~~
+ \frac{256}{9} \psi^{\prime}(\third) \zeta_3
+ \frac{352}{9} \psi^{\prime}(\third) \zeta_3 \alpha
+ \frac{448}{27} \pi^2
+ \frac{616}{27} \pi^2 \alpha
- 48 \zeta_3 \alpha
- 14 \alpha^2
\right. \nonumber \\
&& \left. ~~~
+ 16 \zeta_3 \alpha^2
+ 42 \alpha
\right] a^2
\nonumber \\
&&
+ \left[
\frac{25291531}{540} \zeta_3
- \frac{640705981}{12960}
- \frac{9335291}{3240} \psi^{\prime}(\third) \alpha
- \frac{5544823}{4860} \pi^2
\right. \nonumber \\
&& \left. ~~~
- \frac{1402261}{7290} \pi^4 \alpha
- \frac{448672}{1215} \psi^{\prime}(\third) \zeta_3 \pi^2 \alpha
- \frac{224123}{135} \psi^{\prime}(\third) \zeta_3
\right. \nonumber \\
&& \left. ~~~
- \frac{169595}{972} \pi^2 \alpha^2
- \frac{151669}{30} \zeta_3 \alpha
- \frac{112718}{45} \zeta_3 \pi^2 \alpha
- \frac{98147}{405} \psi^{\prime}(\third)^2 \alpha
\right. \nonumber \\
&& \left. ~~~
- \frac{22883}{36} \alpha^2
- \frac{14080}{9} \psi^{\prime}(\third) \zeta_5 \alpha
- \frac{11896}{45} \zeta_3 \pi^4
- \frac{10240}{9} \psi^{\prime}(\third) \zeta_5
\right. \nonumber \\
&& \left. ~~~
- \frac{9799}{27} \psi^{\prime}(\third) \zeta_3 \alpha^2
- \frac{8155}{243} \pi^4 \alpha^2
- \frac{6944}{81} \psi^{\prime}(\third) \zeta_3 \pi^2 \alpha^2
- \frac{5135}{216} \psi^{\prime}(\third) \alpha^3
\right. \nonumber \\
&& \left. ~~~
- \frac{2560}{27} \zeta_5 \pi^2 \alpha^2
- \frac{1940}{729} \zeta_3 \pi^4 \alpha^3
- \frac{1810}{81} \zeta_3 \pi^2 \alpha^3
- \frac{1792}{243} \psi^{\prime}(\third) \pi^2 \alpha^3
\right. \nonumber \\
&& \left. ~~~
- \frac{1519}{27} \psi^{\prime}(\third)^2 \alpha^2
- \frac{987}{32} \alpha^4
- \frac{944}{15} \psi^{\prime}(\third) \zeta_3 \pi^2
- \frac{749}{8} \psi^{\prime\prime\prime}(\third)
\right. \nonumber \\
&& \left. ~~~
- \frac{653}{18} \psi^{\prime\prime\prime}(\third) \zeta_3 \alpha
- \frac{512}{81} \psi^{\prime}(\third)^2 \zeta_3 \alpha^3
- \frac{451}{48} \alpha^3
- \frac{413}{10} \psi^{\prime}(\third)^2
\right. \nonumber \\
&& \left. ~~~
- \frac{112}{81} \psi^{\prime}(\third) \zeta_3 \pi^2 \alpha^4
- \frac{98}{243} \pi^4 \alpha^4
- \frac{49}{54} \psi^{\prime}(\third)^2 \alpha^4
- \frac{11}{3} \psi^{\prime\prime\prime}(\third) \zeta_3 \alpha^2
\right. \nonumber \\
&& \left. ~~~
- \frac{1}{18} \psi^{\prime\prime\prime}(\third) \zeta_3 \alpha^3
+ \frac{7}{144} \psi^{\prime\prime\prime}(\third) \alpha^3
+ \frac{17}{12} \zeta_3 \alpha^3
+ \frac{21}{2} \psi^{\prime}(\third) \alpha^4
\right. \nonumber \\
&& \left. ~~~
+ \frac{28}{27} \psi^{\prime}(\third)^2 \zeta_3 \alpha^4
+ \frac{77}{24} \psi^{\prime\prime\prime}(\third) \alpha^2
+ \frac{98}{81} \psi^{\prime}(\third) \pi^2 \alpha^4
+ \frac{112}{243} \zeta_3 \pi^4 \alpha^4
\right. \nonumber \\
&& \left. ~~~
+ \frac{183}{4} \zeta_3 \alpha^4
+ \frac{236}{5} \psi^{\prime}(\third)^2 \zeta_3
+ \frac{448}{81} \psi^{\prime}(\third)^2 \alpha^3
+ \frac{826}{15} \psi^{\prime}(\third) \pi^2
\right. \nonumber \\
&& \left. ~~~
+ \frac{905}{27} \psi^{\prime}(\third) \zeta_3 \alpha^3
+ \frac{1280}{9} \psi^{\prime}(\third) \zeta_5 \alpha^2
+ \frac{1736}{27} \psi^{\prime}(\third)^2 \zeta_3 \alpha^2
\right. \nonumber \\
&& \left. ~~~
+ \frac{2048}{243} \psi^{\prime}(\third) \zeta_3 \pi^2 \alpha^3
+ \frac{3395}{1458} \pi^4 \alpha^3
+ \frac{4571}{144} \psi^{\prime\prime\prime}(\third) \alpha
+ \frac{5135}{324} \pi^2 \alpha^3
\right. \nonumber \\
&& \left. ~~~
+ \frac{5311}{12} \zeta_3 \alpha^2
+ \frac{6076}{81} \psi^{\prime}(\third) \pi^2 \alpha^2
+ \frac{9320}{243} \zeta_3 \pi^4 \alpha^2
+ \frac{10409}{45} \pi^4
\right. \nonumber \\
&& \left. ~~~
+ \frac{19598}{81} \zeta_3 \pi^2 \alpha^2
+ \frac{20480}{27} \zeta_5 \pi^2
+ \frac{22400}{3} \zeta_7
+ \frac{28160}{27} \zeta_5 \pi^2 \alpha
\right. \nonumber \\
&& \left. ~~~
+ \frac{56359}{15} \psi^{\prime}(\third) \zeta_3 \alpha
+ \frac{112168}{405} \psi^{\prime}(\third)^2 \zeta_3 \alpha
+ \frac{169595}{648} \psi^{\prime}(\third) \alpha^2
\right. \nonumber \\
&& \left. ~~~
+ \frac{392588}{1215} \psi^{\prime}(\third) \pi^2 \alpha
+ \frac{448246}{405} \zeta_3 \pi^2
+ \frac{485200}{27} \zeta_5
+ \frac{695693}{240} \alpha
\right. \nonumber \\
&& \left. ~~~
+ \frac{801292}{3645} \zeta_3 \pi^4 \alpha
+ \frac{5544823}{3240} \psi^{\prime}(\third)
+ \frac{9335291}{4860} \pi^2 \alpha
- 24900 \zeta_3^2
\right. \nonumber \\
&& \left. ~~~
- 640 \zeta_5 \alpha^2
- 22 \zeta_3^2 \alpha^3
- 12 \zeta_3^2 \alpha^4
- 12 \psi^{\prime}(\third) \zeta_3 \alpha^4
- 7 \pi^2 \alpha^4
+ 8 \zeta_3 \pi^2 \alpha^4
\right. \nonumber \\
&& \left. ~~~
+ 107 \psi^{\prime\prime\prime}(\third) \zeta_3
+ 522 \zeta_3^2 \alpha^2
+ 1576 \zeta_3^2 \alpha
+ 1920 \zeta_5 \alpha
\right] a^3
\end{eqnarray}
and
\begin{eqnarray}
\left. K_{\alpha\,\MAGs}^{\MOMgs}(a,\alpha) \right|^{SU(3)}_{\Nf=0} &=&
\left[
21
- \frac{352}{27} \zeta_3 \pi^2
- \frac{154}{9} \psi^{\prime}(\third)
- \frac{56}{27} \pi^2 \alpha
- \frac{32}{9} \psi^{\prime}(\third) \zeta_3 \alpha
+ \frac{28}{9} \psi^{\prime}(\third) \alpha
\right. \nonumber \\
&& \left. ~
+ \frac{64}{27} \zeta_3 \pi^2 \alpha
+ \frac{176}{9} \psi^{\prime}(\third) \zeta_3
+ \frac{308}{27} \pi^2
- 24 \zeta_3
- 14 \alpha
+ 16 \zeta_3 \alpha
\right] a^2
\nonumber \\
&&
+ \left[
640 \zeta_5
- \frac{16773443}{9720} \psi^{\prime}(\third)
- \frac{1763246}{1215} \zeta_3 \pi^2
- \frac{775271}{7290} \pi^4
\right. \nonumber \\
&& \left. ~~~
- \frac{325472}{1215} \psi^{\prime}(\third) \zeta_3 \pi^2
- \frac{289513}{972} \psi^{\prime}(\third) \alpha
- \frac{71197}{405} \psi^{\prime}(\third)^2
- \frac{64139}{30} \zeta_3
\right. \nonumber \\
&& \left. ~~~
- \frac{41876}{243} \zeta_3 \pi^2 \alpha
- \frac{14080}{27} \psi^{\prime}(\third) \zeta_5
- \frac{12152}{243} \psi^{\prime}(\third) \pi^2 \alpha
- \frac{9136}{729} \zeta_3 \pi^4 \alpha
\right. \nonumber \\
&& \left. ~~~
- \frac{5120}{81} \zeta_5 \pi^2 \alpha
- \frac{3472}{81} \psi^{\prime}(\third)^2 \zeta_3 \alpha
- \frac{2140}{243} \zeta_3 \pi^4 \alpha^2
- \frac{1904}{81} \psi^{\prime}(\third) \pi^2 \alpha^2
\right. \nonumber \\
&& \left. ~~~
- \frac{1280}{3} \zeta_5 \alpha
- \frac{665}{8} \psi^{\prime}(\third) \alpha^2
- \frac{653}{54} \psi^{\prime\prime\prime}(\third) \zeta_3
- \frac{544}{27} \psi^{\prime}(\third)^2 \zeta_3 \alpha^2
\right. \nonumber \\
&& \left. ~~~
- \frac{448}{243} \psi^{\prime}(\third) \zeta_3 \pi^2 \alpha^3
- \frac{392}{729} \pi^4 \alpha^3
- \frac{329}{8} \alpha^3
- \frac{190}{3} \zeta_3 \pi^2 \alpha^2
- \frac{121}{4} \zeta_3 \alpha^2
\right. \nonumber \\
&& \left. ~~~
- \frac{98}{81} \psi^{\prime}(\third)^2 \alpha^3
- \frac{28}{3} \pi^2 \alpha^3
- \frac{22}{9} \psi^{\prime\prime\prime}(\third) \zeta_3 \alpha
- \frac{1}{18} \psi^{\prime\prime\prime}(\third) \zeta_3 \alpha^2
\right. \nonumber \\
&& \left. ~~~
+ \frac{7}{144} \psi^{\prime\prime\prime}(\third) \alpha^2
+ \frac{32}{3} \zeta_3 \pi^2 \alpha^3
+ \frac{77}{36} \psi^{\prime\prime\prime}(\third) \alpha
+ \frac{112}{81} \psi^{\prime}(\third)^2 \zeta_3 \alpha^3
\right. \nonumber \\
&& \left. ~~~
+ \frac{392}{243} \psi^{\prime}(\third) \pi^2 \alpha^3
+ \frac{448}{729} \zeta_3 \pi^4 \alpha^3
+ \frac{476}{27} \psi^{\prime}(\third)^2 \alpha^2
+ \frac{665}{12} \pi^2 \alpha^2
\right. \nonumber \\
&& \left. ~~~
+ \frac{693}{16} \alpha^2
+ \frac{707}{18} \zeta_3 \alpha
+ \frac{1576}{3} \zeta_3^2
+ \frac{2176}{81} \psi^{\prime}(\third) \zeta_3 \pi^2 \alpha^2
+ \frac{2345}{108} \alpha
\right. \nonumber \\
&& \left. ~~~
+ \frac{2560}{27} \psi^{\prime}(\third) \zeta_5 \alpha
+ \frac{3038}{81} \psi^{\prime}(\third)^2 \alpha
+ \frac{3745}{486} \pi^4 \alpha^2
+ \frac{4571}{432} \psi^{\prime\prime\prime}(\third)
\right. \nonumber \\
&& \left. ~~~
+ \frac{7994}{729} \pi^4 \alpha
+ \frac{13888}{243} \psi^{\prime}(\third) \zeta_3 \pi^2 \alpha
+ \frac{20938}{81} \psi^{\prime}(\third) \zeta_3 \alpha
+ \frac{28160}{81} \zeta_5 \pi^2
\right. \nonumber \\
&& \left. ~~~
- 16 \psi^{\prime}(\third) \zeta_3 \alpha^3
+ \frac{81368}{405} \psi^{\prime}(\third)^2 \zeta_3
+ \frac{284788}{1215} \psi^{\prime}(\third) \pi^2
+ \frac{289513}{1458} \pi^2 \alpha
\right. \nonumber \\
&& \left. ~~~
+ \frac{443012}{3645} \zeta_3 \pi^4
+ \frac{881623}{405} \psi^{\prime}(\third) \zeta_3
+ \frac{1405969}{720}
+ \frac{16773443}{14580} \pi^2
\right. \nonumber \\
&& \left. ~~~
- 22 \zeta_3^2 \alpha^2
- 16 \zeta_3^2 \alpha^3
+ 14 \psi^{\prime}(\third) \alpha^3
+ 61 \zeta_3 \alpha^3
+ 95 \psi^{\prime}(\third) \zeta_3 \alpha^2
\right. \nonumber \\
&& \left. ~~~
+ 348 \zeta_3^2 \alpha
\right] a^3 ~+~ O(a^4) 
\end{eqnarray}
for the same theory. What is evident in both cases is that the leading term of
the $\MOMg$ $K_a$ in both gauges, as well as the $\MOMc$ and $\MOMq$ schemes, 
is the same as that of the linear gauge in all schemes. In other words the 
remaining terms of $K_a$ and all in $K_\alpha$  carry both the scheme and gauge
dependence. Moreover by construction the Crewther relation vanishes at the 
corresponding fixed points of the Curci-Ferrari and MAG renormalization group 
functions. In these two nonlinear gauges $\gamma_\alpha(a,\alpha)$ is linearly 
independent of the gluon anomalous dimension $\gamma_A(a,\alpha)$ unlike the 
linear covariant gauge. In other words the critical points of both gauges are 
different.

What is not apparent from the $SU(3)$ expressions is the connection between the
Curci-Ferrari gauge and MAG which requires the results for an arbitrary
colour group. To assist with seeing how the $\Nda$~$\to$~$0$ limit takes effect
we record the situation for the $O(a^2)$ $K$-functions in detail as the next
order expressions are large but available in the associated data file. In the 
Curci-Ferrari gauge we have
\begin{eqnarray}
K_{a\,\CFs}^{\MOMgs}(a,\alpha) &=& 
\left[
12 \zeta_3 C_F
- \frac{21}{2} C_F
\right] a
\nonumber \\
&&
+ \left[
136 \zeta_3 C_F^2
- \frac{512}{27} \zeta_3 \pi^2 \Nf T_F C_F
- \frac{321}{2} C_F C_A
- \frac{224}{9} \psi^{\prime}(\third) \Nf T_F C_F
\right. \nonumber \\
&& \left. ~~~
- \frac{161}{27} \pi^2 C_F C_A
- \frac{112}{3} \zeta_3 \Nf T_F C_F
- \frac{92}{9} \psi^{\prime}(\third) \zeta_3 C_F C_A
- \frac{21}{2} C_F C_A \alpha^2
\right. \nonumber \\
&& \left. ~~~
- \frac{21}{2} \psi^{\prime}(\third) C_F C_A \alpha
- \frac{14}{9} \pi^2 C_F C_A \alpha^2
- \frac{8}{3} \psi^{\prime}(\third) \zeta_3 C_F C_A \alpha^2
\right. \nonumber \\
&& \left. ~~~
+ \frac{7}{3} \psi^{\prime}(\third) C_F C_A \alpha^2
+ \frac{7}{4} C_F C_A \alpha^3
+ \frac{16}{9} \zeta_3 \pi^2 C_F C_A \alpha^2
+ \frac{63}{4} C_F C_A \alpha
\right. \nonumber \\
&& \left. ~~~
+ \frac{158}{3} \Nf T_F C_F
+ \frac{161}{18} \psi^{\prime}(\third) C_F C_A
+ \frac{184}{27} \zeta_3 \pi^2 C_F C_A
\right. \nonumber \\
&& \left. ~~~
+ \frac{256}{9} \psi^{\prime}(\third) \zeta_3 \Nf T_F C_F
+ \frac{356}{3} \zeta_3 C_F C_A
+ \frac{397}{6} C_F^2
+ \frac{448}{27} \pi^2 \Nf T_F C_F
\right. \nonumber \\
&& \left. ~~~
- 240 \zeta_5 C_F^2
- 18 \zeta_3 C_F C_A \alpha
- 8 \zeta_3 \pi^2 C_F C_A \alpha
- 2 \zeta_3 C_F C_A \alpha^3
\right. \nonumber \\
&& \left. ~~~
+ 7 \pi^2 C_F C_A \alpha
+ 12 \zeta_3 C_F C_A \alpha^2
+ 12 \psi^{\prime}(\third) \zeta_3 C_F C_A \alpha
\right] a^2 ~+~ O(a^3)
\label{kafuncfgen}
\end{eqnarray}
and
\begin{eqnarray}
K_{\alpha\,\CFs}^{\MOMgs}(a,\alpha) &=& 
\left[
6 \psi^{\prime}(\third) \zeta_3 C_F C_A
+ 12 \zeta_3 C_F C_A \alpha
- \frac{21}{2} C_F C_A \alpha
- \frac{21}{4} \psi^{\prime}(\third) C_F C_A
\right. \nonumber \\
&& \left. ~
- \frac{14}{9} \pi^2 C_F C_A \alpha
- \frac{8}{3} \psi^{\prime}(\third) \zeta_3 C_F C_A \alpha
+ \frac{7}{2} \pi^2 C_F C_A
+ \frac{7}{3} \psi^{\prime}(\third) C_F C_A \alpha
\right. \nonumber \\
&& \left. ~
+ \frac{16}{9} \zeta_3 \pi^2 C_F C_A \alpha
+ \frac{21}{8} C_F C_A \alpha^2
+ \frac{63}{8} C_F C_A
- 9 \zeta_3 C_F C_A
- 4 \zeta_3 \pi^2 C_F C_A
\right. \nonumber \\
&& \left. ~
- 3 \zeta_3 C_F C_A \alpha^2
\right] a^2 ~+~ O(a^3)
\label{kalfuncfgen}
\end{eqnarray}
while the MAG partners are
\begin{eqnarray}
K_{a\,\MAGs}^{\MOMgs}(a,\alpha) &=& 
\left[
12 \zeta_3 C_F
- \frac{21}{2} C_F
\right] a
\nonumber \\
&&
+ \left[
\frac{7}{3} \psi^{\prime}(\third) C_F C_A \alpha^2
- \frac{512}{27} \zeta_3 \pi^2 \Nf T_F C_F
- \frac{321}{2} C_F C_A
- \frac{224}{9} \psi^{\prime}(\third) \Nf T_F C_F
\right. \nonumber \\
&& \left. ~~~
- \frac{161}{27} \pi^2 C_F C_A
- \frac{112}{3} \zeta_3 \Nf T_F C_F
- \frac{104}{3} \zeta_3 \pi^2 \frac{\Nda}{\Noda} C_F C_A
\right. \nonumber \\
&& \left. ~~~
- \frac{92}{9} \psi^{\prime}(\third) \zeta_3 C_F C_A
- \frac{91}{2} \psi^{\prime}(\third) \frac{\Nda}{\Noda} C_F C_A
- \frac{63}{4} \frac{\Nda}{\Noda} C_F C_A \alpha
\right. \nonumber \\
&& \left. ~~~
- \frac{35}{9} \pi^2 \frac{\Nda}{\Noda} C_F C_A \alpha
- \frac{32}{9} \zeta_3 \pi^2 \frac{\Nda}{\Noda} C_F C_A \alpha^2
- \frac{21}{2} C_F C_A \alpha^2
\right. \nonumber \\
&& \left. ~~~
- \frac{21}{2} \psi^{\prime}(\third) C_F C_A \alpha
- \frac{21}{4} \frac{\Nda}{\Noda} C_F C_A \alpha^3
- \frac{20}{3} \psi^{\prime}(\third) \zeta_3 \frac{\Nda}{\Noda} C_F C_A \alpha
\right. \nonumber \\
&& \left. ~~~
- \frac{14}{3} \psi^{\prime}(\third) \frac{\Nda}{\Noda} C_F C_A \alpha^2
- \frac{14}{9} \pi^2 C_F C_A \alpha^2
- \frac{8}{3} \psi^{\prime}(\third) \zeta_3 C_F C_A \alpha^2
\right. \nonumber \\
&& \left. ~~~
+ \frac{7}{4} C_F C_A \alpha^3
+ \frac{16}{3} \psi^{\prime}(\third) \zeta_3 \frac{\Nda}{\Noda} C_F C_A \alpha^2
+ \frac{16}{9} \zeta_3 \pi^2 C_F C_A \alpha^2
\right. \nonumber \\
&& \left. ~~~
+ \frac{28}{9} \pi^2 \frac{\Nda}{\Noda} C_F C_A \alpha^2
+ \frac{35}{6} \psi^{\prime}(\third) \frac{\Nda}{\Noda} C_F C_A \alpha
+ \frac{40}{9} \zeta_3 \pi^2 \frac{\Nda}{\Noda} C_F C_A \alpha
\right. \nonumber \\
&& \left. ~~~
+ \frac{63}{4} C_F C_A \alpha
+ \frac{91}{3} \pi^2 \frac{\Nda}{\Noda} C_F C_A
+ \frac{158}{3} \Nf T_F C_F
+ \frac{161}{18} \psi^{\prime}(\third) C_F C_A
\right. \nonumber \\
&& \left. ~~~
+ \frac{184}{27} \zeta_3 \pi^2 C_F C_A
+ \frac{256}{9} \psi^{\prime}(\third) \zeta_3 \Nf T_F C_F
+ \frac{356}{3} \zeta_3 C_F C_A
+ \frac{397}{6} C_F^2
\right. \nonumber \\
&& \left. ~~~
+ \frac{448}{27} \pi^2 \Nf T_F C_F
- 240 \zeta_5 C_F^2
- 24 \zeta_3 \frac{\Nda}{\Noda} C_F C_A \alpha^2
- 18 \zeta_3 C_F C_A \alpha
\right. \nonumber \\
&& \left. ~~~
- 8 \zeta_3 \pi^2 C_F C_A \alpha
- 2 \zeta_3 C_F C_A \alpha^3
+ 6 \zeta_3 \frac{\Nda}{\Noda} C_F C_A \alpha^3
+ 7 \pi^2 C_F C_A \alpha
\right. \nonumber \\
&& \left. ~~~
+ 12 \zeta_3 C_F C_A \alpha^2
+ 12 \psi^{\prime}(\third) \zeta_3 C_F C_A \alpha
+ 18 \zeta_3 \frac{\Nda}{\Noda} C_F C_A \alpha
\right. \nonumber \\
&& \left. ~~~
+ 21 \frac{\Nda}{\Noda} C_F C_A \alpha^2
+ 52 \psi^{\prime}(\third) \zeta_3 \frac{\Nda}{\Noda} C_F C_A
+ 136 \zeta_3 C_F^2
\right] a^2
~+~ O(a^3)
\label{kafunmaggen}
\end{eqnarray}
and
\begin{eqnarray}
K_{\alpha\,\MAGs}^{\MOMgs}(a,\alpha) &=&
\left[
\frac{7}{2} \pi^2 C_F C_A
- \frac{63}{8} \frac{\Nda}{\Noda} C_F C_A
- \frac{63}{8} \frac{\Nda}{\Noda} C_F C_A \alpha^2
- \frac{35}{18} \pi^2 \frac{\Nda}{\Noda} C_F C_A
\right. \nonumber \\
&& \left. ~
- \frac{32}{9} \zeta_3 \pi^2 \frac{\Nda}{\Noda} C_F C_A \alpha
- \frac{21}{2} C_F C_A \alpha
- \frac{21}{4} \psi^{\prime}(\third) C_F C_A
\right. \nonumber \\
&& \left. ~
- \frac{14}{3} \psi^{\prime}(\third) \frac{\Nda}{\Noda} C_F C_A \alpha
- \frac{14}{9} \pi^2 C_F C_A \alpha
- \frac{10}{3} \psi^{\prime}(\third) \zeta_3 \frac{\Nda}{\Noda} C_F C_A
\right. \nonumber \\
&& \left. ~
- \frac{8}{3} \psi^{\prime}(\third) \zeta_3 C_F C_A \alpha
+ \frac{7}{3} \psi^{\prime}(\third) C_F C_A \alpha
+ \frac{16}{3} \psi^{\prime}(\third) \zeta_3 \frac{\Nda}{\Noda} C_F C_A \alpha
\right. \nonumber \\
&& \left. ~
+ \frac{16}{9} \zeta_3 \pi^2 C_F C_A \alpha
+ \frac{20}{9} \zeta_3 \pi^2 \frac{\Nda}{\Noda} C_F C_A
+ \frac{21}{8} C_F C_A \alpha^2
\right. \nonumber \\
&& \left. ~
+ \frac{28}{9} \pi^2 \frac{\Nda}{\Noda} C_F C_A \alpha
+ \frac{35}{12} \psi^{\prime}(\third) \frac{\Nda}{\Noda} C_F C_A
+ \frac{63}{8} C_F C_A
\right. \nonumber \\
&& \left. ~
- 24 \zeta_3 \frac{\Nda}{\Noda} C_F C_A \alpha
- 9 \zeta_3 C_F C_A
- 4 \zeta_3 \pi^2 C_F C_A
- 3 \zeta_3 C_F C_A \alpha^2
\right. \nonumber \\
&& \left. ~
+ 6 \psi^{\prime}(\third) \zeta_3 C_F C_A
+ 9 \zeta_3 \frac{\Nda}{\Noda} C_F C_A
+ 9 \zeta_3 \frac{\Nda}{\Noda} C_F C_A \alpha^2
+ 12 \zeta_3 C_F C_A \alpha
\right. \nonumber \\
&& \left. ~
+ 21 \frac{\Nda}{\Noda} C_F C_A \alpha
\right] a^2 ~+~ O(a^3) ~.
\label{kalfunmaggen}
\end{eqnarray}
Deleting the terms with a $\Nda/\Noda$ factor in (\ref{kafunmaggen}) and
(\ref{kalfunmaggen}) reproduces (\ref{kafuncfgen}) and (\ref{kalfuncfgen})
respectively. We have checked that this is also the case for the $O(a^3)$ 
terms.

\sect{Other schemes.}

One question of interest concerns whether or not there are any schemes other 
than the $\MSbar$ one where $\Delta_{\mbox{\footnotesize{csb}}}(a,\alpha)$ 
involves only the $\beta$-function. It turns out in fact that, aside from the
$V$ scheme of \cite{13,14} whose $\beta$-function is independent of the gauge
parameter by construction, there is at least one other such scheme which is the
$\RI$ scheme, \cite{35,36}. It was introduced in relation to lattice 
regularized QCD calculations and has a continuum spacetime analogue which has 
been used to renormalize the theory to high loop order \cite{59,60,61,62}. The 
prescription requires that the wave function renormalization constants of the 
gluon, ghost and quark fields are defined so that there are no $O(a)$ 
corrections to their $2$-point functions at the subtraction point. In this 
sense it incorporates the approach of the MOM schemes of \cite{37,38}. The 
$\RI$ scheme differs however in respect of the coupling constant 
renormalization. For each vertex function the coupling renormalization constant
is defined according to the $\MSbar$ prescription at the subtraction point. So 
the $\RI$ scheme can be regarded as a halfway house between the $\MSbar$ and 
the suite of MOM schemes of \cite{37,38}. In terms of the resulting 
renormalization group functions the relevant issue for this article is that the
$\RI$ $\beta$-function is formally equivalent to that of the $\MSbar$ 
$\beta$-function. By this we mean that although the coupling constant will be 
the variable in the respective schemes the actual coefficients of the coupling 
constant polynomial of the $\RI$ scheme $\beta$-function are identically the 
same as their $\MSbar$ counterparts. We have checked explicitly that the same 
reasoning applies to the structure of $K_a^{\RIs}(a)$ whose coefficients are 
precisely the same as those of $K_a^{\MSbars}(a)$ with in addition now
\begin{equation}
K_\alpha^{\RIs}(a,\alpha) ~=~ 0 ~.
\end{equation}
In other words the ansatz of (\ref{crewms}) holds primarily as a direct result 
of the $\RI$ renormalization prescription of the vertex function.

Although this means that there is a scheme other than $\MSbar$ that takes the
same form as (\ref{deltms}) it is a relatively trivial observation. However the
nature of the $\RI$ scheme gives a more important insight into some of the 
underlying issues. In essence it indicates that the scheme the fields are 
renormalized in plays no role in whether a $K_\alpha(a,\alpha)$ function is 
needed. Instead it is the prescription for the coupling constant 
renormalization that is key. This is understandable since the $\MSbar$ Crewther
construction is $\alpha$ independent. In light of this, one question that is of 
interest is what occurs if one considers a scheme that is opposite to the 
prescription of the $\RI$ scheme. By this we mean one where the fields, gauge 
parameter and quark mass are renormalized in an $\MSbar$ way but the vertex 
functions, and thereby the coupling constant, are renormalized in a 
non-$\MSbar$ fashion. It is possible to explore this possibility given the four
loop data that is available in \cite{18,62}. In this new scheme the gluon, 
ghost and quark $2$-point functions as well as the quark $2$-point function 
with the quark mass operator inserted, \cite{62}, are renormalized in the 
linear covariant gauge by including only the poles in $\epsilon$ into the 
respective renormalization constants. To determine the coupling constant 
renormalization we have chosen to use the ghost-gluon vertex in the 
configuration where the momentum of one ghost field is zero. This is the setup 
that is used to define the $\mMOM$ renormalization. The reason for choosing 
this vertex is that in the Landau gauge the vertex is finite according to 
Taylor's observation, \cite{16}. Therefore in this new scheme the 
renormalization group functions ought to differ from the $\MSbar$ ones in a 
minimal way. We have labelled this scheme as $\RIc$ reflecting that it is based
on the ghost-gluon vertex, being a variation of the $\RI$ scheme, and 
constructed the renormalization group equations to five loops. Expressions for 
them in an arbitrary gauge for a general colour group are given in the 
associated data file. To appreciate their structure we have recorded the 
$SU(3)$ expressions for the $\beta$-function and gauge parameter anomalous
dimension in Appendix B as these are relevant to the current discussion.
However it is worth noting that our expectation that the $\RIc$ scheme has a
minimal effect on the $\beta$-function structure is met in that the
coefficients of the coupling constant in $\beta^{\RIcs}(a,0)$ are precisely
the same as those of $\beta^{\MSbars}(a)$ to five loops.  

Equipped with the $\RIc$ renormalization group functions we have determined the
relation between the $\MSbar$ and $\RIc$ coupling constant and gauge paremeter
that allows us to examine what 
$\Delta_{\mbox{\footnotesize{csb}}}^{\RIcs}(a,\alpha)$ is. By following the 
procedure used for other schemes it transpires that 
$K_\alpha^{\RIcs}(a,\alpha)$ has to be non-zero. More specifically we found 
\begin{eqnarray}
K_a^{\RIcs}(a,\alpha) &=& 
\left[
12 \zeta_3 C_F
- \frac{21}{2} C_F
\right] a
\nonumber \\
&&
+ \left[
\frac{326}{3} \Nf T_F C_F
+ \frac{397}{6} C_F^2
+ \frac{884}{3} \zeta_3 C_F C_A
- \frac{629}{2} C_F C_A
- \frac{304}{3} \zeta_3 \Nf T_F C_F
\right. \nonumber \\
&& \left. ~~~
- 240 \zeta_5 C_F^2
- 24 \zeta_3 C_F C_A \alpha
+ 21 C_F C_A \alpha
+ 136 \zeta_3 C_F^2
\right] a^2
\nonumber \\
&&
+ \left[
\frac{81}{2} \zeta_3^2 C_F C_A^2 \alpha
+ \frac{85}{16} \zeta_3 C_F C_A^2 \alpha^2
+ \frac{211}{16} C_F C_A^2 \alpha^2
- \frac{406043}{36} C_F C_A^2
\right. \nonumber \\
&& \left. ~~~
- \frac{45517}{48} \zeta_3 C_F C_A^2 \alpha
- \frac{40336}{9} \zeta_3 \Nf T_F C_F C_A
- \frac{24880}{3} \zeta_5 C_F^2 C_A
\right. \nonumber \\
&& \left. ~~~
- \frac{9824}{9} \Nf^2 T_F^2 C_F
- \frac{8000}{3} \zeta_5 \Nf T_F C_F C_A
- \frac{7729}{18} \Nf T_F C_F^2
- \frac{652}{3} \Nf T_F C_F C_A \alpha
\right. \nonumber \\
&& \left. ~~~
- \frac{397}{2} C_F^2 C_A \alpha
- \frac{27}{2} \zeta_3^2 C_F C_A^2 \alpha^2
+ \frac{608}{3} \zeta_3 \Nf T_F C_F C_A \alpha
+ \frac{2471}{12} C_F^3
\right. \nonumber \\
&& \left. ~~~
+ \frac{6496}{9} \zeta_3 \Nf^2 T_F^2 C_F
+ \frac{11900}{3} \zeta_5 C_F C_A^2
+ \frac{16570}{3} \zeta_3 C_F^2 C_A
+ \frac{44939}{48} C_F C_A^2 \alpha
\right. \nonumber \\
&& \left. ~~~
+ \frac{67520}{9} \Nf T_F C_F C_A
+ \frac{72028}{9} \zeta_3 C_F C_A^2
+ \frac{99757}{36} C_F^2 C_A
- 5720 \zeta_5 C_F^3
\right. \nonumber \\
&& \left. ~~~
- 3668 \zeta_3 \Nf T_F C_F^2
- 1232 \zeta_3^2 C_F C_A^2
- 840 \zeta_7 C_F^2 C_A
- 408 \zeta_3 C_F^2 C_A \alpha
\right. \nonumber \\
&& \left. ~~~
- 128 \zeta_3^2 \Nf T_F C_F C_A
+ 320 \zeta_5 \Nf^2 T_F^2 C_F
+ 488 \zeta_3 C_F^3
+ 576 \zeta_3^2 \Nf T_F C_F^2
\right. \nonumber \\
&& \left. ~~~
+ 720 \zeta_5 C_F^2 C_A \alpha
+ 4000 \zeta_5 \Nf T_F C_F^2
+ 5040 \zeta_7 C_F^3
\right] a^3 ~+~ O(a^4)
\end{eqnarray}
and
\begin{eqnarray}
K_\alpha^{\RIcs}(a,\alpha) &=& 
\left[
\frac{21}{2} C_F 
- 12 \zeta_3 C_F 
\right] C_A a^2
\nonumber \\
&& + \left[
\frac{27}{2} \zeta_3^2 C_F C_A
+ \frac{231}{8} \zeta_3 C_F C_A \alpha
+ \frac{3461}{48} C_F C_A
- \frac{1477}{16} \zeta_3 C_F C_A
- \frac{397}{6} C_F^2 
\right. \nonumber \\
&& \left. ~~~
- \frac{147}{8} C_F C_A \alpha
- 136 \zeta_3 C_F^2 
- 9 \zeta_3^2 C_F C_A \alpha
+ 240 \zeta_5 C_F^2
\right] C_A a^3 \,+\, O(a^4) ~.
\end{eqnarray}
Although this exercise did not produce another scheme whose Crewther structure
was the same as those of $\MSbar$ and $\RI$ $K_\alpha^{\RIcs}(a,\alpha)$ does
have a different structure. For instance unlike the other schemes considered
earlier the leading term is $\alpha$ independent and the linear terms in 
$\alpha$ at the next order are proportional to the leading term. Also like 
$K_\alpha^{\mMOMs}(a,\alpha)$ we note that $K_\alpha^{\RIcs}(a,\alpha)$ is
proportional to $C_A$. These properties can be traced directly back to those of
the vertex function that this scheme was based on.  

\sect{Renormalization group connection.}

Having accumulated evidence to support the ansatz (\ref{deltgen}) a natural
question is whether it can be derived directly and in general from some 
renormalization group principle. This is important since our analysis so far 
indicating an extension to (\ref{deltms}) is needed has been concerned with 
values of $\Nf$ in the conformal window. Any generalization should not be 
restricted to this range. It turns out the key lies in the way one maps the 
parameters of a theory in one scheme to the corresponding parameters in another
scheme as well as the relation between the renormalization group functions. For
instance if for the moment we regard $a$ as the $\MSbar$ coupling constant and 
$a_{\cal S}$ and $\alpha_{\cal S}$ as the respective coupling constant and 
gauge parameter in another scheme ${\cal S}$ then we found the Adler 
$D$-function and the Bjorken sum rule using the relations
\begin{equation}
C_{{\mbox{\footnotesize{Bjr}}}}^{\MSbars}(a) ~=~ 
C_{{\mbox{\footnotesize{Bjr}}}}^{\cal S}(a_{\cal S},\alpha_{\cal S}) ~~~,~~~
C_{{\mbox{\footnotesize{Adl}}}}^{\MSbars}(a) ~=~ 
C_{{\mbox{\footnotesize{Adl}}}}^{\cal S}(a_{\cal S},\alpha_{\cal S})
\end{equation}
where
\begin{equation}
a ~\equiv~ a(a_{\cal S},\alpha_{\cal S}) 
\label{amapmss}
\end{equation}
expresses the $\MSbar$ coupling in terms of the coupling and gauge parameter in
scheme ${\cal S}$ and implies
\begin{equation}
\Delta_{\mbox{\footnotesize{csb}}}^{\MSbars}(a) ~=~ 
\Delta_{\mbox{\footnotesize{csb}}}^{\cal S}(a_{\cal S},\alpha_{\cal S}) ~.
\end{equation}
Using properties of the renormalization group equation the $\MSbar$ 
$\beta$-function is related to that in another scheme by
\begin{equation}
\beta^{\MSbars}(a) ~=~ \beta^{\cal S}(a_{\cal S},\alpha_{\cal S})
\frac{\partial}{\partial a_{\cal S}} a(a_{\cal S},\alpha_{\cal S}) ~+~ 
\alpha_{\cal S} \gamma_\alpha^{\cal S}(a_{\cal S},\alpha_{\cal S})
\frac{\partial}{\partial \alpha_{\cal S}} a(a_{\cal S},\alpha_{\cal S}) 
\label{betamap}
\end{equation}
which recalling (\ref{deltms}) implies
$\Delta_{\mbox{\footnotesize{csb}}}^{\cal S}(a_{\cal S},\alpha_{\cal S})$ 
can be written as
\begin{equation}
\Delta_{\mbox{\footnotesize{csb}}}^{\cal S}(a_{\cal S},\alpha_{\cal S}) ~=~ 
\frac{\beta^{\cal S}(a_{\cal S},\alpha_{\cal S})}{a_{\cal S}} 
\bar{K}_a^{\cal S}(a_{\cal S},\alpha_{\cal S}) ~+~ 
\alpha_{\cal S} \gamma_\alpha^{\cal S}(a_{\cal S},\alpha_{\cal S}) 
\bar{K}_\alpha^{\cal S}(a_{\cal S},\alpha_{\cal S}) 
\label{deltgenbar}
\end{equation}
where 
\begin{eqnarray} 
\bar{K}_a^{\cal S}(a_{\cal S},\alpha_{\cal S}) &=& a_{\cal S}
\left. \left[ \frac{1}{a} K_a^{\MSbars}(a) \right] 
\right|_{\MSbars \to {\cal S}} ~~
\frac{\partial}{\partial a_{\cal S}} a(a_{\cal S},\alpha_{\cal S}) \nonumber \\ 
\bar{K}_\alpha^{\cal S}(a_{\cal S},\alpha_{\cal S}) &=&
\left. \left[ \frac{1}{a} K_a^{\MSbars}(a) \right] 
\right|_{\MSbars \to {\cal S}} ~~
\frac{\partial}{\partial \alpha_{\cal S}} a(a_{\cal S},\alpha_{\cal S}) ~.
\label{krgecon}
\end{eqnarray} 
The restriction on each appearance of $K_a^{\MSbars}(a)$ indicates that the
$\MSbar$ coupling needs to be mapped to its ${\cal S}$ scheme counterpart by 
inverting (\ref{amapmss}) so that the variables are in the ${\cal S}$ scheme. 
By setting ${\cal S}$ in (\ref{deltgen}) to be the $\MSbar$ scheme and 
comparing with (\ref{deltms}) clearly implies
\begin{equation}
\bar{K}_\alpha^{\MSbars}(a,\alpha) ~=~ 0 ~.
\end{equation}
However (\ref{krgecon}) demonstrates the consistency of this observation 
trivially. For instance if in the second relation of (\ref{krgecon}) we choose 
${\cal S}$ to be the $\MSbar$ scheme itself then 
$\frac{\partial a}{\partial \alpha}$ vanishes as $a$ and $\alpha$ are 
independent in the $\MSbar$ scheme.

While the use of (\ref{betamap}) formally justifies our ansatz of 
(\ref{deltgen}) it provides another way of constructing explicit expressions
for the $K$-functions. However it turns out that the perturbative construction
of the $K$-functions are not in agreement with those derived from 
(\ref{krgecon}). In other words for each of the schemes and gauges considered 
here we define the quantities
\begin{equation}
\Theta_a^{\cal S}(a,\alpha) ~=~ K_a^{\cal S}(a,\alpha) ~-~ 
\bar{K}_a^{\cal S}(a,\alpha) ~~,~~
\Theta_\alpha^{\cal S}(a,\alpha) ~=~ K_\alpha^{\cal S}(a,\alpha) ~-~
\bar{K}_\alpha^{\cal S}(a,\alpha) 
\end{equation}
to illustrate the lack of uniqueness and note they are non-zero. This 
non-uniqueness of the perturbative expansion of the $K$-functions does not 
invalidate the observation that  
$\Delta_{\mbox{\footnotesize{csb}}}^{\cal S}(a_{\cal S},\alpha_{\cal S})$
vanishes at critical points of the renormalization group equation. This still
follows due to the coefficients of the $K$-functions of (\ref{deltgenbar}) 
being the quantities defining the fixed point locations. By way of example we
record a few expressions for both $\Theta_i^{\cal S}(a,\alpha)$ in various
schemes. First in the linear covariant gauge we have 
\begin{eqnarray}
\Theta_a^{\mMOMs}(a,\alpha) &=&
\left[
\frac{434}{3} \alpha \Nf T_F C_F C_A
- \frac{2821}{12} \alpha C_F C_A^2
+ 182 \alpha \zeta_3 C_F C_A^2
- 112 \alpha \zeta_3 \Nf T_F C_F C_A
\right. \nonumber \\
&& \left. ~
+ \frac{31}{8} \alpha^2 C_F C_A^2
+ 31 \alpha^2 \Nf T_F C_F C_A
- 3 \alpha^2 \zeta_3 C_F C_A^2
- 24 \alpha^2 \zeta_3 \Nf T_F C_F C_A
\right. \nonumber \\
&& \left. ~
+ \frac{93}{8} \alpha^3 C_F C_A^2
- 9 \alpha^3 \zeta_3 C_F C_A^2
\right] a^3 ~+~ O(a^4) \nonumber \\
\Theta_\alpha^{\mMOMs}(a,\alpha) &=&
\left[
\frac{434}{3} \Nf T_F C_F C_A
- \frac{2387}{6} C_F C_A^2
+ 308 \zeta_3 C_F C_A^2
- 112 \zeta_3 \Nf T_F C_F C_A
\right. \nonumber \\
&& \left. ~
- \frac{341}{4} \alpha C_F C_A^2
+ 31 \alpha \Nf T_F C_F C_A
+ 66 \alpha \zeta_3 C_F C_A^2
- 24 \alpha \zeta_3 \Nf T_F C_F C_A
\right] a^3 \nonumber \\
&& +~ O(a^4)
\end{eqnarray}
for the $\mMOM$ scheme. By contrast the partially related $\RIc$ scheme 
expressions are simpler since 
\begin{eqnarray}
\Theta_a^{\RIcs}(a,\alpha) &=&
\left[
\frac{326}{3} \alpha \Nf T_F C_F C_A
- \frac{2119}{12} \alpha C_F C_A^2
+ \frac{494}{3} \alpha \zeta_3 C_F C_A^2
- \frac{304}{3} \alpha \zeta_3 \Nf T_F C_F C_A
\right. \nonumber \\
&& \left. ~
+ \frac{163}{4} \alpha^2 C_F C_A^2
- 38 \alpha^2 \zeta_3 C_F C_A^2
\right] a^3 ~+~ O(a^4) \nonumber \\
\Theta_\alpha^{\RIcs}(a,\alpha) &=&
\left[
\frac{326}{3} \Nf T_F C_F C_A
- \frac{1793}{6} C_F C_A^2
+ \frac{836}{3} \zeta_3 C_F C_A^2
- \frac{304}{3} \zeta_3 \Nf T_F C_F C_A
\right] a^3 \nonumber \\
&& +~ O(a^4) ~.
\end{eqnarray}
For balance we record the $\MOMg$ scheme $SU(3)$ Yang-Mills differences are
\begin{eqnarray}
\left. \Theta_a^{\MOMgs}(a,\alpha) \right|^{SU(3)}_{\Nf=0} &=&
\left[
\frac{98}{3} \pi^2 \alpha^4
+ \frac{411}{8} \alpha^4
- \frac{238303}{108} \psi^\prime(\third) \alpha
- \frac{128596}{81} \zeta_3 \pi^2 \alpha
- \frac{99008}{729} \zeta_3 \pi^4 \alpha
\right. \nonumber \\
&& \left. ~
- \frac{86632}{243} \psi^\prime(\third) \pi^2 \alpha
- \frac{24752}{81} \psi^\prime(\third)^2 \zeta_3 \alpha
- \frac{15680}{243} \zeta_3 \pi^4 \alpha^2
- \frac{15001}{36} \psi^\prime(\third) \alpha^2
\right. \nonumber \\
&& \left. ~
- \frac{13720}{81} \psi^\prime(\third) \pi^2 \alpha^2
- \frac{7372}{27} \zeta_3 \pi^2 \alpha^2
- \frac{3920}{27} \psi^\prime(\third)^2 \zeta_3 \alpha^2
- \frac{3445}{3} \zeta_3 \alpha
\right. \nonumber \\
&& \left. ~
- \frac{3299}{8} \alpha^3
- \frac{1792}{27} \psi^\prime(\third) \zeta_3 \pi^2 \alpha^3
- \frac{1568}{81} \pi^4 \alpha^3
- \frac{392}{9} \psi^\prime(\third)^2 \alpha^3
\right. \nonumber \\
&& \left. ~
- \frac{112}{3} \zeta_3 \pi^2 \alpha^4
+ \frac{448}{9} \psi^\prime(\third)^2 \zeta_3 \alpha^3
+ \frac{928}{3} \zeta_3 \pi^2 \alpha^3
+ \frac{1568}{27} \psi^\prime(\third) \pi^2 \alpha^3
\right. \nonumber \\
&& \left. ~
+ \frac{1792}{81} \zeta_3 \pi^4 \alpha^3
+ \frac{2623}{8} \alpha^2
+ \frac{3430}{27} \psi^\prime(\third)^2 \alpha^2
+ \frac{3686}{9} \psi^\prime(\third) \zeta_3 \alpha^2
\right. \nonumber \\
&& \left. ~
+ \frac{13720}{243} \pi^4 \alpha^2
+ \frac{15001}{54} \pi^2 \alpha^2
+ \frac{15680}{81} \psi^\prime(\third) \zeta_3 \pi^2 \alpha^2
+ \frac{21658}{81} \psi^\prime(\third)^2 \alpha
\right. \nonumber \\
&& \left. ~
+ \frac{51415}{24} \alpha
+ \frac{64298}{27} \psi^\prime(\third) \zeta_3 \alpha
+ \frac{86632}{729} \pi^4 \alpha
+ \frac{99008}{243} \psi^\prime(\third) \zeta_3 \pi^2 \alpha
\right. \nonumber \\
&& \left. ~
+ \frac{238303}{162} \pi^2 \alpha
- 464 \psi^\prime(\third) \zeta_3 \alpha^3
- 284 \pi^2 \alpha^3
- 229 \zeta_3 \alpha^2
- 49 \psi^\prime(\third) \alpha^4
\right. \nonumber \\
&& \left. ~
- 33 \zeta_3 \alpha^4
+ 56 \psi^\prime(\third) \zeta_3 \alpha^4
+ 257 \zeta_3 \alpha^3
+ 426 \psi^\prime(\third) \alpha^3
\right] a^3 ~+~ O(a^4) \nonumber \\
\left. \Theta_\alpha^{\MOMgs}(a,\alpha) \right|^{SU(3)}_{\Nf=0} &=&
\left[
\frac{1078}{3} \psi^\prime(\third) \alpha^2
+ \frac{2464}{9} \zeta_3 \pi^2 \alpha^2
- \frac{217624}{81} \zeta_3 \pi^2
- \frac{201641}{54} \psi^\prime(\third)
\right. \nonumber \\
&& \left. ~
- \frac{167552}{729} \zeta_3 \pi^4
- \frac{146608}{243} \psi^\prime(\third) \pi^2
- \frac{41888}{81} \psi^\prime(\third)^2 \zeta_3
- \frac{39424}{243} \zeta_3 \pi^4 \alpha
\right. \nonumber \\
&& \left. ~
- \frac{34496}{81} \psi^\prime(\third) \pi^2 \alpha
- \frac{29216}{27} \zeta_3 \pi^2 \alpha
- \frac{14102}{9} \psi^\prime(\third) \alpha
- \frac{9856}{27} \psi^\prime(\third)^2 \zeta_3 \alpha
\right. \nonumber \\
&& \left. ~
- \frac{5830}{3} \zeta_3
- \frac{2156}{9} \pi^2 \alpha^2
- \frac{1507}{4} \alpha^2
- \frac{1232}{3} \psi^\prime(\third) \zeta_3 \alpha^2
+ \frac{2783}{2} \alpha
\right. \nonumber \\
&& \left. ~
+ \frac{8624}{27} \psi^\prime(\third)^2 \alpha
+ \frac{14608}{9} \psi^\prime(\third) \zeta_3 \alpha
+ \frac{28204}{27} \pi^2 \alpha
+ \frac{34496}{243} \pi^4 \alpha
\right. \nonumber \\
&& \left. ~
+ \frac{36652}{81} \psi^\prime(\third)^2
+ \frac{39424}{81} \psi^\prime(\third) \zeta_3 \pi^2 \alpha
+ \frac{43505}{12}
+ \frac{108812}{27} \psi^\prime(\third) \zeta_3
\right. \nonumber \\
&& \left. ~
+ \frac{146608}{729} \pi^4
+ \frac{167552}{243} \psi^\prime(\third) \zeta_3 \pi^2
+ \frac{201641}{81} \pi^2
- 836 \zeta_3 \alpha
+ 242 \zeta_3 \alpha^2
\right] a^3 \nonumber \\
&& +~ O(a^4)
\end{eqnarray}
to illustrate the situation for a kinematic scheme. While the above linear
covariant gauge expressions for $\Theta_a^{\cal S}(a,\alpha)$ each vanish when 
$\alpha$~$=$~$0$ this is not the case for the MAG as the $\alpha$ dependence of
the respective $\gamma_\alpha(a,\alpha)$ leading terms differ, \cite{57}. For 
instance in the $\MOMc$ scheme in that gauge we find
\begin{eqnarray}
\Theta_{a\,\MAGs}^{\MOMcs}(a,\alpha) &=&
\left[
\frac{2}{3} \pi^2 C_F C_A^2 \alpha^2
+ \frac{2}{3} \psi^\prime(\third) \zeta_3 C_F C_A^2 \alpha^2
+ \frac{8}{9} \zeta_3 \pi^2 \frac{\Nda}{\Noda} C_F C_A^2 \alpha^2
\right. \nonumber \\
&& \left. ~
- \frac{3751}{3} \frac{\Noda}{[\Noda+2\Nda]} C_F C_A^2 \alpha
- \frac{1425}{4} \frac{\Nda}{\Noda} C_F C_A^2
- \frac{1395}{8} \frac{\Nda}{\Noda} C_F C_A^2 \alpha
\right. \nonumber \\
&& \left. ~
- \frac{1323}{8} \frac{\Nda^2}{\Noda^2} C_F C_A^2 \alpha^2
- \frac{1179}{4} \frac{\Nda^2}{\Noda^2} C_F C_A^2 \alpha
- \frac{1023}{2} \frac{\Noda}{[\Noda+2\Nda]} C_F C_A^2
\right. \nonumber \\
&& \left. ~
- \frac{775}{18} \psi^\prime(\third) \frac{\Nda}{\Noda} C_F C_A^2 \alpha
- \frac{621}{4} \frac{\Nda^2}{\Noda^2} C_F C_A^2
- \frac{560}{27} \zeta_3 \pi^2 \frac{\Nda}{\Noda} C_F C_A^2 \alpha
\right. \nonumber \\
&& \left. ~
- \frac{475}{3} \Nf T_F C_F C_A \alpha
- \frac{344}{27} \pi^2 \Nf \frac{\Nda}{\Noda} T_F C_F C_A \alpha
- \frac{153}{2} \zeta_3 \frac{\Nda}{\Noda} C_F C_A^2 \alpha^2
\right. \nonumber \\
&& \left. ~
- \frac{128}{9} \psi^\prime(\third) \zeta_3 \Nf \frac{\Nda}{\Noda} T_F C_F C_A \alpha
- \frac{93}{4} \frac{\Nda^2}{\Noda^2} C_F C_A^2 \alpha^3
- \frac{86}{3} \pi^2 \frac{\Nda^2}{\Noda^2} C_F C_A^2
\right. \nonumber \\
&& \left. ~
- \frac{86}{3} \pi^2 \frac{\Nda^2}{\Noda^2} C_F C_A^2 \alpha
- \frac{64}{27} \zeta_3 \pi^2 \Nf T_F C_F C_A \alpha
- \frac{52}{9} \pi^2 C_F C_A^2 \alpha
\right. \nonumber \\
&& \left. ~
- \frac{52}{9} \psi^\prime(\third) \zeta_3 C_F C_A^2 \alpha
- \frac{43}{9} \pi^2 \frac{\Nda^2}{\Noda^2} C_F C_A^2 \alpha^2
- \frac{19}{18} \pi^2 \frac{\Nda}{\Noda} C_F C_A^2 \alpha^2
\right. \nonumber \\
&& \left. ~
- \frac{16}{3} \zeta_3 \pi^2 \frac{\Nda}{\Noda} C_F C_A^2
- \frac{16}{3} \psi^\prime(\third) \Nf T_F C_F C_A \alpha
- \frac{16}{3} \psi^\prime(\third) \zeta_3 \frac{\Nda^2}{\Noda^2} C_F C_A^2 \alpha^2
\right. \nonumber \\
&& \left. ~
- \frac{9}{2} \zeta_3 C_F C_A^2 \alpha^2
- \frac{9}{2} \zeta_3 C_F C_A^2 \alpha^3
- \frac{4}{3} \psi^\prime(\third) \zeta_3 \frac{\Nda}{\Noda} C_F C_A^2 \alpha^2
\right. \nonumber \\
&& \left. ~
- \frac{4}{9} \zeta_3 \pi^2 C_F C_A^2 \alpha^2
+ \frac{19}{12} \psi^\prime(\third) \frac{\Nda}{\Noda} C_F C_A^2 \alpha^2
+ \frac{26}{3} \psi^\prime(\third) C_F C_A^2 \alpha
\right. \nonumber \\
&& \left. ~
+ \frac{32}{9} \pi^2 \Nf T_F C_F C_A \alpha
+ \frac{32}{9} \zeta_3 \pi^2 \frac{\Nda^2}{\Noda^2} C_F C_A^2 \alpha^2
+ \frac{32}{9} \psi^\prime(\third) \zeta_3 \Nf T_F C_F C_A \alpha
\right. \nonumber \\
&& \left. ~
+ \frac{43}{6} \psi^\prime(\third) \frac{\Nda^2}{\Noda^2} C_F C_A^2 \alpha^2
+ \frac{64}{3} \zeta_3 \pi^2 \frac{\Nda^2}{\Noda^2} C_F C_A^2
+ \frac{64}{3} \zeta_3 \pi^2 \frac{\Nda^2}{\Noda^2} C_F C_A^2 \alpha
\right. \nonumber \\
&& \left. ~
+ \frac{83}{16} C_F C_A^2 \alpha^2
+ \frac{93}{16} C_F C_A^2 \alpha^3
+ \frac{104}{27} \zeta_3 \pi^2 C_F C_A^2 \alpha
\right. \nonumber \\
&& \left. ~
+ \frac{172}{9} \psi^\prime(\third) \Nf \frac{\Nda}{\Noda} T_F C_F C_A \alpha
+ \frac{256}{27} \zeta_3 \pi^2 \Nf \frac{\Nda}{\Noda} T_F C_F C_A \alpha
\right. \nonumber \\
&& \left. ~
+ \frac{280}{9} \psi^\prime(\third) \zeta_3 \frac{\Nda}{\Noda} C_F C_A^2 \alpha
+ \frac{775}{27} \pi^2 \frac{\Nda}{\Noda} C_F C_A^2 \alpha
+ \frac{1023}{2} C_F C_A^2
\right. \nonumber \\
&& \left. ~
+ \frac{1364}{3} \Nf \frac{\Noda}{[\Noda+2\Nda]} T_F C_F C_A \alpha
+ \frac{1571}{16} \frac{\Nda}{\Noda} C_F C_A^2 \alpha^2
+ \frac{18451}{24} C_F C_A^2 \alpha
\right. \nonumber \\
&& \left. ~
- 591 \zeta_3 C_F C_A^2 \alpha
- 396 \zeta_3 C_F C_A^2
- 352 \zeta_3 \Nf \frac{\Noda}{[\Noda+2\Nda]} T_F C_F C_A \alpha
\right. \nonumber \\
&& \left. ~
- 69 \Nf \frac{\Nda}{\Noda} T_F C_F C_A \alpha
- 62 \Nf \frac{\Nda}{\Noda} T_F C_F C_A \alpha^2
- 32 \psi^\prime(\third) \zeta_3 \frac{\Nda^2}{\Noda^2} C_F C_A^2
\right. \nonumber \\
&& \left. ~
- 32 \psi^\prime(\third) \zeta_3 \frac{\Nda^2}{\Noda^2} C_F C_A^2 \alpha
- 24 \zeta_3 \Nf T_F C_F C_A \alpha^2
- 12 \psi^\prime(\third) \frac{\Nda}{\Noda} C_F C_A^2
\right. \nonumber \\
&& \left. ~
- \psi^\prime(\third) C_F C_A^2 \alpha^2
+ 8 \pi^2 \frac{\Nda}{\Noda} C_F C_A^2
+ 8 \psi^\prime(\third) \zeta_3 \frac{\Nda}{\Noda} C_F C_A^2
\right. \nonumber \\
&& \left. ~
+ 18 \zeta_3 \frac{\Nda^2}{\Noda^2} C_F C_A^2 \alpha^3
+ 31 \Nf T_F C_F C_A \alpha^2
+ 43 \psi^\prime(\third) \frac{\Nda^2}{\Noda^2} C_F C_A^2
\right. \nonumber \\
&& \left. ~
+ 43 \psi^\prime(\third) \frac{\Nda^2}{\Noda^2} C_F C_A^2 \alpha
+ 48 \zeta_3 \Nf \frac{\Nda}{\Noda} T_F C_F C_A \alpha^2
+ 56 \zeta_3 \Nf \frac{\Nda}{\Noda} T_F C_F C_A \alpha
\right. \nonumber \\
&& \left. ~
+ 120 \zeta_3 \Nf T_F C_F C_A \alpha
+ 125 \zeta_3 \frac{\Nda}{\Noda} C_F C_A^2 \alpha
+ 126 \zeta_3 \frac{\Nda^2}{\Noda^2} C_F C_A^2
\right. \nonumber \\
&& \left. ~
+ 129 \zeta_3 \frac{\Nda^2}{\Noda^2} C_F C_A^2 \alpha^2
+ 234 \zeta_3 \frac{\Nda^2}{\Noda^2} C_F C_A^2 \alpha
+ 270 \zeta_3 \frac{\Nda}{\Noda} C_F C_A^2
\right. \nonumber \\
&& \left. ~
+ 396 \zeta_3 \frac{\Noda}{[\Noda+2\Nda]} C_F C_A^2
+ 968 \zeta_3 \frac{\Noda}{[\Noda+2\Nda]} C_F C_A^2 \alpha
\right] a^3 ~+~ O(a^4) \nonumber \\
\Theta_{\alpha\,\MAGs}^{\MOMcs}(a,\alpha) &=&
\left[
\frac{32}{9} \pi^2 \Nf T_F C_F C_A
+ \frac{32}{9} \psi^\prime(\third) \zeta_3 \Nf T_F C_F C_A
+ \frac{44}{3} \psi^\prime(\third) C_F C_A^2
\right. \nonumber \\
&& \left. ~
- \frac{3751}{3} \frac{\Noda}{[\Noda+2\Nda]} C_F C_A^2
- \frac{704}{27} \zeta_3 \pi^2 \frac{\Nda}{\Noda} C_F C_A^2
- \frac{475}{3} \Nf T_F C_F C_A
\right. \nonumber \\
&& \left. ~
- \frac{473}{9} \psi^\prime(\third) \frac{\Nda}{\Noda} C_F C_A^2
- \frac{344}{27} \pi^2 \Nf \frac{\Nda}{\Noda} T_F C_F C_A
- \frac{341}{4} C_F C_A^2 \alpha
\right. \nonumber \\
&& \left. ~
- \frac{128}{9} \psi^\prime(\third) \zeta_3 \Nf \frac{\Nda}{\Noda} T_F C_F C_A
- \frac{88}{9} \pi^2 C_F C_A^2
- \frac{88}{9} \psi^\prime(\third) \zeta_3 C_F C_A^2
\right. \nonumber \\
&& \left. ~
- \frac{64}{27} \zeta_3 \pi^2 \Nf T_F C_F C_A
- \frac{16}{3} \psi^\prime(\third) \Nf T_F C_F C_A
+ \frac{172}{9} \psi^\prime(\third) \Nf \frac{\Nda}{\Noda} T_F C_F C_A
\right. \nonumber \\
&& \left. ~
+ \frac{176}{27} \zeta_3 \pi^2 C_F C_A^2
+ \frac{256}{27} \zeta_3 \pi^2 \Nf \frac{\Nda}{\Noda} T_F C_F C_A
+ \frac{341}{2} \frac{\Nda}{\Noda} C_F C_A^2 \alpha
\right. \nonumber \\
&& \left. ~
+ \frac{352}{9} \psi^\prime(\third) \zeta_3 \frac{\Nda}{\Noda} C_F C_A^2
+ \frac{759}{4} \frac{\Nda}{\Noda} C_F C_A^2
+ \frac{946}{27} \pi^2 \frac{\Nda}{\Noda} C_F C_A^2
\right. \nonumber \\
&& \left. ~
+ \frac{1364}{3} \Nf \frac{\Noda}{[\Noda+2\Nda]} T_F C_F C_A
+ \frac{5225}{12} C_F C_A^2
\right. \nonumber \\
&& \left. ~
- 352 \zeta_3 \Nf \frac{\Noda}{[\Noda+2\Nda]} T_F C_F C_A
- 330 \zeta_3 C_F C_A^2
- 154 \zeta_3 \frac{\Nda}{\Noda} C_F C_A^2
\right. \nonumber \\
&& \left. ~
- 132 \zeta_3 \frac{\Nda}{\Noda} C_F C_A^2 \alpha
- 69 \Nf \frac{\Nda}{\Noda} T_F C_F C_A
- 62 \Nf \frac{\Nda}{\Noda} T_F C_F C_A \alpha
\right. \nonumber \\
&& \left. ~
- 24 \zeta_3 \Nf T_F C_F C_A \alpha
+ 31 \Nf T_F C_F C_A \alpha
+ 48 \zeta_3 \Nf \frac{\Nda}{\Noda} T_F C_F C_A \alpha
\right. \nonumber \\
&& \left. ~
+ 56 \zeta_3 \Nf \frac{\Nda}{\Noda} T_F C_F C_A
+ 66 \zeta_3 C_F C_A^2 \alpha
+ 120 \zeta_3 \Nf T_F C_F C_A
\right. \nonumber \\
&& \left. ~
+ 968 \zeta_3 \frac{\Noda}{[\Noda+2\Nda]} C_F C_A^2
\right] a^3 ~+~ O(a^4) ~. 
\end{eqnarray}
By contrast the parallel expressions in the Curci-Ferrari gauge are
\begin{eqnarray}
\Theta_{a\,\CFs}^{\MOMcs}(a,\alpha) &=&
\left[
31 \Nf T_F C_F C_A \alpha^2
+ 377 \zeta_3 C_F C_A^2 \alpha
- \frac{11557}{24} C_F C_A^2 \alpha
- \frac{64}{27} \zeta_3 \pi^2 \Nf T_F C_F C_A \alpha
\right. \nonumber \\
&& \left. ~
- \frac{52}{9} \pi^2 C_F C_A^2 \alpha
- \frac{52}{9} \psi^\prime(\third) \zeta_3 C_F C_A^2 \alpha
- \frac{16}{3} \psi^\prime(\third) \Nf T_F C_F C_A \alpha
- \frac{9}{2} \zeta_3 C_F C_A^2 \alpha^2
\right. \nonumber \\
&& \left. ~
- \frac{9}{2} \zeta_3 C_F C_A^2 \alpha^3
- \frac{4}{9} \zeta_3 \pi^2 C_F C_A^2 \alpha^2
+ \frac{2}{3} \pi^2 C_F C_A^2 \alpha^2
+ \frac{2}{3} \psi^\prime(\third) \zeta_3 C_F C_A^2 \alpha^2
\right. \nonumber \\
&& \left. ~
+ \frac{26}{3} \psi^\prime(\third) C_F C_A^2 \alpha
+ \frac{32}{9} \pi^2 \Nf T_F C_F C_A \alpha
+ \frac{32}{9} \psi^\prime(\third) \zeta_3 \Nf T_F C_F C_A \alpha
\right. \nonumber \\
&& \left. ~
+ \frac{83}{16} C_F C_A^2 \alpha^2
+ \frac{93}{16} C_F C_A^2 \alpha^3
+ \frac{104}{27} \zeta_3 \pi^2 C_F C_A^2 \alpha
+ \frac{889}{3} \Nf T_F C_F C_A \alpha
\right. \nonumber \\
&& \left. ~
- 232 \zeta_3 \Nf T_F C_F C_A \alpha
- 24 \zeta_3 \Nf T_F C_F C_A \alpha^2
- \psi^\prime(\third) C_F C_A^2 \alpha^2
\right] a^3 ~+~ O(a^4) \nonumber \\
\Theta_{\alpha\,\CFs}^{\MOMcs}(a,\alpha) &=&
\left[
31 \Nf T_F C_F C_A \alpha
+ 66 \zeta_3 C_F C_A^2 \alpha
+ 638 \zeta_3 C_F C_A^2
- \frac{9779}{12} C_F C_A^2
- \frac{341}{4} C_F C_A^2 \alpha
\right. \nonumber \\
&& \left. ~
- \frac{88}{9} \pi^2 C_F C_A^2
- \frac{88}{9} \psi^\prime(\third) \zeta_3 C_F C_A^2
- \frac{64}{27} \zeta_3 \pi^2 \Nf T_F C_F C_A
\right. \nonumber \\
&& \left. ~
- \frac{16}{3} \psi^\prime(\third) \Nf T_F C_F C_A
+ \frac{32}{9} \pi^2 \Nf T_F C_F C_A
+ \frac{32}{9} \psi^\prime(\third) \zeta_3 \Nf T_F C_F C_A
\right. \nonumber \\
&& \left. ~
+ \frac{44}{3} \psi^\prime(\third) C_F C_A^2
+ \frac{176}{27} \zeta_3 \pi^2 C_F C_A^2
+ \frac{889}{3} \Nf T_F C_F C_A
- 232 \zeta_3 \Nf T_F C_F C_A
\right. \nonumber \\
&& \left. ~
- 24 \zeta_3 \Nf T_F C_F C_A \alpha
\right] a^3 ~+~ O(a^4) ~. 
\end{eqnarray}
Again when $\alpha$~$=$~$0$ for this and each of the other MOM schemes then 
$\Theta_{a\,\CFs}^{\cal S}(a,\alpha)$ vanishes. 

\sect{Another angle on Crewther's relation.}

While the relation (\ref{deltgen}) appears to hold in all the schemes and
gauges that we have considered so far it is instructive to return to Crewther's
original relation (\ref{deltms}) and consider gauges for which it is valid 
especially in light of the earlier analysis. In other words to determine some 
gauge $\bar{\alpha}$ for which
\begin{equation}
\Delta_{\mbox{\footnotesize{csb}}}(a,\bar{\alpha}) ~=~ 
\frac{\beta(a,\bar{\alpha})}{a} \hat{K}_a(a,\bar{\alpha}) 
\label{deltgauge}
\end{equation}
holds. This was the approach taken in \cite{12} but here rather than assume the 
gauge parameter is a pure constant we will incorporate loop corrections. The
motivation to do so is that various observations noted earlier had properties 
requiring a fixed value of $\alpha$~$=$~$-$~$3$ at a particular loop order
which turned out to be inconsistent when higher order effects were included. An
approach in a similar vein to explore schemes where (\ref{deltgauge}) holds was
given recently in \cite{63}. In particular the $V$ scheme of \cite{13,14} was 
again studied since the coupling constant in that scheme is by construction 
gauge parameter independent being based on the static quark potential. The four
loop renormalization group functions are known \cite{64} and we note that the 
coefficients of the $V$ scheme $\beta$-function are independent of the gauge 
parameter. The first step to achieve the form (\ref{deltgauge}) is to recognize
that for a gauge dependent scheme we have a two dimensional space parameterized
by $a$ and $\bar{\alpha}$. We can attempt to find a gauge for which 
(\ref{deltgauge}) holds by considering curves 
$\bar{\alpha}$~$=$~$\bar{\alpha}(a)$ which are defined by some relation 
$F(a,\bar{\alpha})$~$=$~$0$. In particular, we are interested in curves that 
contain fixed points of the running coupling constant. A natural choice for 
this is the set of curves satisfying
\begin{equation}
\gamma_\alpha(a,\bar{\alpha}) ~+~ S\gamma_a(a,\bar{\alpha}) ~=~ 0
\label{alcond}
\end{equation}
where
\begin{equation}
\gamma_a(a,\alpha) ~\equiv~ \frac{\beta(a,\alpha)}{a} 
\end{equation}
since outside the Landau gauge our fixed points are defined by the requirement
that $\gamma_\alpha(a,\alpha)$~$=$~$0$ and $\gamma_a(a,\alpha)$~$=$~$0$. The
parameter $S$ will be tuned later to examine various scenarios. Under these 
constraints (\ref{deltgen}) becomes
\begin{equation}
\Delta_{\mbox{\footnotesize{csb}}}(a,\bar{\alpha}) ~=~
\left[ K_a(a,\bar{\alpha}) ~-~ SK_\alpha(a,\bar{\alpha}) \right]
\gamma_a(a,\bar{\alpha})
\end{equation}
meaning that
\begin{equation}
\hat{K}_a(a,\bar{\alpha}) ~=~ \left[ K_a(a,\bar{\alpha}) ~-~ 
SK_\alpha(a,\bar{\alpha}) \right] 
\end{equation}
when comparing with (\ref{deltgauge}).

In keeping with our perturbative approach we will expand $\bar{\alpha}$ as a
function of $a$ via
\begin{equation}
\bar{\alpha}(a) ~=~ \sum_{n=0}^\infty \alpha^{(n)} a^n
\label{alexp}
\end{equation}
as well as taking a similar expansion for the two renormalization group
functions
\begin{eqnarray}
\gamma_a(a,\alpha) &=& -~ \beta_0 a ~-~ 
\sum_{n=1}^\infty \left( \sum_{m=0}^{P_m} \beta_n^{(m)} \alpha^m \right) 
a^{n+1} \nonumber \\
\gamma_\alpha(a,\alpha) &=& \sum_{n=1}^\infty \left( \sum_{m=0}^{Q_m}
\gamma_n^{(m)} \alpha^m \right) a^n 
\end{eqnarray}
where $P_1$~$=$~$4$, $P_2$~$=$~$7$, $P_3$~$=$~$5$, $P_4$~$=$~$6$, 
$Q_1$~$=$~$1$, $Q_2$~$=$~$4$, $Q_3$~$=$~$7$, $Q_4$~$=$~$5$ and $Q_5$~$=$~$6$
cover all the schemes considered here. The specific values of $P_m$ and $Q_m$ 
were deduced from the $\alpha$ dependence in the $\MOMg$ scheme with the four 
and five loop ones corresponding to the $\mMOM$ scheme. With these two 
expressions and (\ref{alexp}) solving (\ref{alcond}) produces equations that 
determine $\alpha^{(m)}$. We find the formal solution is
\begin{eqnarray}
\alpha^{(0)}\gamma_1^{(1)} &=& S\beta_0-\gamma_1^{(0)} \nonumber \\
\alpha^{(1)}\gamma_1^{(1)} &=& 
S \left[ \beta_1^{(4)}{\alpha^{(0)}}^4+\beta_1^{(3)}{\alpha^{(0)}}^3
+\beta_1^{(2)}{\alpha^{(0)}}^2+\beta_1^{(1)}{\alpha^{(0)}}
+\beta_1^{(0)} \right] \nonumber\\
&&- \left[ \gamma_2^{(4)}{\alpha^{(0)}}^4+\gamma_2^{(3)}{\alpha^{(0)}}^3
+\gamma_2^{(2)}{\alpha^{(0)}}^2+\gamma_2^{(1)}{\alpha^{(0)}}+\gamma_2^{(0)}
\right] \nonumber \\
\alpha^{(2)}\gamma_1^{(1)} &=& 
S \left[ \beta_2^{(7)}{\alpha^{(0)}}^7+\beta_2^{(6)}{\alpha^{(0)}}^6
+\beta_2^{(5)}{\alpha^{(0)}}^5+\beta_2^{(4)}{\alpha^{(0)}}^4
\right. \nonumber \\
&& \left. ~~~
+\beta_2^{(3)}{\alpha^{(0)}}^3+\beta_2^{(2)}{\alpha^{(0)}}^2
+\beta_2^{(1)}{\alpha^{(0)}}+\beta_2^{(0)} \right] 
\nonumber \\
&& - \left[ \gamma_3^{(7)}{\alpha^{(0)}}^7+\gamma_3^{(6)}{\alpha^{(0)}}^6
+\gamma_3^{(5)}{\alpha^{(0)}}^5+\gamma_3^{(4)}{\alpha^{(0)}}^4
\right. \nonumber \\
&& \left. ~~~
+\gamma_3^{(3)}{\alpha^{(0)}}^3+\gamma_3^{(2)}{\alpha^{(0)}}^2
+\gamma_3^{(1)}{\alpha^{(0)}} + \gamma_3^{(0)} \right] \nonumber\\
&&+~ \alpha^{(1)} \left[ S \left[ 4\beta_1^{(4)}{\alpha^{(0)}}^3
+3\beta_1^{(3)}{\alpha^{(0)}}^2+2\beta_1^{(2)}{\alpha^{(0)}}+\beta_1^{(1)} 
\right] \right. \nonumber\\
&& \left. ~~~~~~~~~- \left[ 4\gamma_2^{(4)}{\alpha^{(0)}}^3
+3\gamma_2^{(3)}{\alpha^{(0)}}^2+2\gamma_2^{(2)}{\alpha^{(0)}}
+\gamma_2^{(1)} \right] \right] \nonumber \\
\alpha^{(3)}\gamma_1^{(1)} &=& 
\sum_{n=0}^{P_4} S\beta_3^{(n)}{\alpha^{(0)}}^n
- \sum_{n=0}^{Q_4} \gamma_4^{(n)}{\alpha^{(0)}}^n
+\alpha^{(1)}\sum_{n=1}^{P_3} nS\beta_2^{(n)}{\alpha^{(0)}}^{n-1}
\nonumber \\
&& -~ \sum_{n=0}^{Q_3} n\gamma_3^{(n)}{\alpha^{(0)}}^{n-1} 
+ 2{\alpha^{(0)}}^{2} \left[ 2{\alpha^{(0)}}{\alpha^{(2)}}
+ 3{\alpha^{(1)}}^{2} \right] 
\left[ S\beta_1^{(4)}-\gamma_2^{(4)} \right]
\nonumber \\
&& +~ 3{\alpha^{(0)}} \left[ {\alpha^{(0)}}{\alpha^{(2)}}+{\alpha^{(1)}}^2 
\right] 
\left[ S\beta_1^{(3)}-\gamma_2^{(3)} \right] \nonumber\\
&& +~ \left[ 2{\alpha^{(0)}}{\alpha^{(2)}}+{\alpha^{(1)}}^2 \right] 
\left[ S\beta_1^{(2)}-\gamma_2^{(2)} \right] 
+{\alpha^{(2)}} \left[ S\beta_1^{(1)}-\gamma_2^{(1)} \right] 
\end{eqnarray}
for the first few orders. We recall that although our first interest at the
moment is on the Crewther relation by finding the expression for $\bar{\alpha}$
that solves (\ref{alcond}) the second component is to find the value of
$a_\infty$ that is the solution to 
$\gamma_a(a_\infty,\bar{\alpha}(a_\infty))$~$=$~$0$. In this way we will
arrive at the fixed points of the running parameters of the theory. Taking this
approach avoids the Banks-Zaks fixed point and will return a value of
$\alpha$~$\approx$~$-$~$3$. This particular integer value was discussed
previously in \cite{20,21,22,23,24,25,26,27,28,29,30,31,32}. In fact, varying 
the defining equation 
\begin{equation}
F(a,\alpha,S) ~\equiv~ \gamma_\alpha(a,\alpha) ~+~ S\gamma_a(a,\alpha)
\end{equation}
with respect to $S$ results in 
\begin{equation}
\frac{dF(a,\alpha,S)}{dS} ~=~ \gamma_a(a,\alpha)
\end{equation}
which should evaluate to zero at a fixed point. So the positions of the fixed 
points determined in this way should be independent of $S$ up to truncation. We
will produce numerical data as evidence of this. In particular choosing 
$S$~$=$~$0$ results in a procedure analogous to the typical process for finding
fixed points where the curves are defined such that the gauge parameter is 
stationary and the value of the coupling constant can then be chosen so that it
is also stationary.

Before considering the details for a specific scheme it is instructive to
examine the leading order equation for the various gauges. First while 
$\beta_0$ will be the same for all gauges since
\begin{equation}
\beta_0 ~=~ -~ \frac{4}{3} \Nf T_F ~+~ \frac{11}{3}C_A 
\end{equation}
$\gamma_1$ is different in each of the three gauges. In a linear covariant 
gauge fixing we have 
\begin{equation}
\gamma_1 ~=~ -~ \frac{1}{2} \alpha C_A ~+~ \frac{13}{6} C_A ~-~
\frac{4}{3} \Nf T_F 
\end{equation}
which means we can write
\begin{equation}
\alpha^{(0)} ~=~ \frac{2}{C_A} 
\left[ S \left[ \frac{4}{3} \Nf T_F ~-~ \frac{11}{3} C_A \right] ~-~ 
\frac{4}{3} \Nf T_F ~+~ \frac{13}{6} C_A \right] ~.
\end{equation}
As this would be singular in the abelian limit it is best to redefine the
parameter $S$ as $S$~$=$~$1$~$+$~$3\sigma C_A$ which gives
\begin{equation}
\alpha^{(0)} ~=~ -~ 3 ~+~ \left[ 8 \Nf T_F ~-~ 22 C_A \right] \sigma ~.
\end{equation}
So $S$~$=$~$1$ produces $\alpha^{(0)}$~$=$~$-$~$3$ corresponding to the 
solution pointed out in \cite{12}. The higher order corrections will be scheme 
dependent. Repeating this leading order exercise produces 
\begin{equation}
\alpha^{(0)}_{\CFs} ~=~ -~ 6 ~+~ \left[ 16 T_F N_f - 44 C_a \right] \sigma
\end{equation}
for the Curci-Ferrari gauge. To analyse the MAG the following modification has 
to be made to the defining equation 
\begin{equation}
F_{\MAGs}(a,\alpha,S) ~=~ \bar{\alpha} \gamma_\alpha(a,\alpha) ~+~ 
\bar{\alpha} S\gamma_a(a,\alpha)
\end{equation}
because the leading order term in $\gamma_\alpha^{\MAGs}(a,\alpha)$ is 
proportional to $\alpha^{-1}$. So the equations for $\alpha^{(i)}$ then need to
be modified in this case. The resulting leading order equation produces two 
solutions. For $S$~$=$~$1$ the numerical values are 
$\alpha^{(0)}_{\MAGs}$~$=$~$-$~$0.430953$ and $-$~$5.569047$. The first of
these solutions corresponds to that termed as Banks-Zaks one found in 
\cite{45}. For each of the three gauges the $S$~$=$~$1$ solutions are 
independent of $\Nf$ unlike the $S$~$=$~$0$ case.

This leading order analysis is scheme independent but this ceases to be the
case when corrections are included. For instance, it was shown in \cite{12} 
that for the linear covariant gauge the form of Crewther's original relation, 
given in (\ref{deltms}) but in the $\mMOM$ scheme, was satisfied at next order 
for particular values of $\alpha$ which were $0$, $-$~$1$ and $-$~$3$. For 
$\alpha$~$=$~$0$ this followed from the $\alpha$ dependence in 
$\gamma_\alpha(a,\alpha)$ whereas $\alpha$~$=$~$-$~$1$ emerged as a special 
case since $K_\alpha \gamma_\alpha(a,\alpha)$ was proportional to $(\alpha+1)$ 
at second order. The situation where $\alpha$~$=$~$-$~$3$ is that value that 
solves $\gamma_a(a,\alpha)$~$=$~$-$~$\gamma_\alpha(a,\alpha)$ at leading order 
which is suggestive of the $\alpha$~$\approx$~$-$~$3$ infrared stable fixed 
point discussed earlier. It is worth examining whether this point of view has 
any credibility at higher order. Therefore we calculated higher order terms in 
the expansion in (\ref{alexp}) for a variety of schemes and gauges. The full 
set of expressions for $\alpha^{(i)}$ are provided in the associated data file 
but we record the $SU(3)$ values for the first few terms are
\begin{eqnarray}
\left. \alpha^{(0)}_{\mMOMs} \right|^{SU(3)} &=& 
-~ 3 + 4\sigma \Nf - 66\sigma \nonumber \\
\left. \alpha^{(1)}_{\mMOMs} \right|^{SU(3)} &=&
725274 {\sigma}^{3} + 41283 {\sigma}^{2} 
+ 3881196 {\sigma}^{4} - \frac{51}{2} - 96 {\sigma}^{3}\Nf^{3} 
+ 5832 {\sigma}^{3}\Nf^{2} \nonumber \\
&& -~ 114048 {\sigma}^{3}\Nf 
+ 132 {\sigma}^{2}\Nf^{2} - 4680 {\sigma}^{2}\Nf 
- 864 {\sigma}^{4}\Nf^{3} + 42768 {\sigma}^{4}\Nf^{2} 
\nonumber \\
&& -~ 705672 {\sigma}^{4}\Nf + 31 \sigma \Nf 
+ \Nf + \frac {261}{2} \sigma \nonumber\\
\left. \alpha^{(2)}_{\mMOMs} \right|^{SU(3)} &=& 
-~ 6562749600 {\sigma}^{6}\Nf^{2} 
+ 59382298800 {\sigma}^{6}\Nf + 103680 {\sigma}^{6}\Nf^{5} 
- 9720000 {\sigma}^{6}\Nf^{4} \nonumber \\
&& +~ 359251200 {\sigma}^ {6}\Nf^{3} 
- \frac{542619}{2} \zeta_3 {\sigma}^{2}
- \frac{1485}{2} \zeta_3 \sigma - 4498659 {\zeta_3} {\sigma}^{3}
\nonumber \\
&& -~ 25150150080 {\sigma}^{7}\Nf^{2} 
+ 207488738160 {\sigma}^{7}\Nf + 559872 {\sigma}^{7}\Nf^{5} 
- 46189440 {\sigma}^{7}\Nf^{4} 
\nonumber \\
&& +~ 1524251520 {\sigma}^{7}\Nf^{3} 
- 17465382 \zeta_3 {\sigma}^{4} - 684712835928 {\sigma}^{7} 
- 27604036251 {\sigma}^{5} 
\nonumber \\
&& -~ 213252314220 {\sigma}^{6} 
+ 6841137204 {\sigma}^{5}\Nf - 12384 {\sigma}^{4}\Nf^{4} 
- 662256 {\sigma}^{5}\Nf^{4} 
\nonumber \\
&& +~ 30458592 {\sigma}^{5}\Nf^{3} 
- 658776888 {\sigma}^{5}\Nf^{2} + 5184 {\sigma}^{5}\Nf^{5} 
- \frac{78139809}{2} {\sigma}^{3} 
\nonumber \\
&& +~ \frac{2146905}{4} {\sigma}^{2}
- \frac{3436534431}{2} {\sigma}^{4} 
+ 3408 {\sigma}^{3}\Nf^{3} - 233370 {\sigma}^{3}\Nf^{2} 
+ 5290650 {\sigma}^{3}\Nf 
\nonumber \\
&& +~ 7242 {\sigma}^{2}\Nf^{2} 
- \frac{271143}{2} {\sigma}^{2}\Nf 
- 72 {\sigma}^{2}\Nf^{3} - \frac{782}{9} \sigma \Nf^{2} 
+ 954396 {\sigma}^{4}\Nf^{3} 
\nonumber \\
&& -~ 27691794 {\sigma}^{4}\Nf^{2} 
+ 356848173 {\sigma}^{4}\Nf 
- 6 {\zeta_3} \Nf - \frac{10}{3} \Nf^{2} - \frac{9405}{8}
- 702 \zeta_3 {\sigma}^{2}\Nf^{2} 
\nonumber \\
&& +~ 28026 \zeta_3 {\sigma}^{2}\Nf 
+ \frac{20}{3} \zeta_3 \sigma \Nf^{2} 
- 65 \zeta_3 \sigma \Nf + 432 \zeta_3 {\sigma}^{3}\Nf^{3} 
- 30780 \zeta_3 {\sigma}^{3}\Nf^{2} 
\nonumber \\
&& +~ 662904 \zeta_3 {\sigma}^{3}\Nf 
+ 3888 \zeta_3 {\sigma}^{4}\Nf^{3} 
- 192456 \zeta_3 {\sigma}^{4}\Nf^{2} 
+ 3175524 \zeta_3 {\sigma}^{4}\Nf 
\nonumber \\
&& +~ \frac{6517}{4} \sigma \Nf
+ \frac{551}{4} \Nf 
+ 153 {\zeta_3} + \frac{909}{8} \sigma 
\end{eqnarray}
in the $\mMOM$ scheme. Up to the truncation order this gives the solution to 
$\bar{\gamma}_a(a)$~$=$~$\bar{\gamma}_\alpha(a)$ allowing us to find a fixed 
point $a_\infty$ with $\bar{\gamma}_a(a_\infty)$~$=$~$0$. When $\Nf$~$=$~$16$ 
and $S$~$=$~$1$ we have 
\begin{eqnarray}
\left. \alpha^{(0)}_{\mMOMs} \right|_{\Nf=16} &=& -~ 3 \nonumber \\
\left. \alpha^{(1)}_{\mMOMs} \right|_{\Nf=16} &=& -~ 9.5 \nonumber \\
\left. \alpha^{(2)}_{\mMOMs} \right|_{\Nf=16} &=& 243.558910
\end{eqnarray}
which gives
\begin{eqnarray}
\gamma_a(a,\bar{\alpha}) &=& -~ 0.333333 a ~+~ 103.666667 a^2 ~+~ 
152.000000 a^3 ~+~ 1060.437500 a^4 \nonumber \\ 
&& +~ 1929.093750 a^5 ~+~ O(a^6)
\end{eqnarray}
to the two loop level. By this we mean the two loop solution of the equations
$\bar{\gamma}_a(a)$~$=$~$0$ and $\bar{\gamma}_\alpha(a)$~$=$~$0$. These have a
positive fixed point at $a_\infty$~$=$~$0.0032000819$ implying 
$\alpha_\infty$~$=$~$-$~$3.030400777$ at this approximation order. The
respective two loop $\mMOM$ fixed point values from \cite{45} are 
$0.0032001941$ and $-$~$3.0301823312$ which shows our $\Nf$~$=$~$16$ solution 
is in reasonable correspondence. We note that by solving this way we have 
introduced an additional truncation in the series for $\alpha$. So it is not 
unexpected that the fixed point found in this way does not result in fully the 
same value. However by including higher order corrections we would expect that 
there would be a degree of convergence. Repeating the exercise at the three 
loop level we find two real fixed point solutions on the $(a,\alpha)$ plane 
which are $(0.0031380752,-3.0274132638)$ and $(0.1624143964,1.8817663994)$. The
first of these has a clear counterpart in the fixed point analysis of \cite{45}
which is $(0.0031380724,-3.0274210489)$ and is an infrared stable fixed point. 
The second solution is not as straightforward to place with a known fixed point
in \cite{45} but may map to the stable fixed point 
$(0.1279084064,1.9051106246)$. 

\begin{table}
\begin{center}
\begin{tabular}{ |c|c||c|c||c|c||c|c| }
 \hline 
& &\multicolumn{2}{|c||}{$\bar{\alpha}$}&\multicolumn{2}{|c||}{\cite{45}}&\multicolumn{2}{|c|}{Difference}  \\
\hline 
 $\Nf$ & Loop & $a_{\infty}$ & $\alpha_{\infty}$& $a_{\infty}$ & $\alpha_{\infty}$ &$\delta_a$&$\delta_\alpha$ \\ 
\hline 
 \hline 
8&2& $0.049337$ & $-3.863406$ &$0.052219$&$-3.795565$&$-0.002882$&$-0.067841$\\ 
 \hline 
 &4&$0.064323$  & $-3.359243$   &$0.060959$&$-3.485534$&$~0.003364$&$~0.126291$\\ 
 \hline 
 &5& $0.046971$& $-2.951933$  &$0.047937$&$-3.192836$&$-0.000966$&$~0.240903$ \\ 
 \hline 
 \hline 
9&2&$0.046323$ & $-3.764323$ &$0.048504$&$-3.705259$&$-0.002181$&$ -0.059064$\\ 
 \hline 
 &4& $0.056983$ & $-2.993633$   &$0.055612$&$-3.212828$&$~0.001371$&$~0.219195 $\\ 
 \hline 
 &5& $0.042375$ & $-2.887326$  &$0.043123$&$-3.003521$&$-0.000748$&$~0.116195 $ \\ 
 \hline 
 \hline 
10&2& $0.042400$ & $-3.657192$  &$0.043898$&$-3.608171$&$-0.001498$&$-0.049021 $\\ 
  \hline 
  &4& $0.046825$ & $-3.006516$  &$0.046629$&$-3.124397$&$~0.000196$&$~ 0.117881$\\ 
 \hline 
   &5&$0.038413$ & $-2.930885$&$0.039060$&$-2.975110$&$-0.000647$&$~0.044225 $\\
 \hline
 \hline 
11&2& $0.037453$ & $-3.543073$   &$0.038351$&$-3.505155$&$-0.000898$&$-0.037918 $\\ 
 \hline 
  &3&  $0.043328$ & $-3.551534$  &$0.042908$&$-3.513882$&$~0.000420$&$-0.037652 $ \\ 
 \hline 
  &4& $0.038483$ & $-3.055520$  &$0.038443$&$-3.114044$&$~0.000040$&$~0.058524 $\\ 
 \hline 
  &5& $0.034366$ & $-3.013171$  &$0.034890$&$-3.025124$&$-0.000524$&$~0.011953 $\\ 
 \hline 
 \hline 
12&2& $0.031475$ & $-3.424915$  &$0.031919$&$-3.398484$&$-0.000444$&$-0.026431 $\\ 
 \hline 
  &3& $0.031275$ & $-3.329545$ &$0.031269$&$-3.323533$&$~0.000006$&$-0.006012 $\\ 
 \hline 
  &4& $0.030877$ & $-3.101381$ &$0.030859$&$-3.126726$&$~0.000018$&$~0.025345 $ \\ 
 \hline 
  &5& $0.029525$ & $-3.092376$  &$0.029873$&$-3.093637$&$-0.000348$&$~0.001261 $\\ 
 \hline 
 \hline 
13&2& $0.024644$ & $-3.308050$  &$0.024812$&$-3.292241$&$-0.000168$&$-0.015809 $\\ 
 \hline 
  &3& $0.023188$ & $-3.213527$  &$0.023188$&$-3.213214$&$~0.000000$&$-0.000313 $\\ 
 \hline 
  &4& $0.023552$ & $-3.122593$  &$0.023543$&$-3.131342$&$~0.000009$&$~0.008749 $\\ 
 \hline 
  &5& $0.023430$ & $-3.131575$  &$0.023575$&$-3.130751$&$-0.000145$&$-0.000824 $ \\ 
 \hline 
 \hline 
14&2& $0.017346$ & $-3.199477$  &$0.017389$&$-3.191979$&$-0.000043$&$-0.007498 $ \\ 
 \hline 
  &3& $0.016146$ & $-3.138109$  &$0.016146$&$-3.138534$&$~0.000000$&$~0.000425 $ \\ 
 \hline 
  &4& $0.016447$ & $-3.112174$  &$0.016445$&$-3.114255$&$~0.000002$&$~0.002081 $\\ 
 \hline 
  &5& $0.016525$ & $-3.119407$  &$0.016555$&$-3.118834$&$-0.000030$&$-0.000573 $\\ 
 \hline 
 \hline 
15&2& $0.010064$ & $-3.105672$  &$0.010070$&$-3.103310$&$-0.000006$&$-0.002362 $\\ 
 \hline 
  &3& $0.009537$ & $-3.080460$  &$0.009537$&$-3.080626$&$~0.000000$&$~0.000166 $\\ 
 \hline 
  &4& $0.009636$ & $-3.076696$  &$0.009636$&$-3.076936$&$~0.000000$&$~0.000240 $\\ 
 \hline 
  &5& $0.009658$ & $-3.078272$  &$0.009660$&$-3.078166$&$-0.000002$&$-0.000106 $\\ 
 \hline 
 \hline 
16&2& $0.003200$ & $-3.030401$   &$0.003200$&$-3.030182$&$~0.000000$&$-0.000219 $\\ 
 \hline 
  &3& $0.003138$ & $-3.027413$  &$0.003138$&$-3.027421$&$~0.000000$&$ ~0.000008$\\ 
 \hline 
  &4& $0.003143$ & $-3.027352$  &$0.003143$&$-3.027354$&$~0.000000$&$~0.000002 $\\ 
 \hline 
  &5& $0.003143$ & $-3.027378$  &$0.003143$&$-3.027377$&$~0.000000$&$-0.000001$\\ 
 \hline 
\end{tabular}
\end{center}
\begin{center}
\caption{Fixed points solutions for $a_\infty$ and $\alpha_\infty$ to the 
curve $\gamma_a(a,\bar{\alpha})$~$=$~$-$~$\gamma_\alpha(a,\bar{\alpha})$, in
the columns headed with $\bar{\alpha}$ and the corresponding solutions for the
fixed points in the $(a,\alpha)$ plane given in \cite{45} together with the 
difference $\delta$ in values in the final pair of columns up to the five loop
level in the $\mMOM$ scheme.}
\label{albarmmom}
\end{center}
\end{table}

While this gives a flavour of the situation at the first few orders for 
$\Nf$~$=$~$16$ we have analysed the remaining values of $\Nf$ in the conformal 
window and the results are recorded in Table \ref{albarmmom}. In that table
columns $3$ and $4$ record the solutions for all values of $\Nf$ in the
conformal window for the $SU(3)$ group. The next pair of columns records the 
values of the critical point solutions of (\ref{betai}) from \cite{45} for 
comparison. The final two columns record the difference $\delta_a$ and
$\delta_\alpha$ of the respective values of $a_\infty$ and $\alpha_\infty$. At 
the upper end of the conformal window, where perturbation theory is more 
reliable, there is a clear indication that the solution for $\bar{\alpha}$ is 
in good agreement with the fixed point of \cite{45}. This continues to be the 
case for lower values of $\Nf$ but the agreement becomes less pronounced except
at high loop order. Below around $\Nf$~$=$~$10$ one strays into territory where
the critical coupling value ceases to be small and perturbation theory may be 
less reliable for drawing a precise conclusion. However in the region of 
validity it does support the premise that the Crewther relation reflects
 conformal properties. Indeed to complete this $\mMOM$ analysis we have 
calculated the value of $C_{{\mbox{\footnotesize{Bjr}}}}^{\mMOMs}(a,\alpha) 
C_{{\mbox{\footnotesize{Adl}}}}^{\mMOMs}(a,\alpha)$ for $\Nf$~$=$~$16$ at
successive loop orders and present the outcome in Table \ref{csbalbarmmom}. 
Again we observe that in effect the product evaluates to $3$ similar to the
earlier cases.

\begin{table}[ht]
\begin{center}
 \begin{tabular}{ |c|c||c|c|}
 \hline \multicolumn{4}{|c|}{$\mMOM$ 2 loop} \\
 \hline \hline
\rule{0pt}{12pt}
 $a_{\infty}$ & $\alpha_{\infty}$&$O(a^3)$&$O(a^4)$\\ \hline
$0.0032000819$ & $-3.0304007777$ & $2.9999982454$ & $3.0000012447$ \\ \hline
 \hline \multicolumn{4}{|c|}{$\mMOM$ 3 loop} \\
 \hline \hline
\rule{0pt}{12pt}
 $a_{\infty}$ & $\alpha_{\infty}$&$O(a^3)$&$O(a^4)$\\ \hline
$0.0031380752$ & $-3.0274132638$ & $2.9999973440$ & $3.0000001217$ \\ \hline
$0.1624143964$ & $1.8817663994$ & $9.7789795843$ & $20.2267682549$ \\ \hline
 \hline \multicolumn{4}{|c|}{$\mMOM$ 4 loop} \\
 \hline \hline
\rule{0pt}{12pt}
 $a_{\infty}$ & $\alpha_{\infty}$&$O(a^3)$&$O(a^4)$\\ \hline
 $0.0031430140$ & $-3.0273515306$ & 
$2.9999974127$ & $3.0000002081$ \\ \hline
$0.0922743518$ & $0.7554173647$ & $4.1917263936$ & $6.5855502329$ \\ \hline
 \hline \multicolumn{4}{|c|}{$\mMOM$ 5 loop} \\
 \hline \hline
\rule{0pt}{12pt}
 $a_{\infty}$ & $\alpha_{\infty}$&$O(a^3)$&$O(a^4)$\\ \hline
  $0.0031434057$ &$-3.0273781422$ & $2.9999974182$ & $3.0000002149$\\ \hline
$0.0498699873$ & $-3.9546945229$ &  $3.1876456375$ &$3.2609125084$ \\ \hline
\end{tabular}
 \end{center}
\begin{center}
\caption{Values of $C_{{\mbox{\footnotesize{Bjr}}}}^{\mMOMs}(a,\alpha) 
C_{{\mbox{\footnotesize{Adl}}}}^{\mMOMs}(a,\alpha)$ from the solution for 
$\bar{\alpha}$ at successive loop orders for $\Nf$~$=$~$16$.}
\label{csbalbarmmom}
\end{center}
\end{table}

\begin{table}
\begin{center}
\small{
\begin{tabular}{|c|c|c|c|c|}
\hline
Loop& $a_\infty$ & $\bar{\alpha}(a_\infty)$&$S=1$& \cite{45} \\  \hline
\hline 
2&0.003311 &$-2.952784 - 0.000061 S - 0.000245 S^2$&$-2.953090$&$-2.952985$\\ \hline
3&0.003162 &$-2.945674 - 0.000400 S + 0.000218 S^2$&$-2.945856$&$-2.945839$\\
&&$ - 6.171139\times10^{-8} S^3$ &&\\\hline
4&0.003170 &$-2.945920 - 0.000033 S + 0.000016 S^2 $&$-2.945937$&$-2.945935$\\
&&$+1.111287 \times10^{-7} S^3 - 2.848273\times10^{-10} S^4$ &&\\\hline
 &0.091700 &$-3.819444 + 8.65019 S - 4.55718 S^2 $&$0.2776970$&$-4.407403$\\
&& $+ 0.00413990 S^3 - 6.895175\times 10^{-6} S^4$&&\\\hline
5&0.003171 &$-2.945966 + 1.06321\times10^{-6} S - 8.15566\times 10^{-7} S^2  $&$-2.945966$&$-2.945965$ \\
&&$- 3.72127\times 10^{-8} S^3+ 8.257949\times 10^{-10} S^4$ &&\\
&&$- 3.041220\times 10^{-14} S^5$ &&\\\hline
 &0.046999 &$-4.691390 + 2.754610 S - 1.377941 S^2  $&$-3.321283$& $-3.682521$ \\
&&$- 0.006615 S^3+0.000053 S^4 - 1.467961\times10^{-9} S^5$ &&\\
\hline
\end{tabular}}
\end{center}
\begin{center}
\caption{Values of $\bar{\alpha}(a_\infty)$ for $\Nf$~$=$~$16$ in the $\MSbar$
scheme as a function of $S$.}
\label{albarmsS}
\end{center}
\end{table}

\begin{table}
\begin{center}
\small{
\begin{tabular}{|c|c|c|c|c|}
\hline
Loop& $a_\infty$ & $\bar{\alpha}(a_\infty)$&$S=1$& \cite{45} \\  \hline
\hline 
2&0.003311 &$-3.032990 - 0.005566S - 0.000559S^2$&$-3.039142$&$-3.040393$\\
&&$ -0.000027 S^3$&&\\ \hline
3&0.003162 &$-3.028600 -0.001502 S +0.000424 S^2$&$-3.029633$&$-3.029389$\\
&&$ +0.000045 S^3 -4.220754\times10^{-7} S^4 -1.523738\times 10^{-8} S^5$&&\\ \hline
4&0.003170 &$-3.027400  -0.000072 S + 0.000095 S^2$&$-3.027390$&$-3.027317$\\
&&$ -0.000014 S^3 + 1.097536\times 10^{-6} S^4 +5.156952\times 10^{-8} S^5$&&\\
&& $-4.450271\times 10^{-10} S^6 -1.186739\times 10^{-11} S^7$&& \\ \hline
 &0.091700 &$29.170852 + 37.509021 S  -7.270245 S^2$&58.084230&$-7.001486$\\
&&$  -1.363470 S^3 + 0.036484 S^4 +0.001606 S^5$&&\\
&&$ -0.000011 S^6 -2.872889\times 10^{-7} S^7$&&\\ \hline
5&0.003171 &$-3.027362 + 7.648957\times 10^{-6} S + 0.000011 S^2$&$-3.027348$&$-3.027350$\\
&&$  -3.795632\times 10^{-6}S^3  -7.923119\times 10^{-7} S^4 $ && \\
&&$ -5.731136\times 10^{-8} S^5 +1.945960\times 10^{-9} S^6 $ && \\
&&$ -5.788853\times 10^{-11} S^7-5.007868\times 10^{-13} S^8$&&\\
&&$  -1.052508\times 10^{-14} S^9$&&\\\hline
 &0.046999 &$3.423566 + 9.178962 S -4.964152 S^2$&7.849534&$-6.454812$\\
&&$ + 0.302490 S^3  -0.086406 S^4 -0.005043 S^5$&&\\
&&$ +0.000114 S^6 +3.328853\times 10^{-6} S^7-2.417239\times 10^{-8} S^8  $&&\\
&&$- 5.080334\times 10^{-10} S^9$&&\\ \hline
\end{tabular}}
\end{center}
\begin{center}
\caption{Values of $\bar{\alpha}(a_\infty)$ for $\Nf$~$=$~$16$ in the $\RI$
scheme as a function of $S$.}
\label{albarriS}
\end{center}
\end{table}

One question that arises is whether the choice of $S$ affects the analysis and
if so to what extent. While we could consider the numerical fixed points for an
array of values of $S$, instead we will consider the fixed points in the 
$\MSbar$ and $\RI$ schemes since the running of the coupling constant is
$\alpha$ independent in both these schemes. This will allow us to calculate the 
fixed points as a function of $S$. The procedure is the same as for the
$\mMOM$ scheme except that $\gamma_a(a_\infty)$~$=$~$0$ immediately determines
$a_\infty$. So solving 
$\gamma_\alpha(a,\alpha)$~$+$~$S\gamma_a(a,\alpha)$~$=$~$0$ means the $S$ 
dependence appears solely in $\bar{\alpha}(a_\infty)$ and allows us to study 
the relation to the loop order. Repeating the earlier example for
$\Nf$~$=$~$16$ we find
\begin{eqnarray}
\alpha_{\MSbars}^{(0)}&=& -~ 2.777778 ~-~ 0.222222 S \nonumber \\
\alpha_{\MSbars}^{(1)}&=& -~ 52.851852 ~+~ 67.092593 S ~-~ 
0.074074S^2 \nonumber\\
\alpha_{\MSbars}^{(2)}&=& -~ 78.713043 ~+~ 968.858403 S ~+~ 45.225583 S^2 ~-~
0.006173 S^3 \nonumber\\
\alpha_{\MSbars}^{(3)}&=& 5792.714293  ~-~ 7038.053127 S ~-~ 
6394.380328 S^2 ~+~ 5.436159 S^3 ~-~ 0.008942 S^4 \nonumber\\
\alpha_{\MSbars}^{(4)}&=& 29279.884317 ~-~ 325028.480677 S ~-~ 
166111.075407 S^2 ~-~ 1468.595029 S^3 \nonumber \\
&& +~ 10.989272 S^4 ~-~ 0.000301 S^5
\end{eqnarray}
which lead to expressions for $\bar{\alpha}(a_\infty)$ given in Table 
\ref{albarmsS} at successive loop order as a function of $S$. Included in the
table are the evaluation of $\bar{\alpha}(a_\infty)$ at $S$~$=$~$1$ and the
critical point value from \cite{45} to compare with. Clearly in the first 
instance the $S$~$=$~$0$ values are an accurate estimate of the eventual
$S$~$=$~$1$ value. Equally as the loop order increases the dependence on $S$
washes out quickly if one uses the coefficients of $S$ itself as a guide. At 
four and five loops additional fixed points emerged. In each case the 
coefficients of $S$ in the polynomial are relatively large for low powers which
is due to the large critical coupling constant value. 

\begin{table}
\begin{center}  
\begin{tabular}{ |c|c||c|c||c|c| }
 \hline 
 \multicolumn{6}{|c|}{Curci-Ferrari Gauge} \\ 
 \hline 
 \hline & &\multicolumn{2}{|c||}{$\bar{\alpha}$}&\multicolumn{2}{|c|}{\cite{45}} \\
\hline 
 $N_f$ & Loop & $a_{\infty}$ & $\alpha_{\infty}$& $a_{\infty}$ & $\alpha_{\infty}$  \\ 
\hline 
 \hline 
8&3& $0.116505$ & $5.425313$ &$0.116505$&$-0.525277$\\ 
 \hline 
 \hline 
9&2& $0.416667$ & $7.333333$ &$0.416667$&$-2.387888$\\ 
 \hline 
 &3& $0.081803$ & $1.386711$ &$0.081803$&$-1.239118$\\ 
 \hline 
 \hline 
10&2& $0.175676$ & $0.207207$ &$0.175676$&$-2.981627$\\ 
 \hline 
  &3& $0.060824$ & $-0.762250$ &$0.060824$&$-1.988405$\\ 
 \hline 
 \hline 
11&2& $0.098214$ & $-2.202381$ &$0.098214$&$-3.548192$\\ 
 \hline 
  &3& $0.046039$ & $-2.180305$ &$0.046039$&$-2.744498$\\ 
 \hline 
 \hline 
12&2& $0.060000$ & $-3.480000$ &$0.060000$&$-4.078489$\\ 
 \hline 
  &3& $0.034607$ & $-3.235417$ &$0.034607$&$-3.479125$\\ 
 \hline 
 \hline 
13&2& $0.037234$ & $-4.312057$&$0.037234$&$-4.569040$ \\ 
 \hline 
  &3& $0.025191$ & $-4.075904$ &$0.025191$&$-4.169158$\\ 
 \hline 
 \hline 
14&2& $0.022124$ & $-4.923304$ &$0.022124$&$-5.020489$\\ 
 \hline 
  &3& $0.017070$ & $-4.768441$ &$0.017070$&$-4.796907$\\ 
 \hline 
 \hline 
15&2& $0.011364$ & $-5.409091$ &$0.011364$&$-5.435844$\\ 
 \hline 
  &3& $0.009818$ & $-5.342638$ &$0.009818$&$-5.347954$\\ 
 \hline 
 \hline 
16&2& $0.003311$ & $-5.816777$ &$0.003311$&$-5.819105$\\ 
 \hline 
  &3& $0.003162$ & $-5.808281$ &$0.003162$&$-5.808458$\\ 
 \hline \hline
 \multicolumn{6}{|c|}{MAG} \\ 
 \hline 
8&3& $0.116505$ & $-14.454928$ &$0.116505$&$-14.732061$\\ 
 \hline 
 \hline 
9&2& $0.416667$ & $-5.748610$ &$0.416667$&$-1.775685$\\ 
 \hline 
 &3& $0.081803$ & $-9.342567$ &$0.081803$&$-6.679483$\\ 
 \hline 
 \hline 
10&2& $0.175676$ & $-5.644755$ &$0.175676$&$-2.041475$\\ 
 \hline 
  &3& $0.060824$ & $-7.302268$ &$0.060824$&$-9.374058$\\ 
 \hline 
 \hline 
11&2& $0.098214$ & $-5.611372$ &$0.098214$&$-2.385628$\\ 
 \hline 
  &3& $0.046039$ & $-6.350328$ &$0.046039$&$-1.943195$\\ 
 \hline 
 \hline 
12&2& $0.060000$ & $-5.594904$  &$0.060000$&$-2.822900$\\ 
 \hline 
  &3& $0.034607$ & $-5.885931$ &$0.034607$&$-2.691103$\\ 
 \hline 
 \hline 
13&2& $0.037234$ & $-5.585093$ &$0.037234$&$-3.356225$\\ 
 \hline 
  &3& $0.025191$ & $-5.668820$ &$0.025191$&$-3.414438$\\ 
 \hline 
 \hline 
14&2& $0.022124$ & $-5.578581$ &$0.022124$&$-3.968278$\\ 
 \hline 
  &3& $0.017070$ & $-5.583155$ &$0.017070$&$-4.074876$\\ 
 \hline 
 \hline 
15&2& $0.011364$ & $-5.573944$ &$0.011364$&$-4.620179$\\ 
 \hline 
  &3& $0.009818$ & $-5.563763$ &$0.009818$&$-4.684705$\\ 
 \hline 
 \hline 
16&2& $0.003311$ & $-5.570474$ &$0.003311$&$-5.263519$\\ 
 \hline 
  &3& $0.003162$ & $-5.568155$ &$0.003162$&$-5.273175$\\ 
 \hline 
\end{tabular}
\end{center}
\begin{center}
\caption{Fixed points solutions for $a_\infty$ and $\alpha_\infty$ to the curve
$\gamma_a(a,\bar{\alpha})$~$=$~$-$~$\gamma_\alpha(a,\bar{\alpha})$ in the
Curci-Ferrari gauge and MAG in the $\MSbar$ scheme.}
\label{albarmsScfmag}
\end{center}
\end{table}

While this analysis is in the $\MSbar$ scheme, similar to that of the original
Crewther study \cite{5}, it is worth repeating the exercise in the $\RI$ scheme
to ascertain whether the same picture emerges. Therefore taking $\Nf$~$=$~$16$ 
again we find 
\begin{eqnarray}
\alpha_{\RIs}^{(0)} &=& -~ 2.777778 ~-~ 0.222222 S \nonumber \\
\alpha_{\RIs}^{(1)} &=& -~ 77.074074 ~+~ 65.430041 S ~-~ 0.168724S^2 ~-~ 
0.008230 S^3 
\nonumber\\
\alpha_{\RIs}^{(2)} &=& -~ 712.810800 ~+~ 1384.485718 S ~+~ 95.815127 S^2 ~+~ 
7.080171 S^3 ~-~ 0.042219 S^4 \nonumber \\
&& -~ 0.001524 S^5 \nonumber\\
\alpha_{\RIs}^{(3)} &=& 58371.788435 ~+~  26052.906414 S ~-~ 
10453.261109 S^2 ~-~ 1844.453717 S^3 \nonumber \\
&& +~ 47.775386 S^4 ~+~ 2.099823 S^5 ~-~ 0.013971 S^6 ~-~ 0.000373 S^7 
\nonumber \\
\alpha_{\RIs}^{(4)} &=& 1094061.824587 ~+~  115396.980509 S ~-~ 
836720.663333 S^2 ~+~ 98112.588891 S^3 \nonumber \\
&& -~ 18705.923877 S^4 ~-~ 1077.589340 S^5~+~ 23.656656 S^6 ~+~ 
0.690162 S^7 ~-~ 0.004954 S^8 \nonumber \\
&& -~ 0.000104 S^9
\end{eqnarray}
with the subsequent implications for $\alpha(a_\infty)$ recorded in Table
\ref{albarriS}. We recall that as the $\RI$ $\beta$-function has the same
coefficients as those of the $\MSbar$ scheme the critical couplings will be the
same and so there will also be four and five loop solutions. However comparing
the solutions for each coefficient $\alpha_{\RIs}^{(i)}$ with its $\MSbar$
partner we note that in the $\RI$ scheme the degree of the polynomial solution
in $S$ at each loop order is higher. Despite this the same outcome emerges when
$S$ is set to unity in that the numerical value is in very good agreement with 
the fixed point values given in \cite{45}. Again the convergence to a 
practically $S$ independent solution for $\alpha(a_\infty)$ transpires. This is
indicative of the idea argued earlier that the condition 
$\gamma_a(a,\alpha)$~$+$~$S\gamma_\alpha(a,\alpha)$~$=$~$0$ can be used to find
the stable infrared fixed point and also that the value of $S$ is unimportant. 
This means that while the value of $S$~$=$~$1$ has properties of interest 
specifically to studies of the Crewther relation the relation of this 
particular curve to the stable infrared fixed point is no more important than 
any similar curve generated from any other value of $S$.

To this point we have concentrated on the linear covariant gauge with this
approach. However we have also examined the two nonlinear gauges in depth and
for the kinematic schemes of \cite{37,38} although we are limited to three
loops in this case. The outcome is in essence the same as the linear gauge
case except that the leading value of $\bar{\alpha}$ is in the neighbourhood
of $\alpha_{\CFs}$~$=$~$-$~$6$ for the Curci-Ferrari gauge. Due to the nature 
of $\gamma_\alpha(a,\alpha)$ for the MAG the leading value of $\alpha$ is
$\alpha_{\MAGs}$~$=$~$-$~$5.56$. For both gauges the same property is present
as the linear gauge in that over the range of the conformal window the higher
order corrections do not unduly alter these respective $\alpha$ values.  
Equally for values of $\Nf$ down to around $12$ the solutions for 
$\bar{\alpha}$ are in keeping with the fixed point solutions of \cite{45}. We 
would expect this to be improved for lower values of $\Nf$ when higher order 
corrections become available. This viewpoint can be seen in Table 
\ref{albarmsScfmag}. Similar tables are also available in the data file 
associated with this article but they are more comprehensive in that they 
include extra fixed points which have a mapping to the $\bar{\alpha}$ solution.
In addition we include the results for all the schemes and gauges we have 
discussed in the article. Examining these wider tables it is evident that for 
certain schemes and gauges the connection with the fixed points of \cite{45} 
and the solutions found by the method of this section is remarkably stable even
at the lower end of the conformal window. Although for the MOM schemes of 
\cite{37,38} only a few perturbative orders are available. We note that in 
several schemes there were no fixed point solutions for various $\Nf$ values in
\cite{45}. In that instance there are no entries but we did find solutions for 
$\bar{\alpha}$. It is not clear whether this is indicating that there will be 
solutions for fixed points in these cases when higher order perturbative 
results become available in those schemes. 

\sect{Wider perspective.}

It is worth taking stock of the wider perspective that our investigations have
arrived at. We begin by putting the choice of the left hand side of 
(\ref{alcond}) in a field theory context. Clearly the combined renormalization
group function
\begin{equation}
\hat{\gamma}_\alpha(a,\alpha) ~=~ \gamma_\alpha(a,\alpha) ~+~ 
\gamma_a(a,\alpha)
\label{modgammaalpha}
\end{equation}
plays an important role in the analysis of the Crewther relation extension as
well as being connected to the special choice of $\alpha$~$=$~$-$~$3$ in the
linear covariant gauge as illustrated in the previous section. To understand 
the origin of (\ref{modgammaalpha}) we recall the purely gluonic sector of the 
Lagrangian of a non-abelian gauge theory is
\begin{equation}
L^{\mbox{\footnotesize{gluonic}}} ~=~ 
-~ \frac{1}{4} G^a_{\mu\nu} G^{a \, \mu\nu} ~-~
\frac{1}{2\alpha} \left( \partial^\mu A^a_\mu \right)^2 
\label{laggluon}
\end{equation}
for a linear covariant gauge fixing. The field strength tensor is given by
\begin{equation}
G^a_{\mu\nu} ~=~ \partial_\mu A^a_\nu ~-~ \partial_\nu A^a_\mu ~+~
g f^{abc} A^b_\mu A^c_\nu
\label{fsdef}
\end{equation}
where $g$ is the coupling and $f^{abc}$ are the gauge group structure
constants. While this is invariably the usual presentation of the gluonic 
sector and the one used for perturbative computations one can always redefine
the fields and variables without altering any physical predictions. Therefore 
defining the rescaling
\begin{equation}
\hat{A}^a_\mu ~=~ g A^a_\mu
\end{equation}
we have 
\begin{equation}
L^{\mbox{\footnotesize{gluonic}}} ~=~ 
-~ \frac{1}{4a} \hat{G}^a_{\mu\nu} \hat{G}^{a \, \mu\nu} ~-~
\frac{1}{2\alpha a} \left( \partial^\mu \hat{A}^a_\mu \right)^2
\label{laggluonhat}
\end{equation}
where $a$~$=$~$g^2/(16\pi^2)$ and $\hat{G}^a_{\mu\nu}$ corresponds to 
(\ref{fsdef}) at the formal value of $g$~$=$~$1$. Formulating the gluonic 
sector in this way identifies the renormalization of the dimension four 
operator $\hat{G}^a_{\mu\nu} \hat{G}^{a \, \mu\nu}$ with its associated 
coupling which is $1/a$ and whose renormalization group function is related to 
the $\beta$-function. One can view the second term of (\ref{laggluonhat}) in 
the same light and regard the coupling associated with the linear gauge fixing 
term as $\alpha a$. Clearly both operators are independent which can be seen by 
focussing on the quadratic part of (\ref{laggluonhat}). The field strength term
is transverse while the other is purely longitudinal. Therefore the running of 
the associated couplings will be different with that of the latter operator
being governed by (\ref{modgammaalpha}). At a more formal level the treatment
of the transverse and longitudinal components of the gauge field as separate
entities within loop computations has been recognized earlier. See, for
instance, \cite{16,65,66,67,68,69}. Indeed in \cite{68,69} the concept of 
invariant charges of the two operators was coined and formalized within the 
Hopf algebra construct of renormalization theory. Therefore from the point of
view of (\ref{laggluonhat}) a more appropriate formulation would be to treat 
$\alpha a$ as a second coupling constant in the renormalization group running 
in a gauge theory. Moreover while we concentrated on the linear covariant gauge
repeating the rescaling argument for the Curci-Ferrari gauge and MAG will 
produce the same observation with regard to the definition of a second coupling
from the two independent operators of the quadratic sector of 
$L^{\mbox{\footnotesize{gluonic}}}$. For completeness we note the first few 
terms of (\ref{modgammaalpha}) for the three gauges are
\begin{eqnarray}
\hat{\gamma}_\alpha^{\lins , \MSbars}(a,\alpha) &=& 
-~ \left[ \alpha + 3 \right] \frac{C_A a}{2} ~+~
\left[ 40 \Nf T_F - 6 \alpha^2 C_A - 33 \alpha C_A - 95 C_A \right] 
\frac{C_A a^2}{24} ~+~ O(a^3) \nonumber \\
\hat{\gamma}_\alpha^{\CFs , \MSbars}(a,\alpha) &=& 
-~ \left[ \alpha + 6 \right] \frac{C_A a}{4} ~+~
\left[ 80 \Nf T_F - 3 \alpha^2 C_A - 51 \alpha C_A - 190 C_A \right]
\frac{C_A a^2}{48} ~+~ O(a^3) \nonumber \\
\hat{\gamma}_\alpha^{\MAGs , \MSbars}(a,\alpha) &=& 
-~ \left[ 2 \alpha^2 \Nda + \alpha^2 \Noda + 12 \alpha \Nda + 6 \alpha \Noda
+ 12 \Nda \right] \frac{C_A a}{4 \alpha \Noda} \nonumber \\
&& +~ \left[ 512 \Nda \Noda \Nf T_F + 80 \alpha \Noda^2 \Nf T_F
- 30 \alpha^3 C_A \Nda^2 - 27 \alpha^3 C_A \Nda \Noda 
\right. \nonumber \\
&& \left. ~~~~
- 3 \alpha^3 C_A \Noda^2 - 366 \alpha^2 C_A \Nda^2 
- 339 \alpha^2 C_A \Nda \Noda - 51 \alpha^2 C_A \Noda^2 
\right. \nonumber \\
&& \left. ~~~~
+ 294 \alpha C_A \Nda^2 - 647 \alpha C_A \Nda \Noda - 190 \alpha C_A \Noda^2 
+ 160 \alpha \Nda \Noda \Nf T_F 
\right. \nonumber \\
&& \left. ~~~~
+ 2016 C_A \Nda^2 - 928 C_A \Nda \Noda \right] 
\frac{C_A a^2}{48 \alpha \Noda^2} ~+~ O(a^3) ~.
\end{eqnarray}
Clearly the leading terms of the first two gauges indicate the special gauge
choices that emerged in the Crewther analysis. For the MAG the leading term
vanishes at two values of $\alpha$ which are
\begin{equation}
\alpha_{\MAGs} ~=~ -~ 3 ~\pm~ \sqrt{\frac{3[2\Nda+3\Noda]}{[2\Nda+\Noda]}} 
\end{equation}
and reproduces the Curci-Ferrari values in the $\Nda$~$\to$~$0$ limit. Also
for example
\begin{equation}
\left. \alpha_{\MAGs} \right|^{SU(\Nc)} ~=~ 
-~ 3 ~\pm~ \sqrt{\frac{3[3\Nc+2]}{[\Nc+2]}} ~.
\end{equation}
For $SU(3)$ the two values are $-$~$3$~$\pm$~$\sqrt{\frac{33}{5}}$ which equate
to $-$~$0.43095348$ and $-$~$5.56904651$ respectively. It is important to note
that these special gauge parameter values are constructed from the scheme 
independent terms of $\hat{\gamma}(a,\alpha)$ and therefore will be the 
foundation for the infrared stable fixed points in all schemes at higher
orders. 

It is clear the Crewther relation has provided another example where a special 
value of the gauge parameter plays a crucial role in a deeper property of QCD
as observed in \cite{12} in the $\mMOM$ scheme. The other situations where this
occurred were in a variety of different problems and it was worth pausing to
understand if there is any underlying connections as to when this arises. 
Moreover by doing so would provide a signpost where to expect the special case 
to occur in other computations. Examining the various articles where 
$\alpha$~$=$~$-$~$3$ has been singled out, 
\cite{20,21,22,23,24,25,26,27,28,29,30,31,32},
several common themes emerge. For instance, in \cite{23,24,25,26} the quark
$2$-point function was the main topic. In \cite{23} this involved solving the
Schwinger-Dyson equation defining the quark $2$-point function to examine the
quark mass gap to ascertain the condition when chiral symmetry was broken. 
Although \cite{24,25} dealt with mesonic quark bound states in a gauge 
invariant setup the path integral and worldline approach effectively trade off
gauge invariance for path dependence. This resulted in what was termed the
connector having a renormalization that at one loop depended on $(\alpha+3)$.
The special choice appeared to remove residual dependence on path and gauge at
one loop. For the approaches in \cite{25,26,30} the common thread was the
resummation of a class of Feynman diagrams which invariably were one loop
bubble graphs. In \cite{25,26} this naive non-abelianization was used to study 
the evolution of quark bilinear operators that underpin deep inelastic 
scattering processes. As the treatment of these operators was by insertion in a
quark $2$-point function the Green's function is $\alpha$ dependent. What was
interesting is that the choice of $\alpha$~$=$~$-$~$3$ provided a reliable 
estimate of the one loop $\Nf$ independent part of the operator anomalous 
dimension for a range of moments. This accuracy was not as reliable for the
two loop case. Similarly in \cite{30} a quark based Green's function was 
studied from which an effective coupling was constructed. Its partial 
resummation indicated that asymptotic freedom was conditional on the special
gauge value. In \cite{31} the same gauge parameter choice showed a remarkable
connection between a set of gluonic bubble insertions in a quark current
correlation function and the one loop gluonic contribution to the $V$ scheme
coupling constant derived from the static quark potential. By contrast the
emergence of $\alpha$~$=$~$-$~$3$ was not restricted to quark based quantities.
In \cite{27,28,29} a dimension six operator correction to the Yang-Mills
action was used to probe infrared gluon dynamics to determine what conditions
if any led to the gluon propagator having a double pole in the $p^2$ where
$p$ is the momentum. Such a behaviour would lead to a linear confining
potential. Solving the Schwinger-Dyson equation for the ghost propagator with
a massless double pole gluon propagator, that derived from the dimension six
operator, the solution required $\alpha$~$=$~$-$~$3$. 

From an overall point of view the studies that have identified 
$\alpha$~$=$~$-$~$3$ as having particular significance share several similar
features. Aside from all using the linear covariant gauge the calculations were
all carried out using the $\MSbar$ scheme. In addition most involved 
calculations of Green's functions which would introduce gauge parameter
dependence. For instance, although the twist-$2$ operators used in the operator
product expansion for deep inelastic scattering that were the focus of
\cite{25,26} are gauge invariant, the quark $2$-point function where they were 
inserted is gauge variant. Therefore the expression will be $\alpha$ dependent
but the operator renormalization constant extracted from this Green's function 
will be gauge parameter independent in the $\MSbar$ scheme. By contrast if the 
$2$-point correlation function for the twist-$2$ operator was evaluated then it
would be gauge parameter independent in the $\MSbar$ scheme since the operator 
is gauge invariant. In other words the same result would ensue for the operator
correlation function if it was evaluated in the Curci-Ferrari gauge or MAG in 
the $\MSbar$ scheme. However if the operator correlation function was evaluated
in a scheme such as $\mMOM$ or the MOM schemes of \cite{37,38} then it would 
depend on the gauge parameter. This is similar to the situation of the 
anomalous dimension of a gauge invariant operator which depends on the gauge 
parameter in general but not in a scheme such as $\MSbar$. In other words the 
correlation function of gauge invariant operators is in general gauge parameter
dependent but independent of it in a subset of renormalization schemes. 

In distilling the essence of previous observations of a special gauge parameter 
value it is evident now where the Crewther relation rests within this 
framework. The two ingredients of the relation are correlation functions of 
gauge invariant operators which are the vector and axial vector quark currents. 
Therefore in the $\MSbar$ scheme each correlation function will be gauge 
parameter independent meaning the Crewther relation will be too. Equally 
applying the formalism to effect a change of scheme to say $\mMOM$ will produce
gauge parameter dependence in the separate correlation functions and hence 
produce a similar dependence in $\Delta_{\mbox{\footnotesize{csb}}}(a,\alpha)$.
This is evident in \cite{12}. We have shown how that can be accommodated 
consistently using the renormalization group formalism in (\ref{deltgenbar}). A
similar mapping that established (\ref{deltgenbar}) for the Crewther case could
equally well be applied to the quantities computed in 
\cite{22,23,24,25,26,30,31}. What is important to remember is that the gauge 
parameter dependence of any operator correlation function does not affect any 
prediction from its evaluation. The two underlying parameters, $a$ and 
$\alpha$, both run with respect to the renormalization mass scale $\mu$ in a 
way derived from their definition in the renormalization group equation. The 
formal dependence of each on $\mu$ will be different in different schemes but 
ultimately within the domain of perturbative applicability the behaviour of the
correlation function with respect to $\mu$ will be very similar since the 
scheme and gauge dependence will wash out as the loop order increases. 

\sect{Discussion.}

We have examined the Crewther relation at high loop order from a variety of
different angles to first ascertain whether it is possible to consistently 
extend the observation of \cite{5} to schemes other than $\MSbar$ such as 
$\mMOM$ and several kinematic schemes \cite{37,38} in different non-Lorenz 
covariant gauge fixings. In other words to see if the multiplicative form that 
the relation has in the $\MSbar$ and $V$ schemes is retained universally in 
other schemes and gauges. While we have added the $\RI$ scheme to that list,
albeit for trivial reasons, the main conclusion is that the multiplicative form 
is not universal. In schemes where the $\beta$-function depends on the gauge 
parameter, which is not the case for the $\MSbar$, $V$ and $\RI$ schemes, the 
Crewther relation is already known to  depend on the gauge parameter. This is 
despite the fact that the two correlation functions that lead to the relation 
involve gauge invariant operators. Instead in this situation the relation has 
to be extended to accommodate the gauge parameter dependence. Moreover we have
extensively shown that this is fully consistent with quantum field theoretic 
considerations and primarily the properties of the renormalization group 
equation. Essentially the gauge parameter acts as a second coupling and 
therefore its associated $\beta$-function has to be in included as indicated in
(\ref{deltgen}). While we deduced this initially by probing the product of the 
Bjorken sum rule and Adler $D$-function using the fixed point properties of 
QCD, as the $\beta$-function will always be zero at such points, we showed how 
to effect scheme changes on the coupling and gauge parameters in general to 
verify (\ref{deltgen}) formally. Subsequently the robustness of this structure 
was probed from a different angle when we posed the problem of trying to find 
what conditions on the gauge parameter would preserve the multiplicative 
structure akin to that of (\ref{deltgen}). In doing so we showed that the 
infrared stable critical point value of the gauge parameter resulted, thereby 
completing the circle of our analysis. Equally by doing so we provided a 
concrete example of the invariant charge concept defined by the couplings 
associated with the transverse and longitudinal components of the gluon field. 
It would seem that this is not restricted to the Crewther relation since the 
gauge variant quantities studied in \cite{22,23,24,25,26,30,31} could equally 
well be recast in the two coupling language whence the reasons for the 
emergence of a special gauge parameter value underlying key properties of each 
quantity would be apparent. If one took a wider viewpoint concerning where this
study appears to point it might be that caution is advised when trying to 
adduce general properties of seemingly gauge independent quantities from one 
specific renormalization scheme such as $\MSbar$. The minimal way the 
underlying divergences are subtracted in divergent Green's function would 
appear to cloud more general considerations.

\vspace{1cm}
\noindent
{\bf Acknowledgements.} This work was carried out with the support of the STFC
Consolidated Grant ST/T000988/1 (JAG) and an EPSRC Studentship EP/R513271/1 
(RHM). We thank A.L. Kataev for several discussions and S.V. Mikhailov for 
comments. For the purpose of open access, the authors have applied a Creative 
Commons Attribution (CC-BY) licence to any Author Accepted Manuscript version 
arising. The data representing the main results here are accessible in 
electronic form from the arXiv ancillary directory associated with the article.

\appendix

\sect{Results for a general colour group.}

To illustrate aspects of the $K$-functions for an arbitrary colour group we
record them for the $\MOMg$ scheme in a linear covariant by way of an example.
We have
\begin{eqnarray}
K_a^{\MOMgs}(a,\alpha) &=& 
\left[
12 \zeta_3 C_F
- \frac{21}{2} C_F
\right] a
\nonumber \\
&&
+ \left[
\frac{7}{3} \psi^{\prime}(\third) C_F C_A \alpha^2
- \frac{512}{27} \zeta_3 \pi^2 \Nf T_F C_F
- \frac{321}{2} C_F C_A
- \frac{224}{9} \psi^{\prime}(\third) \Nf T_F C_F
\right. \nonumber \\
&& \left. ~~~
- \frac{161}{27} \pi^2 C_F C_A
- \frac{112}{3} \zeta_3 \Nf T_F C_F
- \frac{92}{9} \psi^{\prime}(\third) \zeta_3 C_F C_A
- \frac{21}{2} C_F C_A \alpha^2
\right. \nonumber \\
&& \left. ~~~
- \frac{21}{2} \psi^{\prime}(\third) C_F C_A \alpha
- \frac{14}{9} \pi^2 C_F C_A \alpha^2
- \frac{8}{3} \psi^{\prime}(\third) \zeta_3 C_F C_A \alpha^2
+ \frac{7}{4} C_F C_A \alpha^3
\right. \nonumber \\
&& \left. ~~~
+ \frac{16}{9} \zeta_3 \pi^2 C_F C_A \alpha^2
+ \frac{63}{4} C_F C_A \alpha
+ \frac{158}{3} \Nf T_F C_F
+ \frac{161}{18} \psi^{\prime}(\third) C_F C_A
\right. \nonumber \\
&& \left. ~~~
+ \frac{184}{27} \zeta_3 \pi^2 C_F C_A
+ \frac{256}{9} \psi^{\prime}(\third) \zeta_3 \Nf T_F C_F
+ \frac{356}{3} \zeta_3 C_F C_A
+ \frac{397}{6} C_F^2
\right. \nonumber \\
&& \left. ~~~
+ \frac{448}{27} \pi^2 \Nf T_F C_F
- 240 \zeta_5 C_F^2
- 18 \zeta_3 C_F C_A \alpha
- 8 \zeta_3 \pi^2 C_F C_A \alpha
\right. \nonumber \\
&& \left. ~~~
- 2 \zeta_3 C_F C_A \alpha^3
+ 7 \pi^2 C_F C_A \alpha
+ 12 \zeta_3 C_F C_A \alpha^2
+ 12 \psi^{\prime}(\third) \zeta_3 C_F C_A \alpha
\right. \nonumber \\
&& \left. ~~~
+ 136 \zeta_3 C_F^2
\right] a^2
\nonumber \\
&&
+ \left[
\frac{1}{48} \psi^{\prime\prime\prime}(\third) \zeta_3 C_F C_A^2 \alpha^3
- \frac{5457913}{1152} C_F C_A^2
- \frac{336299}{648} \psi^{\prime}(\third) C_F C_A^2 \alpha
\right. \nonumber \\
&& \left. ~~~
- \frac{177919}{2916} \psi^{\prime}(\third) \pi^2 C_F C_A^2 \alpha
- \frac{87808}{729} \psi^{\prime}(\third) \zeta_3 \pi^2 \Nf T_F C_F C_A \alpha
\right. \nonumber \\
&& \left. ~~~
- \frac{78529}{1296} \pi^2 C_F C_A^2 \alpha^2
- \frac{78454}{243} \zeta_3 \pi^2 C_F C_A^2 \alpha
- \frac{76832}{2187} \pi^4 \Nf T_F C_F C_A \alpha
\right. \nonumber \\
&& \left. ~~~
- \frac{66269}{384} C_F C_A^2 \alpha^2
- \frac{47950}{243} \pi^2 \Nf T_F C_F C_A \alpha
- \frac{35689}{54} \psi^{\prime}(\third) \Nf T_F C_F C_A
\right. \nonumber \\
&& \left. ~~~
- \frac{33338}{2187} \zeta_3 \pi^4 C_F C_A^2 \alpha
- \frac{28672}{243} \psi^{\prime}(\third) \zeta_3 \pi^2 \Nf^2 T_F^2 C_F
- \frac{27181}{1458} \zeta_3 \pi^4 C_F C_A^2
\right. \nonumber \\
&& \left. ~~~
- \frac{26776}{81} \zeta_3 \pi^2 \Nf T_F C_F C_A
- \frac{25921}{2592} \psi^{\prime}(\third)^2 C_F C_A^2
- \frac{25088}{729} \pi^4 \Nf^2 T_F^2 C_F
\right. \nonumber \\
&& \left. ~~~
- \frac{25417}{486} \psi^{\prime}(\third)^2 \zeta_3 C_F C_A^2 \alpha
- \frac{24040}{81} \psi^{\prime}(\third) \zeta_3 \Nf T_F C_F C_A \alpha
\right. \nonumber \\
&& \left. ~~~
- \frac{21805}{2592} \psi^{\prime}(\third)^2 C_F C_A^2 \alpha^2
- \frac{19208}{243} \psi^{\prime}(\third)^2 \Nf T_F C_F C_A \alpha
\right. \nonumber \\
&& \left. ~~~
- \frac{18032}{243} \psi^{\prime}(\third) \pi^2 \Nf T_F C_F C_A
- \frac{14257}{54} \pi^2 C_F C_A^2
- \frac{11008}{27} \zeta_3 \pi^2 \Nf T_F C_F^2
\right. \nonumber \\
&& \left. ~~~
- \frac{10240}{729} \zeta_3 \pi^4 \Nf T_F C_F C_A
- \frac{9986}{27} \psi^{\prime}(\third) \zeta_3 C_F C_A^2
- \frac{9850}{9} \zeta_3 \Nf T_F C_F C_A
\right. \nonumber \\
&& \left. ~~~
- \frac{9809}{18} \zeta_3 C_F C_A^2 \alpha
- \frac{9487}{108} \psi^{\prime}(\third) \zeta_3 C_F C_A^2 \alpha^2
- \frac{9131}{108} \psi^{\prime}(\third) C_F^2 C_A
\right. \nonumber \\
&& \left. ~~~
- \frac{9040}{3} \zeta_5 C_F^2 C_A
- \frac{8768}{81} \pi^2 \Nf^2 T_F^2 C_F
- \frac{8000}{3} \zeta_5 \Nf T_F C_F C_A
\right. \nonumber \\
&& \left. ~~~
- \frac{7775}{4} \zeta_3^2 C_F C_A^2
- \frac{6656}{81} \pi^2 \Nf T_F C_F^2
- \frac{6272}{81} \psi^{\prime}(\third)^2 \Nf^2 T_F^2 C_F
\right. \nonumber \\
&& \left. ~~~
- \frac{5152}{81} \psi^{\prime}(\third)^2 \zeta_3 \Nf T_F C_F C_A
- \frac{4459}{1458} \pi^4 C_F C_A^2 \alpha^2
- \frac{4058}{9} \Nf^2 T_F^2 C_F
\right. \nonumber \\
&& \left. ~~~
- \frac{3920}{9} \Nf T_F C_F^2
- \frac{3703}{243} \psi^{\prime}(\third) \zeta_3 \pi^2 C_F C_A^2
- \frac{3115}{243} \psi^{\prime}(\third) \zeta_3 \pi^2 C_F C_A^2 \alpha^2
\right. \nonumber \\
&& \left. ~~~
- \frac{2989}{384} \psi^{\prime\prime\prime}(\third) C_F C_A^2
- \frac{2816}{27} \psi^{\prime}(\third) \zeta_3 \Nf^2 T_F^2 C_F
- \frac{2560}{3} \psi^{\prime}(\third) \zeta_5 \Nf T_F C_F^2
\right. \nonumber \\
&& \left. ~~~
- \frac{2197}{9} \Nf T_F C_F C_A \alpha
- \frac{1840}{9} \zeta_5 \pi^2 C_F^2 C_A
- \frac{1792}{729} \zeta_3 \pi^4 \Nf T_F C_F C_A \alpha^2
\right. \nonumber \\
&& \left. ~~~
- \frac{1783}{128} C_F C_A^2 \alpha^4
- \frac{1568}{243} \psi^{\prime}(\third) \pi^2 \Nf T_F C_F C_A \alpha^2
- \frac{1564}{9} \psi^{\prime}(\third) \zeta_3 C_F^2 C_A
\right. \nonumber \\
&& \left. ~~~
- \frac{1328}{81} \zeta_3 \pi^2 \Nf T_F C_F C_A \alpha^2
- \frac{1191}{8} C_F^2 C_A \alpha
- \frac{1115}{216} \pi^2 C_F C_A^2 \alpha^3
\right. \nonumber \\
&& \left. ~~~
- \frac{763}{144} \pi^2 C_F C_A^2 \alpha^4
- \frac{643}{486} \zeta_3 \pi^4 C_F C_A^2 \alpha^3
- \frac{641}{27} \psi^{\prime}(\third) \Nf T_F C_F C_A \alpha^2
\right. \nonumber \\
&& \left. ~~~
- \frac{539}{162} \psi^{\prime}(\third) \pi^2 C_F C_A^2 \alpha^3
- \frac{448}{81} \psi^{\prime}(\third)^2 \zeta_3 \Nf T_F C_F C_A \alpha^2
- \frac{397}{6} \pi^2 C_F^2 C_A \alpha
\right. \nonumber \\
&& \left. ~~~
- \frac{397}{18} \psi^{\prime}(\third) C_F^2 C_A \alpha^2
- \frac{397}{24} C_F^2 C_A \alpha^3
- \frac{160}{3} \zeta_5 \pi^2 C_F^2 C_A \alpha^2
\right. \nonumber \\
&& \left. ~~~
- \frac{136}{3} \psi^{\prime}(\third) \zeta_3 C_F^2 C_A \alpha^2
- \frac{125}{18} \psi^{\prime}(\third) \zeta_3 C_F C_A^2 \alpha^3
- \frac{109}{12} \psi^{\prime}(\third) \zeta_3 C_F C_A^2 \alpha^4
\right. \nonumber \\
&& \left. ~~~
- \frac{77}{27} \psi^{\prime}(\third)^2 \zeta_3 C_F C_A^2 \alpha^3
- \frac{51}{8} \zeta_3 C_F C_A^2 \alpha^5
- \frac{49}{48} \psi^{\prime}(\third) C_F C_A^2 \alpha^5
\right. \nonumber \\
&& \left. ~~~
- \frac{49}{72} \psi^{\prime}(\third)^2 C_F C_A^2 \alpha^4
- \frac{49}{128} C_F C_A^2 \alpha^6
- \frac{49}{162} \pi^4 C_F C_A^2 \alpha^4
\right. \nonumber \\
&& \left. ~~~
- \frac{49}{192} \psi^{\prime\prime\prime}(\third) C_F C_A^2 \alpha^2
- \frac{38}{3} \zeta_3 \Nf T_F C_F C_A \alpha^2
- \frac{28}{27} \psi^{\prime}(\third) \zeta_3 \pi^2 C_F C_A^2 \alpha^4
\right. \nonumber \\
&& \left. ~~~
- \frac{16}{3} \psi^{\prime\prime\prime}(\third) \zeta_3 \Nf T_F C_F C_A
- \frac{13}{8} \zeta_3 C_F C_A^2 \alpha^3
- \frac{7}{9} \zeta_3 \pi^2 C_F C_A^2 \alpha^5
\right. \nonumber \\
&& \left. ~~~
- \frac{7}{384} \psi^{\prime\prime\prime}(\third) C_F C_A^2 \alpha^3
+ \frac{3}{4} \zeta_3^2 C_F C_A^2 \alpha^2
+ \frac{7}{6} \psi^{\prime}(\third) \zeta_3 C_F C_A^2 \alpha^5
\right. \nonumber \\
&& \left. ~~~
+ \frac{7}{9} \psi^{\prime}(\third)^2 \zeta_3 C_F C_A^2 \alpha^4
+ \frac{7}{16} \zeta_3 C_F C_A^2 \alpha^6
+ \frac{7}{24} \psi^{\prime\prime\prime}(\third) \zeta_3 C_F C_A^2 \alpha^2
\right. \nonumber \\
&& \left. ~~~
+ \frac{14}{3} \psi^{\prime\prime\prime}(\third) \Nf T_F C_F C_A
+ \frac{21}{8} \psi^{\prime\prime\prime}(\third) C_F C_A^2 \alpha
+ \frac{28}{81} \zeta_3 \pi^4 C_F C_A^2 \alpha^4
\right. \nonumber \\
&& \left. ~~~
+ \frac{49}{54} \psi^{\prime}(\third) \pi^2 C_F C_A^2 \alpha^4
+ \frac{49}{72} \pi^2 C_F C_A^2 \alpha^5
+ \frac{109}{18} \zeta_3 \pi^2 C_F C_A^2 \alpha^4
\right. \nonumber \\
&& \left. ~~~
+ \frac{125}{27} \zeta_3 \pi^2 C_F C_A^2 \alpha^3
+ \frac{217}{96} C_F C_A^2 \alpha^3
+ \frac{253}{12} \Nf T_F C_F C_A \alpha^2
\right. \nonumber \\
&& \left. ~~~
+ \frac{272}{9} \zeta_3 \pi^2 C_F^2 C_A \alpha^2
+ \frac{289}{16} \zeta_3 C_F C_A^2 \alpha^4
+ \frac{308}{81} \psi^{\prime}(\third) \zeta_3 \pi^2 C_F C_A^2 \alpha^3
\right. \nonumber \\
&& \left. ~~~
+ \frac{357}{64} C_F C_A^2 \alpha^5
+ \frac{392}{81} \psi^{\prime}(\third)^2 \Nf T_F C_F C_A \alpha^2
+ \frac{397}{4} C_F^2 C_A \alpha^2
\right. \nonumber \\
&& \left. ~~~
+ \frac{397}{4} \psi^{\prime}(\third) C_F^2 C_A \alpha
+ \frac{397}{27} \pi^2 C_F^2 C_A \alpha^2
+ \frac{427}{48} \psi^{\prime\prime\prime}(\third) \zeta_3 C_F C_A^2
\right. \nonumber \\
&& \left. ~~~
+ \frac{539}{216} \psi^{\prime}(\third)^2 C_F C_A^2 \alpha^3
+ \frac{664}{27} \psi^{\prime}(\third) \zeta_3 \Nf T_F C_F C_A \alpha^2
\right. \nonumber \\
&& \left. ~~~
+ \frac{763}{96} \psi^{\prime}(\third) C_F C_A^2 \alpha^4
+ \frac{920}{3} \psi^{\prime}(\third) \zeta_5 C_F^2 C_A
+ \frac{1115}{144} \psi^{\prime}(\third) C_F C_A^2 \alpha^3
\right. \nonumber \\
&& \left. ~~~
+ \frac{1282}{81} \pi^2 \Nf T_F C_F C_A \alpha^2
+ \frac{1448}{9} \zeta_3 \Nf T_F C_F C_A \alpha
+ \frac{1552}{9} \zeta_3 \Nf^2 T_F^2 C_F
\right. \nonumber \\
&& \left. ~~~
+ \frac{1568}{729} \pi^4 \Nf T_F C_F C_A \alpha^2
+ \frac{1792}{243} \psi^{\prime}(\third) \zeta_3 \pi^2 \Nf T_F C_F C_A \alpha^2
+ \frac{2471}{12} C_F^3
\right. \nonumber \\
&& \left. ~~~
+ \frac{2548}{729} \zeta_3 \pi^4 C_F C_A^2 \alpha^2
+ \frac{3115}{324} \psi^{\prime}(\third)^2 \zeta_3 C_F C_A^2 \alpha^2
+ \frac{3128}{27} \zeta_3 \pi^2 C_F^2 C_A
\right. \nonumber \\
&& \left. ~~~
+ \frac{3328}{27} \psi^{\prime}(\third) \Nf T_F C_F^2
+ \frac{3703}{324} \psi^{\prime}(\third)^2 \zeta_3 C_F C_A^2
+ \frac{4384}{27} \psi^{\prime}(\third) \Nf^2 T_F^2 C_F
\right. \nonumber \\
&& \left. ~~~
+ \frac{4501}{3888} \pi^4 C_F C_A^2 \alpha^3
+ \frac{4508}{81} \psi^{\prime}(\third)^2 \Nf T_F C_F C_A
+ \frac{5120}{9} \zeta_5 \pi^2 \Nf T_F C_F^2
\right. \nonumber \\
&& \left. ~~~
+ \frac{5504}{9} \psi^{\prime}(\third) \zeta_3 \Nf T_F C_F^2
+ \frac{5632}{81} \zeta_3 \pi^2 \Nf^2 T_F^2 C_F
\right. \nonumber \\
&& \left. ~~~
+ \frac{7168}{81} \psi^{\prime}(\third)^2 \zeta_3 \Nf^2 T_F^2 C_F
+ \frac{7594}{3} \zeta_3 C_F^2 C_A
+ \frac{8960}{729} \pi^4 \Nf T_F C_F C_A
\right. \nonumber \\
&& \left. ~~~
+ \frac{9131}{162} \pi^2 C_F^2 C_A
+ \frac{9487}{162} \zeta_3 \pi^2 C_F C_A^2 \alpha^2
+ \frac{10711}{96} \zeta_3 C_F C_A^2 \alpha^2
\right. \nonumber \\
&& \left. ~~~
+ \frac{11900}{3} \zeta_5 C_F C_A^2
+ \frac{13388}{27} \psi^{\prime}(\third) \zeta_3 \Nf T_F C_F C_A
+ \frac{14257}{36} \psi^{\prime}(\third) C_F C_A^2
\right. \nonumber \\
&& \left. ~~~
+ \frac{19972}{81} \zeta_3 \pi^2 C_F C_A^2
+ \frac{20608}{243} \psi^{\prime}(\third) \zeta_3 \pi^2 \Nf T_F C_F C_A
\right. \nonumber \\
&& \left. ~~~
+ \frac{21805}{1944} \psi^{\prime}(\third) \pi^2 C_F C_A^2 \alpha^2
+ \frac{21952}{243} \psi^{\prime}(\third)^2 \zeta_3 \Nf T_F C_F C_A \alpha
\right. \nonumber \\
&& \left. ~~~
+ \frac{23975}{81} \psi^{\prime}(\third) \Nf T_F C_F C_A \alpha
+ \frac{25088}{243} \psi^{\prime}(\third) \pi^2 \Nf^2 T_F^2 C_F
\right. \nonumber \\
&& \left. ~~~
+ \frac{25921}{1944} \psi^{\prime}(\third) \pi^2 C_F C_A^2
+ \frac{28672}{729} \zeta_3 \pi^4 \Nf^2 T_F^2 C_F
+ \frac{35689}{81} \pi^2 \Nf T_F C_F C_A
\right. \nonumber \\
&& \left. ~~~
+ \frac{39227}{81} \psi^{\prime}(\third) \zeta_3 C_F C_A^2 \alpha
+ \frac{47353}{36} C_F^2 C_A
+ \frac{48080}{243} \zeta_3 \pi^2 \Nf T_F C_F C_A \alpha
\right. \nonumber \\
&& \left. ~~~
+ \frac{50834}{729} \psi^{\prime}(\third) \zeta_3 \pi^2 C_F C_A^2 \alpha
+ \frac{76832}{729} \psi^{\prime}(\third) \pi^2 \Nf T_F C_F C_A \alpha
\right. \nonumber \\
&& \left. ~~~
+ \frac{78529}{864} \psi^{\prime}(\third) C_F C_A^2 \alpha^2
+ \frac{87808}{2187} \zeta_3 \pi^4 \Nf T_F C_F C_A \alpha
\right. \nonumber \\
&& \left. ~~~
+ \frac{110051}{36} \Nf T_F C_F C_A
+ \frac{116683}{8748} \pi^4 C_F C_A^2 \alpha
+ \frac{177919}{3888} \psi^{\prime}(\third)^2 C_F C_A^2 \alpha
\right. \nonumber \\
&& \left. ~~~
+ \frac{190267}{11664} \pi^4 C_F C_A^2
+ \frac{327281}{576} C_F C_A^2 \alpha
+ \frac{336299}{972} \pi^2 C_F C_A^2 \alpha
\right. \nonumber \\
&& \left. ~~~
+ \frac{741599}{288} \zeta_3 C_F C_A^2
- 5720 \zeta_5 C_F^3
- 1464 \zeta_3 \Nf T_F C_F^2
- 840 \zeta_7 C_F^2 C_A
\right. \nonumber \\
&& \left. ~~~
- 360 \zeta_5 C_F^2 C_A \alpha^2
- 360 \psi^{\prime}(\third) \zeta_5 C_F^2 C_A \alpha
- 306 \zeta_3 C_F^2 C_A \alpha
\right. \nonumber \\
&& \left. ~~~
- 136 \zeta_3 \pi^2 C_F^2 C_A \alpha
- 34 \zeta_3 C_F^2 C_A \alpha^3
- 3 \zeta_3^2 C_F C_A^2 \alpha^3
- 3 \psi^{\prime\prime\prime}(\third) \zeta_3 C_F C_A^2 \alpha
\right. \nonumber \\
&& \left. ~~~
+ 60 \zeta_5 C_F^2 C_A \alpha^3
+ 80 \psi^{\prime}(\third) \zeta_5 C_F^2 C_A \alpha^2
+ 171 \zeta_3^2 C_F C_A^2 \alpha
+ 204 \zeta_3 C_F^2 C_A \alpha^2
\right. \nonumber \\
&& \left. ~~~
+ 204 \psi^{\prime}(\third) \zeta_3 C_F^2 C_A \alpha
+ 240 \zeta_5 \pi^2 C_F^2 C_A \alpha
+ 320 \zeta_5 \Nf^2 T_F^2 C_F
+ 488 \zeta_3 C_F^3
\right. \nonumber \\
&& \left. ~~~
+ 540 \zeta_5 C_F^2 C_A \alpha
+ 640 \zeta_3^2 \Nf T_F C_F C_A
+ 2080 \zeta_5 \Nf T_F C_F^2
\right. \nonumber \\
&& \left. ~~~
+ 5040 \zeta_7 C_F^3
\right] a^3 ~+~ O(a^4)
\end{eqnarray}
and
\begin{eqnarray}
K_\alpha^{\MOMgs}(a,\alpha) &=& 
\left[
12 \zeta_3 C_F C_A \alpha
- \frac{21}{2} C_F C_A \alpha
- \frac{21}{4} \psi^{\prime}(\third) C_F C_A
- \frac{14}{9} \pi^2 C_F C_A \alpha
\right. \nonumber \\
&& \left. ~
- \frac{8}{3} \psi^{\prime}(\third) \zeta_3 C_F C_A \alpha
+ \frac{7}{2} \pi^2 C_F C_A
+ \frac{7}{3} \psi^{\prime}(\third) C_F C_A \alpha
+ \frac{16}{9} \zeta_3 \pi^2 C_F C_A \alpha
\right. \nonumber \\
&& \left. ~
+ \frac{21}{8} C_F C_A \alpha^2
+ \frac{63}{8} C_F C_A
- 9 \zeta_3 C_F C_A
- 4 \zeta_3 \pi^2 C_F C_A
- 3 \zeta_3 C_F C_A \alpha^2
\right. \nonumber \\
&& \left. ~
+ 6 \psi^{\prime}(\third) \zeta_3 C_F C_A
\right] a^2
\nonumber \\
&&
+ \left[
\frac{1}{2} \zeta_3^2 C_F C_A^2 \alpha
- \frac{548047}{1296} \psi^{\prime}(\third) C_F C_A^2
- \frac{177037}{2916} \psi^{\prime}(\third) \pi^2 C_F C_A^2
\right. \nonumber \\
&& \left. ~~~
- \frac{69841}{243} \zeta_3 \pi^2 C_F C_A^2
- \frac{44750}{2187} \zeta_3 \pi^4 C_F C_A^2
\right. \nonumber \\
&& \left. ~~~
- \frac{39424}{729} \psi^{\prime}(\third) \zeta_3 \pi^2 \Nf T_F C_F C_A
- \frac{34496}{2187} \pi^4 \Nf T_F C_F C_A
\right. \nonumber \\
&& \left. ~~~
- \frac{28204}{243} \pi^2 \Nf T_F C_F C_A
- \frac{25291}{486} \psi^{\prime}(\third)^2 \zeta_3 C_F C_A^2
\right. \nonumber \\
&& \left. ~~~
- \frac{20231}{432} \psi^{\prime}(\third) C_F C_A^2 \alpha
- \frac{14608}{81} \psi^{\prime}(\third) \zeta_3 \Nf T_F C_F C_A
- \frac{11221}{36} \zeta_3 C_F C_A^2
\right. \nonumber \\
&& \left. ~~~
- \frac{8624}{243} \psi^{\prime}(\third)^2 \Nf T_F C_F C_A
- \frac{3323}{192} C_F C_A^2 \alpha
- \frac{2783}{18} \Nf T_F C_F C_A
\right. \nonumber \\
&& \left. ~~~
- \frac{2009}{108} \psi^{\prime}(\third) \pi^2 C_F C_A^2 \alpha
- \frac{971}{27} \zeta_3 \pi^2 C_F C_A^2 \alpha
- \frac{777}{32} C_F C_A^2 \alpha^3
\right. \nonumber \\
&& \left. ~~~
- \frac{616}{81} \zeta_3 \pi^4 C_F C_A^2 \alpha
- \frac{397}{8} C_F^2 C_A
- \frac{397}{18} \pi^2 C_F^2 C_A
- \frac{397}{24} C_F^2 C_A \alpha^2
\right. \nonumber \\
&& \left. ~~~
- \frac{397}{27} \psi^{\prime}(\third) C_F^2 C_A \alpha
- \frac{385}{36} \pi^2 C_F C_A^2 \alpha^3
- \frac{320}{9} \zeta_5 \pi^2 C_F^2 C_A \alpha
\right. \nonumber \\
&& \left. ~~~
- \frac{287}{18} \psi^{\prime}(\third)^2 \zeta_3 C_F C_A^2 \alpha
- \frac{272}{9} \psi^{\prime}(\third) \zeta_3 C_F^2 C_A \alpha
- \frac{245}{144} \psi^{\prime}(\third) C_F C_A^2 \alpha^4
\right. \nonumber \\
&& \left. ~~~
- \frac{136}{3} \zeta_3 \pi^2 C_F^2 C_A
- \frac{112}{81} \psi^{\prime}(\third) \zeta_3 \pi^2 C_F C_A^2 \alpha^3
- \frac{98}{243} \pi^4 C_F C_A^2 \alpha^3
\right. \nonumber \\
&& \left. ~~~
- \frac{85}{8} \zeta_3 C_F C_A^2 \alpha^4
- \frac{55}{3} \psi^{\prime}(\third) \zeta_3 C_F C_A^2 \alpha^3
- \frac{49}{6} \psi^{\prime}(\third) \pi^2 C_F C_A^2 \alpha^2
\right. \nonumber \\
&& \left. ~~~
- \frac{49}{54} \psi^{\prime}(\third)^2 C_F C_A^2 \alpha^3
- \frac{49}{64} C_F C_A^2 \alpha^5
- \frac{49}{288} \psi^{\prime\prime\prime}(\third) C_F C_A^2 \alpha
\right. \nonumber \\
&& \left. ~~~
- \frac{35}{24} \pi^2 C_F C_A^2 \alpha^2
- \frac{35}{27} \zeta_3 \pi^2 C_F C_A^2 \alpha^4
- \frac{23}{8} \zeta_3 C_F C_A^2 \alpha^2
\right. \nonumber \\
&& \left. ~~~
- \frac{19}{6} \zeta_3 \pi^4 C_F C_A^2 \alpha^2
- \frac{7}{384} \psi^{\prime\prime\prime}(\third) C_F C_A^2 \alpha^2
- \frac{5}{2} \psi^{\prime}(\third) \zeta_3 C_F C_A^2 \alpha^2
\right. \nonumber \\
&& \left. ~~~
+ \frac{1}{48} \psi^{\prime\prime\prime}(\third) \zeta_3 C_F C_A^2 \alpha^2
+ \frac{5}{3} \zeta_3 \pi^2 C_F C_A^2 \alpha^2
+ \frac{7}{8} \zeta_3 C_F C_A^2 \alpha^5
\right. \nonumber \\
&& \left. ~~~
+ \frac{7}{8} \psi^{\prime\prime\prime}(\third) C_F C_A^2
+ \frac{7}{36} \psi^{\prime\prime\prime}(\third) \zeta_3 C_F C_A^2 \alpha
+ \frac{28}{3} \psi^{\prime}(\third) \zeta_3 \pi^2 C_F C_A^2 \alpha^2
\right. \nonumber \\
&& \left. ~~~
+ \frac{28}{27} \psi^{\prime}(\third)^2 \zeta_3 C_F C_A^2 \alpha^3
+ \frac{35}{16} \psi^{\prime}(\third) C_F C_A^2 \alpha^2
+ \frac{35}{18} \psi^{\prime}(\third) \zeta_3 C_F C_A^2 \alpha^4
\right. \nonumber \\
&& \left. ~~~
+ \frac{49}{8} \psi^{\prime}(\third)^2 C_F C_A^2 \alpha^2
+ \frac{98}{81} \psi^{\prime}(\third) \pi^2 C_F C_A^2 \alpha^3
+ \frac{110}{9} \zeta_3 \pi^2 C_F C_A^2 \alpha^3
\right. \nonumber \\
&& \left. ~~~
+ \frac{111}{4} \zeta_3 C_F C_A^2 \alpha^3
+ \frac{112}{243} \zeta_3 \pi^4 C_F C_A^2 \alpha^3
+ \frac{133}{48} \pi^4 C_F C_A^2 \alpha^2
\right. \nonumber \\
&& \left. ~~~
+ \frac{160}{3} \psi^{\prime}(\third) \zeta_5 C_F^2 C_A \alpha
+ \frac{245}{216} \pi^2 C_F C_A^2 \alpha^4
+ \frac{251}{48} C_F C_A^2 \alpha^2
\right. \nonumber \\
&& \left. ~~~
+ \frac{279}{16} \zeta_3 C_F C_A^2 \alpha
+ \frac{385}{24} \psi^{\prime}(\third) C_F C_A^2 \alpha^3
+ \frac{397}{6} C_F^2 C_A \alpha
\right. \nonumber \\
&& \left. ~~~
+ \frac{397}{12} \psi^{\prime}(\third) C_F^2 C_A
+ \frac{539}{81} \pi^4 C_F C_A^2 \alpha
+ \frac{544}{27} \zeta_3 \pi^2 C_F^2 C_A \alpha
\right. \nonumber \\
&& \left. ~~~
+ \frac{574}{27} \psi^{\prime}(\third) \zeta_3 \pi^2 C_F C_A^2 \alpha
+ \frac{595}{64} C_F C_A^2 \alpha^4
+ \frac{794}{81} \pi^2 C_F^2 C_A \alpha
\right. \nonumber \\
&& \left. ~~~
+ \frac{836}{9} \zeta_3 \Nf T_F C_F C_A
+ \frac{971}{18} \psi^{\prime}(\third) \zeta_3 C_F C_A^2 \alpha
+ \frac{2009}{144} \psi^{\prime}(\third)^2 C_F C_A^2 \alpha
\right. \nonumber \\
&& \left. ~~~
+ \frac{9856}{243} \psi^{\prime}(\third)^2 \zeta_3 \Nf T_F C_F C_A
+ \frac{14102}{81} \psi^{\prime}(\third) \Nf T_F C_F C_A
\right. \nonumber \\
&& \left. ~~~
+ \frac{20231}{648} \pi^2 C_F C_A^2 \alpha
+ \frac{29216}{243} \zeta_3 \pi^2 \Nf T_F C_F C_A
\right. \nonumber \\
&& \left. ~~~
+ \frac{34496}{729} \psi^{\prime}(\third) \pi^2 \Nf T_F C_F C_A
+ \frac{39424}{2187} \zeta_3 \pi^4 \Nf T_F C_F C_A
\right. \nonumber \\
&& \left. ~~~
+ \frac{50582}{729} \psi^{\prime}(\third) \zeta_3 \pi^2 C_F C_A^2
+ \frac{69841}{162} \psi^{\prime}(\third) \zeta_3 C_F C_A^2
+ \frac{156625}{8748} \pi^4 C_F C_A^2
\right. \nonumber \\
&& \left. ~~~
+ \frac{177037}{3888} \psi^{\prime}(\third)^2 C_F C_A^2
+ \frac{248837}{576} C_F C_A^2
+ \frac{548047}{1944} \pi^2 C_F C_A^2
\right. \nonumber \\
&& \left. ~~~
- 240 \zeta_5 C_F^2 C_A \alpha
- 120 \psi^{\prime}(\third) \zeta_5 C_F^2 C_A
- 102 \zeta_3 C_F^2 C_A
- 34 \zeta_3 C_F^2 C_A \alpha^2
\right. \nonumber \\
&& \left. ~~~
- 7 \psi^{\prime}(\third)^2 \zeta_3 C_F C_A^2 \alpha^2
- 3 \zeta_3^2 C_F C_A^2 \alpha^2
- \psi^{\prime\prime\prime}(\third) \zeta_3 C_F C_A^2
+ 57 \zeta_3^2 C_F C_A^2
\right. \nonumber \\
&& \left. ~~~
+ 60 \zeta_5 C_F^2 C_A \alpha^2
+ 68 \psi^{\prime}(\third) \zeta_3 C_F^2 C_A
+ 80 \zeta_5 \pi^2 C_F^2 C_A
+ 136 \zeta_3 C_F^2 C_A \alpha
\right. \nonumber \\
&& \left. ~~~
+ 180 \zeta_5 C_F^2 C_A
\right] a^3 ~+~ O(a^4) ~.
\end{eqnarray}
The corresponding MAG expression depends on $\Nda$ and $\Noda$ and is available
in the data file.

\sect{$\RIc$ scheme $\beta$-function.}

We record the $SU(3)$ expressions for the two key renormalization group
equations that are central to the Crewther relation in the $\RIc$ scheme. The
five loop $\beta$-function is
\begin{eqnarray}
\left. \beta^{\RIcs}(a,\alpha) \right|^{SU(3)} &=& 
\left[
\frac{2}{3} \Nf
- 11
\right] a^2
+ \left[
\frac{38}{3} \Nf
+ \frac{39}{2} \alpha
- \frac{9}{2} \alpha^2
- 102
- 2 \Nf \alpha
\right] a^3
\nonumber \\
&&
+ \left[
\frac{21}{4} \Nf \alpha^2
+ \frac{27}{8} \alpha^3
- \frac{2857}{2}
- \frac{2655}{16} \alpha^2
- \frac{325}{54} \Nf^2
- \frac{137}{2} \Nf \alpha
- \frac{81}{8} \zeta_3 \alpha^3
\right. \nonumber \\
&& \left. ~~~
- \frac{9}{4} \zeta_3 \Nf \alpha^2
+ \frac{351}{16} \zeta_3 \alpha^2
+ \frac{729}{16} \zeta_3 \alpha
+ \frac{4599}{16} \alpha
+ \frac{5033}{18} \Nf
\right] a^4
\nonumber \\
&&
+ \left[
\frac{29}{2} \zeta_3 \Nf^2 \alpha
+ \frac{243}{8} \zeta_3 \Nf \alpha^3
- \frac{1172325}{128} \alpha
- \frac{717363}{128} \alpha^2
- \frac{149753}{6}
\right. \nonumber \\
&& \left. ~~~
- \frac{50065}{162} \Nf^2
- \frac{6472}{81} \zeta_3 \Nf^2
- \frac{6321}{16} \Nf \alpha
- \frac{2025}{16} \zeta_5 \alpha^4
- \frac{1725}{4} \zeta_3 \Nf \alpha
\right. \nonumber \\
&& \left. ~~~
- \frac{1323}{2} \zeta_3 \alpha^3
- \frac{1093}{729} \Nf^3
- \frac{621}{16} \zeta_3 \alpha^2
- \frac{567}{8} \alpha^4
- \frac{243}{2} \zeta_3 \Nf \alpha^2
\right. \nonumber \\
&& \left. ~~~
- \frac{177}{8} \Nf \alpha^3
- \frac{75}{4} \zeta_5 \Nf \alpha^3
- \frac{45}{2} \zeta_5 \Nf \alpha
- \frac{13}{8} \Nf^2 \alpha
+ \frac{3465}{16} \zeta_5 \alpha^3
\right. \nonumber \\
&& \left. ~~~
+ \frac{3645}{4} \zeta_5 \alpha^2
+ \frac{4131}{32} \zeta_3 \alpha^4
+ \frac{4185}{8} \zeta_5 \alpha
+ \frac{6508}{27} \zeta_3 \Nf
+ \frac{17169}{32} \Nf \alpha^2
\right. \nonumber \\
&& \left. ~~~
+ \frac{33705}{64} \alpha^3
+ \frac{197307}{32} \zeta_3 \alpha
+ \frac{1078361}{162} \Nf
- 3564 \zeta_3
\right] a^5
\nonumber \\
&&
+ \left[
\frac{152}{81} \zeta_3 \Nf^4
+ \frac{159}{4} \zeta_3^2 \Nf^2 \alpha
- \frac{15970474577}{9216} \alpha
- \frac{372908785}{3072} \alpha^2
\right. \nonumber \\
&& \left. ~~~
- \frac{207136629}{8192} \zeta_7 \alpha^2
- \frac{32567937}{512} \zeta_3 \alpha^2
- \frac{25960913}{1944} \Nf^2
\right. \nonumber \\
&& \left. ~~~
- \frac{14959621}{144} \zeta_3 \Nf \alpha
- \frac{13960107}{8192} \zeta_7 \alpha^3
- \frac{9452457}{256} \zeta_3^2 \alpha
- \frac{8157455}{16}
\right. \nonumber \\
&& \left. ~~~
- \frac{5668191}{1024} \alpha^4
- \frac{3852207}{512} \zeta_5 \alpha^4
- \frac{3388995}{128} \zeta_3 \alpha^3
- \frac{3383613}{256} \zeta_4 \alpha
\right. \nonumber \\
&& \left. ~~~
- \frac{2636597}{216} \Nf^2 \alpha
- \frac{1358995}{27} \zeta_5 \Nf
- \frac{1209579}{1024} \zeta_7 \Nf \alpha
- \frac{941891}{256} \zeta_5 \Nf \alpha^2
\right. \nonumber \\
&& \left. ~~~
- \frac{886581}{128} \zeta_5 \Nf \alpha
- \frac{793143}{512} \zeta_5 \alpha^3
- \frac{698531}{81} \zeta_3 \Nf^2
- \frac{621885}{2} \zeta_3
\right. \nonumber \\
&& \left. ~~~
- \frac{580203}{256} \zeta_4 \alpha^2
- \frac{573237}{2048} \zeta_7 \alpha^5
- \frac{476263}{1728} \Nf^2 \alpha^2
- \frac{404607}{128} \Nf \alpha^3
\right. \nonumber \\
&& \left. ~~~
- \frac{191727}{256} \zeta_4 \alpha^3
- \frac{151875}{128} \zeta_6 \Nf \alpha
- \frac{92097}{256} \zeta_3 \alpha^5
- \frac{63693}{2048} \zeta_7 \Nf \alpha^4
\right. \nonumber \\
&& \left. ~~~
- \frac{43011}{256} \zeta_3^2 \alpha^4
- \frac{40365}{64} \zeta_3^2 \Nf \alpha^2
- \frac{35559}{256} \zeta_4 \alpha^4
- \frac{34845}{32} \zeta_5 \Nf \alpha^3
\right. \nonumber \\
&& \left. ~~~
- \frac{33935}{6} \zeta_4 \Nf
- \frac{16235}{96} \zeta_3 \Nf^2 \alpha^2
- \frac{14175}{256} \zeta_6 \alpha^5
- \frac{14175}{256} \zeta_6 \Nf \alpha^2
\right. \nonumber \\
&& \left. ~~~
- \frac{5895}{32} \zeta_3 \Nf \alpha^4
- \frac{3105}{256} \zeta_5 \alpha^5
- \frac{1809}{16} \zeta_3^2 \Nf \alpha^3
- \frac{1618}{27} \zeta_4 \Nf^3
\right. \nonumber \\
&& \left. ~~~
- \frac{1575}{256} \zeta_6 \Nf \alpha^4
- \frac{1353}{8} \zeta_4 \Nf^2 \alpha
- \frac{1205}{2916} \Nf^4
- \frac{460}{9} \zeta_5 \Nf^3
- \frac{382}{9} \zeta_3 \Nf^3 \alpha
\right. \nonumber \\
&& \left. ~~~
- \frac{153}{16} \zeta_4 \Nf^2 \alpha^2
- \frac{100}{3} \zeta_5 \Nf^3 \alpha
+ \frac{243}{32} \zeta_4 \Nf \alpha^3
+ \frac{459}{128} \zeta_4 \Nf \alpha^4
\right. \nonumber \\
&& \left. ~~~
+ \frac{1701}{64} \zeta_3^2 \Nf \alpha^4
+ \frac{4131}{128} \zeta_4 \alpha^5
+ \frac{5931}{8} \zeta_3^2 \Nf \alpha
+ \frac{6025}{48} \zeta_5 \Nf^2 \alpha^2
\right. \nonumber \\
&& \left. ~~~
+ \frac{6093}{64} \Nf \alpha^4
+ \frac{10526}{9} \zeta_4 \Nf^2
+ \frac{10935}{64} \zeta_3^2 \alpha^5
+ \frac{13273}{16} \zeta_5 \Nf^2 \alpha
+ \frac{16605}{128} \alpha^5
\right. \nonumber \\
&& \left. ~~~
+ \frac{19465}{81} \Nf^3 \alpha
+ \frac{27285}{256} \zeta_5 \Nf \alpha^4
+ \frac{46143}{128} \zeta_4 \Nf \alpha^2
+ \frac{48722}{243} \zeta_3 \Nf^3
\right. \nonumber \\
&& \left. ~~~
+ \frac{88209}{2} \zeta_4
+ \frac{126891}{64} \zeta_4 \Nf \alpha
+ \frac{188529}{64} \zeta_3 \Nf \alpha^3
+ \frac{207225}{1024} \zeta_6 \alpha^4
\right. \nonumber \\
&& \left. ~~~
+ \frac{324577}{128} \zeta_3 \Nf \alpha^2
+ \frac{381760}{81} \zeta_5 \Nf^2
+ \frac{502929}{256} \zeta_3^2 \alpha^3
+ \frac{630559}{5832} \Nf^3
\right. \nonumber \\
&& \left. ~~~
+ \frac{948933}{512} \zeta_5 \alpha
+ \frac{1075275}{1024} \zeta_6 \alpha^3
+ \frac{1816857}{2048} \zeta_7 \Nf \alpha^2
+ \frac{2023353}{256} \zeta_3 \alpha^4
\right. \nonumber \\
&& \left. ~~~
+ \frac{2116639}{432} \zeta_3 \Nf^2 \alpha
+ \frac{2480247}{256} \zeta_3^2 \alpha^2
+ \frac{3067875}{1024} \zeta_6 \alpha^2
+ \frac{3572667}{8192} \zeta_7 \alpha^4
\right. \nonumber \\
&& \left. ~~~
+ \frac{4811164}{81} \zeta_3 \Nf
+ \frac{24148125}{1024} \zeta_6 \alpha
+ \frac{24208323}{512} \zeta_5 \alpha^2
+ \frac{33423111}{1024} \alpha^3
\right. \nonumber \\
&& \left. ~~~
+ \frac{51960893}{2304} \Nf \alpha^2
+ \frac{192323061}{8192} \zeta_7 \alpha
+ \frac{336460813}{1944} \Nf
+ \frac{435326091}{512} \zeta_3 \alpha
\right. \nonumber \\
&& \left. ~~~
+ \frac{1723650635}{6912} \Nf \alpha
+ 6 \zeta_4 \Nf^3 \alpha
+ 15 \zeta_3^2 \Nf^2 \alpha^2
+ 288090 \zeta_5
\right] a^6 
\nonumber \\
&& +~ O(a^7)
\end{eqnarray}
while the gauge parameter anomalous dimension also in the linear covariant
gauge is 
\begin{eqnarray}
\left. \gamma_\alpha^{\RIcs}(a,\alpha) \right|^{SU(3)} &=& 
\left[
\frac{13}{2}
- \frac{3}{2} \alpha
- \frac{2}{3} \Nf
\right] a
+ \left[
+ \frac{9}{4} \alpha^2
+ \frac{531}{8}
+ 2 \Nf \alpha
- \frac{255}{8} \alpha
- \frac{61}{6} \Nf
\right] a^2
\nonumber \\
&&
+ \left[
\frac{9}{4} \zeta_3 \Nf \alpha^2
+ \frac{81}{16} \zeta_3 \alpha^3
+ \frac{215}{27} \Nf^2
- \frac{27315}{32} \alpha
- \frac{8155}{36} \Nf
- \frac{675}{16} \zeta_3 \alpha^2
\right. \nonumber \\
&& \left. ~~~
- \frac{243}{16} \zeta_3
- \frac{135}{32} \alpha^3
- \frac{27}{4} \zeta_3 \Nf \alpha
- \frac{21}{4} \Nf \alpha^2
+ \frac{409}{4} \Nf \alpha
+ \frac{729}{16} \zeta_3 \alpha
\right. \nonumber \\
&& \left. ~~~
+ \frac{5445}{32} \alpha^2
+ \frac{29895}{32}
+ 33 \zeta_3 \Nf
\right] a^3
\nonumber \\
&&
+ \left[
\frac{17}{2} \zeta_3 \Nf^2 \alpha
+ \frac{45}{4} \zeta_4 \Nf \alpha
+ \frac{75}{4} \zeta_5 \Nf \alpha^3
- \frac{23350603}{5184} \Nf
- \frac{21210899}{768} \alpha
\right. \nonumber \\
&& \left. ~~~
- \frac{1012023}{256} \zeta_3
- \frac{184473}{64} \zeta_3 \alpha^2
- \frac{177021}{256} \alpha^3
- \frac{46755}{128} \zeta_5 \alpha^3
- \frac{44331}{64} \Nf \alpha^2
\right. \nonumber \\
&& \left. ~~~
- \frac{13781}{108} \Nf^2 \alpha
- \frac{13149}{256} \zeta_3 \alpha^4
- \frac{12549}{16} \zeta_3 \Nf \alpha
- \frac{8955}{16} \zeta_4 \Nf
- \frac{3355}{2} \zeta_5 \Nf
\right. \nonumber \\
&& \left. ~~~
- \frac{255}{4} \zeta_5 \Nf \alpha^2
- \frac{243}{8} \zeta_3 \Nf \alpha^3
- \frac{45}{2} \zeta_5 \Nf \alpha
- \frac{27}{16} \zeta_4 \Nf \alpha^2
- \frac{8}{3} \zeta_3 \Nf^3
\right. \nonumber \\
&& \left. ~~~
+ \frac{177}{8} \Nf \alpha^3
+ \frac{243}{64} \zeta_4 \alpha^3
+ \frac{405}{8} \zeta_4 \alpha^2
+ \frac{1809}{64} \alpha^4
+ \frac{2017}{81} \zeta_3 \Nf^2
\right. \nonumber \\
&& \left. ~~~
+ \frac{4017}{16} \zeta_3 \Nf \alpha^2
+ \frac{4427}{1458} \Nf^3
+ \frac{4455}{128} \zeta_5 \alpha^4
+ \frac{8019}{32} \zeta_4
+ \frac{16443}{64} \zeta_4 \alpha
\right. \nonumber \\
&& \left. ~~~
+ \frac{40905}{4} \zeta_5
+ \frac{43033}{162} \Nf^2
+ \frac{68625}{128} \zeta_5 \alpha^2
+ \frac{87831}{128} \zeta_3 \alpha^3
+ \frac{198315}{128} \zeta_5 \alpha
\right. \nonumber \\
&& \left. ~~~
+ \frac{311301}{128} \zeta_3 \alpha
+ \frac{387649}{216} \zeta_3 \Nf
+ \frac{712315}{144} \Nf \alpha
+ \frac{2137221}{256} \alpha^2
\right. \nonumber \\
&& \left. ~~~
+ \frac{10596127}{768}
+ 33 \zeta_4 \Nf^2
\right] a^4
\nonumber \\
&&
+ \left[
\frac{3248889939}{4096} \zeta_5
- \frac{40020851929}{36864} \alpha
- \frac{15059970043}{124416} \Nf
\right. \nonumber \\
&& \left. ~~~
- \frac{9151899939}{16384} \zeta_7
- \frac{514366083}{16384} \zeta_7 \alpha
- \frac{322439503}{4608} \zeta_3 \Nf \alpha
\right. \nonumber \\
&& \left. ~~~
- \frac{281043751}{4608} \Nf \alpha^2
- \frac{169349259}{1024} \zeta_3 \alpha^2
- \frac{110282119}{864} \zeta_5 \Nf
\right. \nonumber \\
&& \left. ~~~
- \frac{70747047}{8192} \zeta_7 \alpha^2
- \frac{65529567}{2048} \zeta_3^2
- \frac{46637683}{3456} \Nf^2 \alpha
- \frac{45332901}{2048} \zeta_5 \alpha^3
\right. \nonumber \\
&& \left. ~~~
- \frac{44784819}{2048} \zeta_3^2 \alpha
- \frac{32357475}{2048} \zeta_6 \alpha
- \frac{31012901}{768} \zeta_4 \Nf
- \frac{17999415}{16384} \zeta_7 \alpha^4
\right. \nonumber \\
&& \left. ~~~
- \frac{12215151}{2048} \zeta_3 \alpha^4
- \frac{3696633}{64} \alpha^3
- \frac{3378739}{1296} \zeta_5 \Nf^2
- \frac{2259077}{11664} \Nf^3
\right. \nonumber \\
&& \left. ~~~
- \frac{2249775}{8} \zeta_6
- \frac{1905231}{512} \zeta_3 \Nf \alpha^3
- \frac{1633959}{1024} \zeta_3^2 \alpha^3
- \frac{987899}{256} \zeta_5 \Nf \alpha^2
\right. \nonumber \\
&& \left. ~~~
- \frac{726489}{2048} \zeta_4 \alpha^3
- \frac{393579}{4096} \alpha^5
- \frac{370305}{4096} \zeta_5 \alpha^5
- \frac{257775}{512} \zeta_6 \Nf \alpha
\right. \nonumber \\
&& \left. ~~~
- \frac{231399}{2048} \zeta_7 \Nf \alpha^3
- \frac{123525}{1024} \zeta_6 \alpha^2
- \frac{105921}{128} \zeta_3^2 \Nf \alpha^2
- \frac{94527}{2048} \zeta_3^2 \alpha^5
\right. \nonumber \\
&& \left. ~~~
- \frac{80771}{243} \zeta_3 \Nf^3
- \frac{45441}{4096} \zeta_4 \alpha^5
- \frac{40139}{96} \zeta_3 \Nf^2 \alpha^2
- \frac{28575}{512} \zeta_6 \Nf \alpha^3
\right. \nonumber \\
&& \left. ~~~
- \frac{27285}{256} \zeta_5 \Nf \alpha^4
- \frac{16775}{6} \zeta_6 \Nf^2
- \frac{6093}{64} \Nf \alpha^4
- \frac{2659}{3} \zeta_3^2 \Nf^2
\right. \nonumber \\
&& \left. ~~~
- \frac{1701}{64} \zeta_3^2 \Nf \alpha^4
- \frac{1317}{4} \zeta_4 \Nf^2 \alpha
- \frac{729}{2} \zeta_4 \Nf \alpha^2
- \frac{675}{8} \zeta_6 \Nf \alpha^2
\right. \nonumber \\
&& \left. ~~~
- \frac{459}{128} \zeta_4 \Nf \alpha^4
- \frac{50}{3} \zeta_5 \Nf^3 \alpha
- \frac{32}{81} \zeta_3 \Nf^4
- \frac{8}{3} \zeta_4 \Nf^4
+ \frac{99}{16} \zeta_4 \Nf^2 \alpha^2
\right. \nonumber \\
&& \left. ~~~
+ \frac{139}{9} \zeta_3 \Nf^3 \alpha
+ \frac{159}{8} \zeta_3^2 \Nf^2 \alpha
+ \frac{1321}{27} \zeta_4 \Nf^3
+ \frac{1575}{256} \zeta_6 \Nf \alpha^4
+ \frac{2164}{9} \zeta_5 \Nf^3
\right. \nonumber \\
&& \left. ~~~
+ \frac{2299}{48} \zeta_5 \Nf^2 \alpha^2
+ \frac{4715}{2916} \Nf^4
+ \frac{5895}{32} \zeta_3 \Nf \alpha^4
+ \frac{7425}{1024} \zeta_6 \alpha^4
+ \frac{11045}{144} \Nf^3 \alpha
\right. \nonumber \\
&& \left. ~~~
+ \frac{14851}{32} \zeta_5 \Nf^2 \alpha
+ \frac{16119}{256} \zeta_4 \alpha^4
+ \frac{17469}{512} \zeta_4 \Nf \alpha^3
+ \frac{28305}{256} \zeta_3^2 \Nf \alpha^3
\right. \nonumber \\
&& \left. ~~~
+ \frac{63693}{2048} \zeta_7 \Nf \alpha^4
+ \frac{65817}{16} \Nf \alpha^3
+ \frac{70875}{2048} \zeta_6 \alpha^5
+ \frac{86103}{128} \zeta_4 \alpha^2
\right. \nonumber \\
&& \left. ~~~
+ \frac{114075}{128} \zeta_6 \alpha^3
+ \frac{163139}{72} \zeta_4 \Nf^2
+ \frac{187509}{2048} \zeta_7 \Nf \alpha
+ \frac{440809}{64} \zeta_3^2 \Nf
\right. \nonumber \\
&& \left. ~~~
+ \frac{543079}{432} \Nf^2 \alpha^2
+ \frac{638961}{256} \zeta_3^2 \Nf \alpha
+ \frac{820773}{4096} \zeta_3 \alpha^5
+ \frac{863541}{16384} \zeta_7 \alpha^5
\right. \nonumber \\
&& \left. ~~~
+ \frac{996423}{512} \zeta_5 \Nf \alpha^3
+ \frac{1164213}{2048} \zeta_3^2 \alpha^4
+ \frac{1477789}{864} \zeta_3 \Nf^2 \alpha
+ \frac{1816857}{2048} \zeta_7 \Nf \alpha^2
\right. \nonumber \\
&& \left. ~~~
+ \frac{2670405}{512} \zeta_4 \Nf \alpha
+ \frac{2988559}{128} \zeta_3 \Nf \alpha^2
+ \frac{3662827}{48} \zeta_7 \Nf
+ \frac{4089225}{512} \zeta_5 \Nf \alpha
\right. \nonumber \\
&& \left. ~~~
+ \frac{5660469}{1024} \zeta_3^2 \alpha^2
+ \frac{7740549}{2048} \alpha^4
+ \frac{10890829}{2592} \zeta_3 \Nf^2
+ \frac{17186355}{4096} \zeta_5 \alpha^4
\right. \nonumber \\
&& \left. ~~~
+ \frac{24104115}{8192} \zeta_7 \alpha^3
+ \frac{25440057}{2048} \zeta_3
+ \frac{32297109}{2048} \zeta_5 \alpha^2
+ \frac{39034359}{512} \zeta_4
\right. \nonumber \\
&& \left. ~~~
+ \frac{48001201}{3888} \Nf^2
+ \frac{61806555}{4096} \zeta_5 \alpha
+ \frac{77357835}{4096} \zeta_4 \alpha
+ \frac{97546581}{2048} \zeta_3 \alpha^3
\right. \nonumber \\
&& \left. ~~~
+ \frac{302449841}{20736} \zeta_3 \Nf
+ \frac{612462769}{2048}
+ \frac{676597463}{1536} \alpha^2
+ \frac{877406273}{3456} \Nf \alpha
\right. \nonumber \\
&& \left. ~~~
+ \frac{1067166537}{4096} \zeta_3 \alpha
- 3 \zeta_4 \Nf^3 \alpha
+ 15 \zeta_3^2 \Nf^2 \alpha^2
+ 63175 \zeta_6 \Nf
\right] a^5
\nonumber \\
&& +~ O(a^6) ~.
\end{eqnarray}
The five loop expressions for these as well as the other renormalization group
equations in an arbitrary Lie group are provided in the associated data file.


\begin{thebibliography}{99}
\bibitem{1} R.J. Crewther, Phys. Rev. Lett. {\bf 28} (1972), 1421.
\bibitem{2} S.G. Gorishnii, A.L. Kataev \& S.A. Larin, Phys. Lett. {\bf B259} 
(1991), 144. 
\bibitem{3} S.G. Gorishnii, A.L. Kataev \& S.A. Larin, Pisma Zh. Eksp. Teor.
Fiz. {\bf 53} (1991), 121. 
\bibitem{4} S.A. Larin \& J.A.M. Vermaseren, Phys. Lett. {\bf B259} (1991), 
345. 
\bibitem{5} D.J. Broadhurst \& A.L. Kataev, Phys. Lett. {\bf B315} (1993), 179.
\bibitem{6} G.T. Gabadadze \& A.L. Kataev, JETP Lett. {\bf 61} (1995), 448.
\bibitem{7} R.J. Crewther, Phys. Lett. {\bf B397} (1997), 137.
\bibitem{8} D. M\"{u}ller, Phys. Rev. {\bf D58} (1998), 054005.
\bibitem{9} P.A. Baikov, K.G. Chetyrkin \& J.H. K\"{u}hn, Phys. Rev. Lett. 
{\bf 104} (2010), 132004.
\bibitem{10} A.L. Kataev \& V.S. Molokoedov, Phys. Rev. {\bf D92} (2015),
054008.
\bibitem{11} A.L. Kataev \& V.S. Molokoedov, J. Phys. Conf. Ser. {\bf 938} 
(2017), 012050.
\bibitem{12} A.V. Garkusha, A.L. Kataev \& V.S. Molokoedov, JHEP {\bf 02} 
(2018), 161.
\bibitem{13} M. Peter, Phys. Rev. Lett. {\bf 78} (1997), 602.
\bibitem{14} Y. Schr\"{o}der, Phys. Lett. {\bf B447} (1999), 321.
\bibitem{15} L. von Smekal, K. Maltman \& A. Sternbeck, Phys. Lett. {\bf B681}
(2009), 336.
\bibitem{16} J.C. Taylor, Nucl. Phys. {\bf B33} (1971), 436.
\bibitem{17} J.A. Gracey, J. Phys. {\bf A46} (2013), 225403; J. Phys. {\bf A48}
(2015), 119501(E).
\bibitem{18} B. Ruijl, T. Ueda, J.A.M. Vermaseren \& A. Vogt, JHEP {\bf 06}
(2017), 040.
\bibitem{19} J.A. Gracey \& R.H. Mason, J. Phys. {\bf A56} (2023), 085401.
\bibitem{20} N.S. Craigie \& H. Dorn, Nucl. Phys. {\bf B185} (1981), 204.
\bibitem{21} N.S. Craigie, V.K. Dobrev \& I.T. Todorov, Annals Phys.
{\bf 159} (1985), 411.
\bibitem{22} T. Appelquist, K.D. Lane \& U. Mahanta, Phys. Rev. Lett. {\bf 61} 
(1988), 1553.
\bibitem{23} N.G. Stefanis, Nuovo Cim. {\bf A83} (1984), 205. 
\bibitem{24} N.G. Stefanis, Acta Phys. Polon. Supp. {\bf 6} (2013), 71.
\bibitem{25} S.V. Mikhailov, Phys. Lett. {\bf B431} (1998), 387. 
\bibitem{26} S.V. Mikhailov, Phys. Rev. {\bf D62} (2000), 034002. 
\bibitem{27} A.I. Alekseev, B.A. Arbuzov \& V.A. Baikov, Theor. Math. Phys.
{\bf 52} (1982), 739.
\bibitem{28} B.A. Arbuzov, E.E. Boos \& K.Sh. Turashvili, Z. Phys. {\bf C30}
(1986), 287.
\bibitem{29} B.A. Arbuzov, E.E. Boos \& A.I. Davydychev, Theor. Math. Phys.
{\bf 74} (1988), 103.
\bibitem{30} J.H. Field, Phys. Part. Nucl. {\bf 40} (2009), 353.
\bibitem{31} K.G. Chetyrkin, A.H. Hoang, J.H. K\"{u}hn, M. Steinhauser \& T.
Teubner, Phys. Lett. {\bf B384} (1996), 233.
\bibitem{32} L. Baulieu \& R. Coquereaux, Ann. Phys. {\bf 140} (1982), 163.
\bibitem{33} T.A. Ryttov, Phys. Rev. {\bf D89} (2014), 016013.
\bibitem{34} T. Banks \& A. Zaks, Nucl. Phys. {\bf B196} (1982), 189.
\bibitem{35} G. Martinelli, C. Pittori, C.T. Sachrajda, M. Testa \& A.
Vladikas, Nucl. Phys. {\bf B445} (1995), 81.
\bibitem{36} E. Franco \& V. Lubicz, Nucl. Phys. {\bf B531} (1998), 641.
\bibitem{37} W. Celmaster \& R.J. Gonsalves, Phys. Rev. Lett. {\bf 42} (1979),
1435.
\bibitem{38} W. Celmaster \& R.J. Gonsalves, Phys. Rev. {\bf D20} (1979),
1420.
\bibitem{39} J.-M. Shen, X.-G. Wu, Y. Ma \& S.J. Brodsky, Phys. Lett. 
{\bf B770} (2017), 494.
\bibitem{40} D.J. Gross \& F.J. Wilczek, Phys. Rev. Lett. {\bf 30} (1973),
1343.
\bibitem{41} H.D. Politzer, Phys. Rev. Lett. {\bf 30} (1973), 1346.
\bibitem{42} G. 't Hooft, Nucl. Phys. {\bf B61} (1973), 455.
\bibitem{43} O.V. Tarasov \& D.V. Shirkov, Sov. J. Nucl. Phys. {\bf 51} (1990),
877.
\bibitem{44} W.E. Caswell, Phys. Rev. Lett. {\bf 33} (1974), 244.
\bibitem{45} J.A. Gracey, R.H. Mason, T.A. Ryttov \& R.M. Simms, 
arXiv:2306.09056 [hep-ph].
\bibitem{46} K.G. Chetyrkin \& T. Seidensticker, Phys. Lett. {\bf B495} (2000),
74.
\bibitem{47} J.A. Gracey, Phys. Rev. {\bf D84} (2011), 085011.
\bibitem{48} A. Bednyakov \& A. Pikelner, Phys. Rev. {\bf D101} (2020), 
071502(R).
\bibitem{49} G. Curci \& R. Ferrari, Nuovo Cim. {\bf A32} (1976), 151.
\bibitem{50} G. 't Hooft, Nucl. Phys. {\bf B190} (1981), 455.
\bibitem{51} A.S. Kronfeld, G. Schierholz \& U.J. Wiese, Nucl. Phys. {\bf B293}
(1987), 461.
\bibitem{52} A.S. Kronfeld, M.L. Laursen, G. Schierholz \& U.J. Wiese, Phys.
Lett. {\bf B198} (1987), 516.
\bibitem{53} T. Ueda, B. Ruijl \& J.A.M. Vermaseren, PoS LL2016 (2016), 070.
\bibitem{54} T. Ueda, B. Ruijl \& J.A.M. Vermaseren, Comput. Phys. Commun.
{\bf 253} (2020), 107198.
\bibitem{55} J.A.M. Vermaseren, math-ph/0010025.
\bibitem{56} M. Tentyukov \& J.A.M. Vermaseren, Comput. Phys. Commun. {\bf 181}
(2010), 1419.
\bibitem{57} J.A. Gracey, JHEP {\bf 0504} (2005), 012.
\bibitem{58} J.M. Bell \& J.A. Gracey, Phys. Rev. {\bf D92} (2015), 125001.
\bibitem{59} K.G. Chetyrkin \& A. R\'{e}tey, Nucl. Phys. {\bf B583} (2000), 3.
\bibitem{60} K.G. Chetyrkin \& A. R\'{e}tey, hep-ph/0007088.
\bibitem{61} J.A. Gracey, Nucl. Phys. {\bf B662} (2003), 247.
\bibitem{62} J.A. Gracey, Eur. Phys. J. {\bf C83} (2023), 83.
\bibitem{63} A.L. Kataev \& V.S. Molokoedov, arXiv:2302.03443 [hep-ph].
\bibitem{64} A.L. Kataev \& V.S. Molokoedov, Phys. Rev. {\bf D92} (2015), 
054008.
\bibitem{65} B.E. Lautrup, {\it Of ghoulies and ghosties: an introduction to
QCD}, (1977), NBI-HE-76-14,
{\tt http://www.lautrup.nbi.dk/papers/ghoulies.pdf}.
\bibitem{66} P. Cvitanovic, {\it Field Theory}, NORDITA Lecture Notes (1983),
 RX-1012, \\
{\tt http://chaosbook.org/FieldTheory/}.
\bibitem{67} D. Kreimer, M. Sars \& W. van Suijlekom, Annals Phys. {\bf 336}
(2013), 180.
\bibitem{68} H. Ki\ss{}ler, Annals Phys. {\bf 372} (2016), 159.
\bibitem{69} H. Ki\ss{}ler, {\it Computational and diagrammatic techniques for 
perturbative Quantum Electrodynamics}, Ph.D. thesis (2017), \\
{\small {\tt http://www2.mathematik.hu-berlin.de/$\sim$kreimer/wp-content/uploads/KisslerDiss.pdf}}.
\end{thebibliography}
\end{document}